\documentclass[12pt]{iopart}

\usepackage{hyperref}

\usepackage{amssymb}
\expandafter\let\csname equation*\endcsname\relax
\expandafter\let\csname endequation*\endcsname\relax

\usepackage{bbold}
\usepackage{amsmath}

\usepackage{slashed}
\usepackage{nicefrac}

\usepackage{braket}

\usepackage{hyperref}
\usepackage[nosort]{cite}

\usepackage{etoolbox}

\makeatletter
\def\@mkboth#1#2{}
\newlength\appendixwidth
\preto\appendix{\addtocontents{toc}{\protect\patchl@section}}
\newcommand{\patchl@section}{%
  \settowidth{\appendixwidth}{\textbf{Appendix }}%
  \addtolength{\appendixwidth}{1.5em}%
  \patchcmd{\l@section}{1.5em}{\appendixwidth}{}{\ddt}%
}
\makeatother

\def\be{\begin{equation}}
\def\ee{\end{equation}}
\def\ba{\begin{eqnarray}}
\def\ea{\end{eqnarray}}
\newcommand{\bea}{\begin{eqnarray}}
\newcommand{\eea}{\end{eqnarray}}

\def\Li{\textrm{Li}}
\def\eps{\epsilon}

\newcommand{\fwboxL}[2]{\text{\makebox[#1][l]{$#2$}}}

\def\Disc{{\rm Disc}}
\def\Gcusp{\Gamma_{\rm cusp}}

\newcommand{\cA}{\begin{cal}A\end{cal}}

\newcommand{\cE}{\begin{cal}E\end{cal}}

\newcommand{\cH}{\begin{cal}H\end{cal}}

\newcommand{\cN}{\begin{cal}N\end{cal}}

\newcommand{\cR}{\begin{cal}R\end{cal}}

\newcommand{\cX}{\begin{cal}X\end{cal}}

\newcommand\hns[1]{\text{hns}[#1]}

\newcommand{\la}{\langle}
\newcommand{\ra}{\rangle}
\newcommand{\ab}[1]{\la #1 \ra}

\usepackage[makeroom]{cancel}

\usepackage{tikz}
\usepackage{pgf}
\usepackage{tcolorbox}
\usetikzlibrary{tikzmark}
\usetikzlibrary{snakes}

\usetikzlibrary{decorations.pathmorphing}
\usetikzlibrary{intersections}
  \usetikzlibrary{positioning}
\usetikzlibrary{shapes}
\usetikzlibrary{fit}
\usetikzlibrary{backgrounds}

\usetikzlibrary{decorations.pathreplacing,calc,arrows,decorations.markings}

\tikzset{
	vertex/.style={draw,shape=circle,fill=black,minimum size=0pt,inner sep=0pt},
	vertex2/.style={draw,shape=circle,fill=black,minimum size=4pt,inner sep=0pt},
	polygonEdge/.style={draw, very thick,line cap = round},
	blueNode/.style={circle,fill=blue!20,draw},
	factNode/.style={draw,shape=circle,fill=black,inner sep=2pt},
	grayOut/.style={fill=gray!20}
}

\begin{document}
\bibliographystyle{iopart-num}

%
\newcommand{\eprint}[2][]{\href{https://arxiv.org/abs/#2}{#2}}
%

\begin{flushright}
	SAGEX-22-06\\
	DESY-22-055
\end{flushright}

\title[Analytic Bootstraps for Scattering Amplitudes and Beyond]{The SAGEX Review on Scattering Amplitudes \\ Chapter 5: Analytic Bootstraps for Scattering Amplitudes and Beyond}

\author{Georgios Papathanasiou}

\address{Deutsches Elektronen-Synchrotron DESY, Notkestr. 85, 22607 Hamburg, Germany}
\ead{georgios.papathanasiou@desy.de}
\vspace{10pt}
\begin{indented}
\item[]March 2022
\end{indented}

\begin{abstract}
One of the main challenges in obtaining predictions for collider experiments from perturbative quantum field theory, is the direct evaluation of the Feynman integrals it gives rise to. In this chapter, we review an alternative bootstrap method that instead efficiently constructs physical quantities by exploiting their analytic structure. We present in detail the setting where this method has been originally developed, six- and seven-particle amplitudes in the large-color limit of $\mathcal{N}=4$ super Yang-Mills theory. We discuss the class of functions these amplitudes belong to, and the strong clues mathematical objects known as cluster algebras provide for rendering this function space both finite and of relatively small dimension at each loop order. We then describe how to construct this function space, as well as how to locate the amplitude inside of it with the help of kinematic limits, and apply the general procedure to a concrete example: The determination of the two-loop correction to the first nontrivial six-particle amplitude. We also provide an overview of other areas where the realm of the bootstrap paradigm is expanding, including other scattering amplitudes, form factors and Feynman integrals, and point out the analytic properties of potentially wider applicability that it has revealed.
\end{abstract}

%
%
%
%
%
\newpage
\tableofcontents
\newpage

\section{Introduction}

The idea that analytic properties could be exploited so as to circumvent difficulties in obtaining predictions from the traditional perturbative quantum field theory approach is not new~\cite{White:2000zs}. Yet it is only over the last decade that it has been applied so successfully in order to compute physical quantities for general values of their kinematic parameters, albeit mostly in the large-color limit of the simplest interacting four-dimensional gauge theory, known as $\mathcal{N}=4$ super Yang-Mills (SYM). 

In its modern reincarnation, this perturbative analytic bootstrap has been initiated by Dixon, Drummond and Henn in order to determine the three-loop correction to the essentially first nontrivial six-particle scattering amplitude of the theory~\cite{Dixon:2011pw}, following a remarkable earlier conjecture on the all-loop structure of amplitudes of any multiplicity~\cite{Bern:2005iz}, as well as the exploration of its consequences at strong coupling~\cite{Alday:2007hr,Alday:2009yn,Alday:2009dv,Alday:2010vh} via the gauge/string duality~\cite{Maldacena:1997re}. The cornerstone of the method is the construction of a finite space of functions expected to contain the physical quantity in question, from which the latter may then be uniquely determined from the knowledge of its behavior in special kinematic limits. Its upgrade to higher loops~\cite{Dixon:2011pw,Dixon:2011nj,Dixon:2013eka,Dixon:2014voa,Dixon:2014iba,Dixon:2015iva,Caron-Huot:2016owq,Caron-Huot:2019vjl} and to multiplicity seven~\cite{Drummond:2014ffa,Dixon:2016nkn,Drummond:2018caf,Dixon:2020cnr} has been achieved in parallel with the discovery of new analytic properties~\cite{Drummond:2017ssj,Caron-Huot:2019bsq} that prune the initial space of functions, thereby making the method more efficient.

Perhaps more importantly, the discovered analytic properties have potential for applicability to more general gauge theories~\cite{Abreu:2020jxa,Chicherin:2020umh}. And as a proof of concept that the bootstrap may also work for phenomenologically relevant physical quantities, it has been used to rederive the three-loop soft anomalous dimension in quantum chromodynamics (QCD), up to an overall numerical factor~\cite{Almelid:2017qju}. Further areas expanding the realm of the bootstrap paradigm include amplitudes in simpler kinematics~\cite{Dixon:2012yy,Chestnov:2020ifg,He:2021fwf} and three-particle form factors in $\cN=4$ SYM\cite{Brandhuber:2012vm,Dixon:2020bbt}, as well as individual Feynman integrals~\cite{Dixon:2011nj,Drummond:2017ssj,Chicherin:2017dob,Caron-Huot:2018dsv,Henn:2018cdp,He:2021non,He:2021eec}, or basis sets thereof~\cite{Heller:2019gkq,Heller:2021gun}.

Given that there already exists a six-particle bootstrap review~\cite{Dixon:2014xca}, as well as a more recent amplitude bootstrap review~\cite{Caron-Huot:2020bkp} based on a talk by the author, our goal here will be two-fold: On the one hand, to cover some of the key concepts behind the amplitude bootstrap in more detail, and to show how it works in a concrete example; and on the other hand, to give an overview of the new frontiers of its application. In section~\ref{sec:SymKin} we briefly introduce the $\mathcal{N}=4$ SYM amplitudes of interest, and in section~\ref{sec:ClusterMPLs} we discuss the class of polylogarithmic functions they belong to. We also describe how mathematical objects known as cluster algebras provide strong clues for the singularities of these amplitudes at multiplicities $n=6,7$, as well as for how these singularities can appear consecutively. This is the information that renders the function space finite and of relatively small size, and we explain how to construct it and how to locate the amplitude inside it in section~\ref{sec:HexHepFuns}, also working out all the steps explicitly for the two-loop six-particle case in the simplest helicity configuration. We close with an overview of the new frontiers in section~\ref{sec:NewFrontiers}.

\section{Planar $\mathcal{N}=4$ SYM Amplitudes in a Nutshell}\label{sec:SymKin}

\subsection{Planar limit, color-ordering and discrete symmetries}
For the most part we will be focusing on $\cN=4$ SYM theory~\cite{Brink:1976bc,Gliozzi:1976qd}, the gauge theory containing the maximal amount of supersymmetry, and whose particle content apart from gluons also contains their scalar and fermionic partners, all of which are in the adjoint representation. We will also be restricting to the origin in the moduli space of the theory, where the scalar fields have zero vacuum expectation value, and all particles are massless. Many of the properties of the theory and of its amplitudes are discussed in chapter 1~\cite{Brandhuber:2022qbk} of the SAGEX review~\cite{Travaglini:2022uwo}, so here we will briefly recall some of their main features that we will need later on. Unless otherwise noted we will also be considering 't Hooft's planar limit~\cite{tHooft:1973alw}, where the number of colors $N\to \infty$ while its product $g_{YM}^{2}N$ with the gauge theory coupling  is held fixed. The latter is the only parameter of the theory that survives in the limit, and in our normalization conventions we will be denoting the perturbative expansion of any quantity $F$ with respect to it as
\be\label{eq:gLoopExpansion}
F=\sum_{L=0}^\infty g^{2L} F^{(L)}\,,\quad g^2 = \frac{{g_{YM}^{2}N}}{16 \pi^{2}}\,.
\ee

A major simplification that occurs in the planar limit, is that only a single color structure $\Tr(T^{a_1}\ldots T^{a_n})$ of the gauge group generators $T^{a_i}$ is the leading term in the $1/N$ expansion of the $n$-particle amplitude, as is reviewed in \cite{Brandhuber:2022qbk} or~\cite{Dixon:2011xs}. We may thus restrict our attention to the coefficient of this leading color structure, the \emph{color-ordered amplitude} $A_n$. Due to its relation to the trace, it is evident that it it is invariant under cyclic shifts $i\to i+1$ of the particle labels. What is less obvious, is that $A_n$ is also invariant (up to an overall sign for odd $n$) under reflections $i\to n+1-i$, where the equivalence $n+i\sim i$ is understood. At tree level, this holds both for gluons and quarks in any gauge theory, and is a consequence of the (anti-)symmetry of the (three-) four-point vertex of the color-ordered Feynman rules, see e.g.~\cite{Srednicki:1019751}. As a consequence of supersymmetry, in $\cN=4$ SYM this property also persists at loop level and for all types of external states, after grouping them in a single superfield~\cite{Nair:1988bq}, and similarly combining all component amplitudes in a single superamplitude~\cite{Elvang:2009wd}. Together cyclic permutations and reflections form the discrete \emph{dihedral group}, which is thus a symmetry group of $A_n$.

\subsection{Helicity dependence}
Having factored out the color degrees of freedom from the amplitude in the planar limit, the remaining physical quantities it will depend on are the momenta and helicities of the external particles. We recall that \emph{helicity}, namely the projection of a spin in the direction of momentum is a good quantum number for massless particles, as it cannot be altered by Lorentz transformations. With the help of spinor-helicity variables, which as the name suggests are friendly to the aforementioned quantum number, and which by now have also made their way into quantum field theory textbooks~\cite{Srednicki:1019751,Zee2010}, it is straightforward to show that all gluon amplitudes where all or all but one helicities are positive (for example in conventions where all momenta are outgoing) vanish at tree level. While this vanishing is lifted at loop level in generic gauge theories (see~\cite{Gehrmann:2015bfy} for a relatively recent example in QCD), it does persist in supersymmetric theories such as $\cN=4$ SYM, since it is a consequence of supersymmetric Ward identities~\cite{Grisaru:1977px,Elvang:2009wd}.

Hence the first nontrivial helicity configuration at any loop order corresponds to the so-called \emph{Maximally Helicity Violating} (MHV) amplitudes, with all but two gluons having positive helicity, and similarly amplitudes with all but $(k+2)$ gluons of positive helicity are denoted as N$^k$MHV. As reviewed in e.g.~\cite{Drummond:2010km}, different distributions of the negative helicity states are also simultaneously accounted for in the superamplitude,  which has a natural grading with respect to total helicity, following from  the fact that the entire aforementioned superfield it depends on has well-defined helicity. Also note that helicity degrees $k$ and $\bar k=n-4-k$ are related by a parity or spatial reflection tranformation, which in spinor-helicity variables simply corresponds to complex conjugation of spinors. As a result, we may always restrict $k\le  \lfloor \frac{n-4}{2}{\rfloor}$, where $\lfloor x {\rfloor}$ denotes the integer part of $x$. Summarizing what we have discussed so far, the nontrivial physical quantity encoding scattering in planar $\cN=4$ SYM is the color-ordered $n$-particle, helicity degree-$k$ superamplitude
\be
A_{n,k}(p_1,\ldots p_n)\,,\qquad n\ge4,\,\,\, 0\le k\le  \lfloor \frac{n-4}{2}{\rfloor}\,,
\ee
where $p_i$ denotes the momentum of the $i$-th particle. In other words only $A_{n,0}$ or the MHV superamplitude is needed for $n=4,5$, additionally $A_{n,1}$ or the NMHV superamplitude is needed for $n=6,7$, and so on. Evidently, amplitudes obeying $k=(n-4)/2$, such as $A_{4,0}$ or $A_{6,1}$, will be invariant under parity.

\subsection{The kinematic space: Dual conformal invariance and momentum twistors}\label{sec:DCIKin}
With not only color but also helicity dependence specified as above, we now move on to describe what is the space of kinematics of the amplitude. Remarkably, planar $\cN=4$ SYM possesses \emph{dual conformal symmetry}~\cite{Drummond:2006rz,Alday:2007hr,Drummond:2007aua,Drummond:2007au}, which acts on dual position variables $x_i$ related to the usual momenta by
\be\label{eq:pTox}
p_i \equiv x_{i+1} - x_i\,.
\end{equation}
This symmetry, which is also reviewed in chapter 1 of this review~\cite{Brandhuber:2022qbk}, and combines with the usual conformal symmetry of the theory so as to form an infinite-dimensional Yangian symmetry~\cite{Drummond:2009fd}, implies that the amplitude (in the appropriate normalization, which we will describe at the end of this section) depends on $3n-15$ instead of $3n-10$ kinematic variables\footnote{This counting corresponds to the number of independent components of the lightlike-separated points $x_i$, minus the dimension of the 4D conformal group or Poincar\'e group, for planar $\cN=4$ SYM or a generic massless gauge theory, respectively.} Said differently, instead of an algebraically independent subset of the $n(n-3)/2$ distinct Mandelstam invariants 
\begin{equation}\label{eq:Mandelstamtox}
s_{i,\ldots,j-1}\equiv (p_{i} + p_{i+1} + \ldots + p_{j-1})^2 = (x_i - x_j)^2\equiv x^2_{ij}\,,\qquad j\ge i+1 \mod n\,,
\end{equation}
one must instead pick an algebraically independent subset of their distinct 
$n(n-5)/2$ \emph{conformal cross ratios}
\begin{equation}\label{u_def}
u_{ij}\equiv\frac{x_{ij+1}^2 x_{i+1j}^2}{x_{ij}^2 x_{i+1j+1}^2}\,,\qquad j\ge i+2 \mod n\,.
\end{equation}
The fact that dual conformal invariance reduces the number of independent kinematic variables of the amplitude to $3n-15$ has the following important implication: \emph{Only normalized $n$-particle amplitudes with $n\ge 6$ have nontrivial kinematic dependence}, and will therefore be the focus of this article. For the simplest cases with $n=6,7$ the abbreviated notation
\be\label{eq:udef}
u_i\equiv {u_{i+8-n,i+11-n}}
\ee
for the cross ratios~\eqref{u_def} is also used. 

The algebraic relations among different Mandelstam invariants (cross ratios) are known as (conformal) Gram determinant constraints, and simply encode the fact that the number of independent vectors is bounded in a given spacetime dimension. In the presence of conformal symmetry, these constraints were worked out in~\cite{Eden:2012tu}, and while their solution yields an independent subset of cross ratios that can be used to parametrize the kinematics, in practice this parametrization turns out to be quite complicated.

Instead, massless, planar, dual conformal invariant kinematics may be most conveniently described in terms of \emph{momentum twistors} \cite{Hodges:2009hk}, which are also very nicely reviewed in e.g. \cite{ArkaniHamed:2010gh,Bullimore:2013jma}. Very briefly, one way for obtaining these variables is by representing $x^\mu \in \mathbb{R}^{1,3}$ as a projective null vector  $X^M\in \mathbb{R}^{2,4}$, $X^2=0$, $X\sim \lambda X$. This $SO(2,4)$ vector $X^M$ is also equivalent to an antisymmetric representation $X^{IJ}$ of $SU(2,2)$, since the two algebras are isomorphic (in practice one representation can be converted into the other by six-dimensional analogues of the Pauli matrices, which may similarly be used in order to transform Lorentz vectors to $2 \times 2$ antisymmetric matrices). The antisymmetric representation can in turn be built out of two copies of the fundamental representation $Z^{I}$ of $SU(2,2)$, or, after complexifying, $SL(4,\mathbb{C})$. Momentum twistors precisely correspond to these $Z$'s, and we see that our original point $x^\mu \in \mathbb{R}^{1,3}$ is mapped to a pair of points, i.e. a line in momentum twistor space. As is the case for the vector $X$ they originate from, momentum twistors are also defined up to rescalings $Z\sim t Z$, thus they may be equivalently viewed as homogeneous coordinates on complex projective space $\mathbb{P}^3$. It is then possible to show that the usual Mandelstam invariants \eqref{eq:Mandelstamtox} can be expressed in terms of momentum twistors as
\begin{equation}\label{xToZ}
x^2_{ij} \propto \langle i-1 i j-1 j \rangle\,,
\end{equation}
up to proportionality factors that drop out from conformally invariant quantities, where
\be\label{fourbrak}
\ab{ijkl} \equiv\langle Z_{i} Z_j Z_{k} Z_l \rangle=\det(Z_i Z_j Z_k Z_l)
\ee
is a \emph{four-bracket} of momentum twistors. 

The advantage of momentum twistor variables is that they automatically satisfy both momentum conservation and the constraint on external lightlike momenta, $p_i^2=x_{ii+1}^2=0$: Indeed, from eq.~\eqref{eq:pTox} it is evident that the former already holds for the dual space coordinates giving rise to them, and the latter follows from eqs.~\eqref{xToZ}-\eqref{fourbrak}. Furthermore, conformal transformations of the dual space coordinates $x$ map to $SO(2,4)$ rotations of $X$, and in turn to $SL(4,\mathbb{C})$ transformations of the momentum twistors. Therefore the space of dual conformal invariant kinematics can be written as a $4\times n$ matrix, whose columns are the cyclically ordered momentum twistors/homogeneous $\mathbb{CP}^3$ coordinates defined up to rescalings, and modulo $SL(4,\mathbb{C})$ transformations. Fixing this gauge redundancy, so as to obtain explicit parametrizations of the kinematics in terms of $3n-15$ independent variables is then very straightforward; one such example are the web variables, that may be algorithmically constructed for any $n$~\cite{Speyer2005}, see also~\cite{Drummond:2019cxm,Henke:2021ity} for a simplified reformulation.

Let's see how these types of kinematic parametrizations may be used in practice in the $n=6$ case. The matrix of momentum twistors in terms of web variables is
\be\label{eq:Zweb6}
(Z_1,\ldots,Z_6)=\left(
\begin{array}{cccccc}
 1 & 0 & 0 & 0 & -1 & -1-x_1-x_1 x_2-x_1 x_2 x_3 \\
 0 & 1 & 0 & 0 & 1 & 1+x_1+x_1 x_2 \\
 0 & 0 & 1 & 0 & -1 & -1-x_1 \\
 0 & 0 & 0 & 1 & 1 & 1 \\
\end{array}
\right)\,,
\ee
from which we can compute any four-bracket by choosing the corresponding $4 \times 4$ minor according to eq.~\eqref{fourbrak}, such that e.g.
\be\label{eq:ab1346}
\ab{1346}=1+x_1+x_1 x_2\,,
\ee
and similarly evaluate any other kinematic variable that depends on them. For example due to eq.\eqref{xToZ}, the cross ratios~\eqref{eq:udef} become
\be
u_1=\frac{x_2 x_3}{\left(1+x_1+x_1 x_2\right) \left(1+x_2+x_2 x_3\right)}\,,\quad u_2=\frac{x_1 x_2}{1+x_1+x_1 x_2}\,,\quad u_3=\frac{1}{1+x_2+x_2 x_3}\,.\nonumber
\ee

It is worth noting that the space of kinematics in momentum twistor variables is also very closely related to the \emph{Gra\ss mannian}  $Gr(m,n)$, defined as the space of $m$-dimensional planes going through the origin in $n$-dimensional space: From this definition, it follows that $Gr(m,n)$ may also be realized as an $m\times n$ matrix, this time modulo $GL(m)$ transformations. As was first noted in~\cite{Golden:2013xva}, based on the previously discovered relevance of the Gra\ss mannian for the amplitude integrand~\cite{Arkani-Hamed:2012zlh}, the comparison of their matrix realizations reveals that the space of external momentum kinematics is in fact equivalent to the quotient $Gr(4,n)/(\mathbb{C}^{*})^{n-1}$.

\subsection{Amplitude normalizations}
Let us finally come to address the question of the normalization of the amplitude. In any massless gauge theory, loop amplitudes have infrared divergences, arising from integration regions where the loop momenta become soft or collinear. It can be shown that these divergences exponentiate quite universally~\cite{Sterman:2002qn}, and that particularly in  planar $\cN=4$ SYM they are captured to all loops by the Bern-Dixon-Smirnov (BDS) ansatz \cite{Bern:2005iz}. The latter is essentially the exponential of the one-loop amplitude times the cusp anomalous dimension $\Gcusp$,
\be
\frac14 \Gcusp(g^2)\ =\ g^2 - 2\,\zeta_2\,g^4 + 22\,\zeta_4\,g^6
- \Bigl[ 219\, \zeta_6 + 8 \, (\zeta_3)^2 \Bigr] \, g^8 + \cdots,
\label{Gcusp}
\ee
also known to all loops~\cite{Beisert:2006ez} thanks to the integrability of the theory, as reviewed for example in~\cite{Freyhult:2010kc}\footnote{The integrability of a deformation of $\cN=4$ SYM theory is also described in chapter 9\cite{Chicherin:2022nqq} of the SAGEX review~\cite{Travaglini:2022uwo}.}. So the BDS ansatz by construction satisfies the exponentiation property of infrared divergences, but also includes additional one-loop contributions. 


From the above discussion, it therefore follows that it is possible to obtain an infrared-finite normalized amplitude by dividing out $A_{n,k}$ by the BDS ansatz. It should also make apparent, however, that this normalization is not unique: Similarly to the difference between renormalization schemes in any gauge theory, there is still freedom in the finite, dual conformal invariant terms that the infrared-divergent factor may be chosen to absorb. While this choice ultimately leads to just equivalent representations of the amplitude, it proves advantageous to \emph{tune} it such that the normalized amplitude inherits certain important physical properties of $A_{n,k}$, and hence becomes simpler to compute. Indeed, when $n$ is not a multiple of 4, choosing to factor out the closely related BDS-like ansatz, which naturally appears in the strong-coupling description of the amplitude~\cite{Alday:2009dv}, ensures that the normalized amplitude respects the Steinmann relations, whose significance will be discussed in subsection~\ref{sec:SteinClus}.

In what follows, whenever possible we will will thus focus on BDS-like normalized amplitude, denoted as
\be\label{eq:cE_def}
\cE_{n,k}\equiv\frac{A_{n,k}}{\,A^{(0)}_{n,0}\,A^{\text{BDS-like}}_{n}}\,,\quad n\,\text{mod}\, 4\ne 0\,,
\ee
where for convenience we have also additionally divided by the tree-level MHV superamplitude, $A^{(0)}_{n,0}$. Since the precise form of the BDS and BDS-like ans\"atze will not be important for our purposes, we will refrain from quoting them here, and refer the interested reader to the original references, or e.g. \cite{Alday:2008yw,Yang:2010as}. The ratio of the two ans\"atze is however closely related to the one-loop correction to $\cE_{n,0}$\,,
\be
\frac{A_n^{\text{BDS}}}{A_n^{\text{BDS-like}}}=\exp\left[\frac{\Gamma_{\text{cusp}}}{4} \cE^{(1)}_{n,0}\right],
\label{BDSlikeBDSn}
\ee
where for $n=6,7$ we explicitly have,
\bea
\cE^{(1)}_{6,0} &=& \, \sum_{i=1}^3  \Li_2\left(1-\frac{1}{u_i}\right)\,,\label{E61def}\\
\cE^{(1)}_{7,0} &=& \sum_{i=1}^7 \biggl[ \Li_2\left(1-\frac{1}{u_i}\right)
  + \frac{1}{2} \log \left(\frac{u_{i+2}u_{i{-}2}}{u_{i+3}u_{i}u_{i{-}3}}\right)
               \log u_i \biggr]\,.
\label{E71def}
\eea
Converting between BDS and BDS-like normalizations also follows immediately from eq.~\eqref{BDSlikeBDSn}. For example, the BDS-normalized MHV amplitude, which in the original literature was expressed in terms of an exponentiated \emph{remainder function} $\cR_n$, is related to $\cE_{n,0}$ by
\be\label{eq:R6}
e^{\cR_n}=e^{-\frac{\Gamma_{\text{cusp}}}{4} \cE^{(1)}_{n,0}} \cE_{n,0}\,.
\ee

To recapitulate the main lesson of this section, infrared-normalized, color-ordered superamplitudes in the planar limit of $\cN=4$ SYM only depend on the particle number $n$, the helicity degree $k$, $3n-15$ variables in the space of dual conformally invariant kinematics, and the order $L$ of loops or perturbative corrections.

\section{Cluster Polylogarithmic Functions}\label{sec:ClusterMPLs}

Having reviewed the parameters that normalized amplitudes $\cE^{(L)}_{n,k}$ in planar $\cN=4$ SYM theory depend on, here we will continue to describe the type of functions they evaluate to. As we will recall in subsection~\eqref{sec:MPLs}, the latter fall in the general class of multiple polylogarithms, which are also relevant for  a wide range of Standard Model processes at the forefront of precision phenomenology, especially when mediated by internal particles that can be considered as massless, see for example~\cite{Gehrmann:2015bfy,Papadopoulos:2015jft,Abreu:2020jxa,Canko:2020ylt,Abreu:2021smk}. Within this class, however, there still exists an infinite number of these functions at each loop order, depending on where they are allowed to have singularities. In subsection~\eqref{sec:AlphabetCluster} we will then see that beautiful mathematical objects known as cluster algebras appear to correctly predict these singularities at multiplicity $n=6,7$, thereby making the function spaces expected to contain $\cE^{(L)}_{n,k}$ finite. Furthermore, in subsection~\eqref{sec:SteinClus} we will see that cluster algebras also dictate how these singularities are allowed to appear consecutively, and how these additional restrictions can be physically interpreted as the (extended) Steinmann relations of axiomatic quantum field theory. The finite function spaces that are further reduced by the latter restrictions will be the starting point for bootstrapping the corresponding amplitudes in the most efficient manner, to be discussed in the next section.

\subsection{Multiple polylogarithms and symbols}\label{sec:MPLs}

All explicit calculations to date, as well as an analysis at the level of the integrand~\cite{ArkaniHamed:2012nw} (note however the subtleties pointed out in~\cite{Brown:2020rda}), suggest that at least for $k=0,1$, $\cE_{n,k}^{(L)}$ can be expressed in terms of generalized or Goncharov or \emph{multiple polylogarithms} (MPL) \cite{Chen,G91b,Goncharov:1998kja} of \emph{weight} $m=2L$. These functions are also mentioned in  chapters 3~\cite{Abreu:2022mfk} and 4~\cite{Blumlein:2022qci} of the SAGEX review\cite{Travaglini:2022uwo}, and in addition they are discussed in great detail in the recent textbook~\cite{Weinzierl:2022eaz}. Let us briefly collect here the definitions and properties that will be useful for our purposes.

A function $F_m$ is defined as an MPL
of weight $m$ if its total differential obeys
\be 
dF_m = \sum_{\phi_\beta\in{\Phi}} F_{m-1}^{\phi_\beta}\,\, d\log \phi_\beta\,,
\label{dFPhi}
\ee
such that $F^{\phi_\alpha}_{m-1}$ is an MPL of weight $m-1$,
\be
dF_{m-1}^{\phi_\beta} = \sum_{\phi_\alpha\in{\Phi}} F_{m-2}^{\phi_\alpha,\phi_\beta} d\log \phi_\alpha\label{dFPhiCop}\,,
\ee
and so on, with this recursive definition \eqref{dFPhi} terminating at $m=1$ with the usual logarithms on the left-hand side, and rational numbers as coefficients of the total differentials on the right-hand side. The arguments of the dlogs $\phi_{\alpha_i}$ are algebraic functions of the independent variables of $F_m$ known as the (symbol) \emph{letters}, and similarly their collection $\Phi$ from all steps of the recursion is called the (symbol) \emph{alphabet}. Evidently, it encodes the positions of the possible branch point singularities of $F_m$, which may appear when $\phi_{\alpha_i}=0,\infty$.

The iterative structure we have described forms part of the \emph{coaction} operation $\Delta$ \cite{Gonch3,Gonch2,Brown:2011ik,Duhr:2012fh} (also loosely
referred to as the \emph{coproduct}), which `decomposes' an MPL
of weight $m$ into linear combinations of pairs of MPLs with weight $\{m-m_1,m_1\}$
for $m_1=0,1,\ldots m$. Concretely, the total differential \eqref{dFPhi} is essentially equivalent to the $\{m-1,1\}$ component of $\Delta$,
\be\label{eq:Deltan1}
\Delta_{m-1,1} F_m = \sum_{\phi_{\beta} \in \Phi} F_{m-1}^{\phi_\beta}\otimes
\big[\log\phi_{\beta}\mod\,(i\pi)\big]\,.
\ee
The coaction may be repeatedly applied to either the first or the second factor of the $\{m-m_1,m_1\}$ pair when $m_1>1$, yielding a decomposition of an MPL of weight $m$ into subspaces of MPLs with weight $\{m_1,\ldots,m_r\}$, $\sum_{i=1}^r m_i=m$, that is unique thanks to the coassociativity property of the coaction. Denoting the projection of the coaction on each of these subspaces by $\Delta_{m_1,\ldots,m_r}$, for example the equations \eqref{dFPhi} and \eqref{dFPhiCop} combine to yield the $\{m-2,1,1\}$ coproduct,
\begin{align}
\Delta_{m-2,1,1}F_m &= \sum_{\phi_\alpha,\phi_\beta\in{\Phi}} F_{m-2}^{\phi_\alpha,\phi_\beta} \otimes \log \phi_\alpha \otimes \log \phi_\beta\,,\label{Delta11}
\end{align}
where from eq.~\eqref{Delta11} onwards, identification of $\log \phi$ factors up to $i\pi$ will be implied. Furthermore, the maximal iteration of the coaction defines the  \emph{symbol} \cite{2009arXiv0908.2238G,Goncharov:2010jf},
\be\label{symbol}
S[F_m]=\Delta_{\fwboxL{27pt}{{\underbrace{1,\ldots,1}_{m\,\,\text{times}}}}}F_m=\sum_{\phi_{\alpha_1},\ldots,\phi_{\alpha_m}} F_0^{\phi_{\alpha_1},\ldots,\phi_{\alpha_n}}\,  \left[\log {\phi_{\alpha_1}} \otimes\cdots \otimes\log \phi_{\alpha_m}\right]\,,
\ee
where it is also customary to adopt a more compact notation by replacing $\log \phi_{\alpha_i}\to \phi_{\alpha_i}$. To make the above general definitions more tangible, we will also apply them to a concrete example towards the end of this subsection.

Comparing  \eqref{dFPhi} and \eqref{eq:Deltan1}, we see that derivatives only act on the rightmost factor of the coaction, and the same carries over to the symbol. Similarly, the discontinuities of MPLs can be shown to be encoded in the leftmost factor of their coaction.  For example, at the level of the symbol the discontinuity of $F_m$ when going around a potential branch point $\phi_\beta=0$ with no other letter  vanishing simultaneously is given by
\be\label{symbolDisc}
S[\text{Disc}_{\phi_\beta} (F_m)]=2\pi i \sum_{\phi_{\alpha_1},\ldots,\phi_{\alpha_m}} F_0^{\phi_{\alpha_1},\ldots,\phi_{\alpha_n}}\,  \delta_{\alpha_1\beta }\left[ {\phi_{\alpha_2}} \otimes\cdots \otimes \phi_{\alpha_m}\right]\,,
\ee
in other words it is equivalent to clipping off the first entry.

We now move on to present some further definitions that we will only need when we perform the sample bootstrap computation of $\cE_{6,0}^{(2)}$ in subsection~\ref{sec:R62}. The reader only interested in conceptual aspects may thus choose to skip to the next subsection. Alternatively to the differential definition of MPLs presented above, one may also reverse the direction and define them as iterated integrals. Choosing the integration contour in the simplest possible manner, leads to the ($G$-function) definition
\be\label{eq:Gdef}
G(a_1,\ldots,a_m; z) = \int_0^z\frac{dt}{t-a_1}G(a_2,\ldots,a_m;t)\,,
\ee
where the recursion starts with $G(;z)=1$, and for the special case where all the $a_i$ are zero, we define
\be\label{eq:Gzeros}
G(\underbrace{0,\ldots,0}_{m\textrm{ times}};z)= \frac{1}{m!}\log^mz\,.
\ee
Indeed, by differentiating eq.~\eqref{eq:Gdef}, applying the identity
\be
\frac{\partial}{\partial a}\frac{1}{t-a}=-\frac{\partial}{\partial t}\frac{1}{t-a}
\ee
and partial fractioning, it can be shown that $dG$ takes the general form of eq.~\eqref{dFPhi}. 

In the integral definition~\eqref{eq:Gdef}, we see that the weight $m$ corresponds to the number of iterated integrations. The single-variable case with $a_i \in\{-1,0,1\}$ has also independently appeared in the physics literature under the name of \emph{harmonic polylogarithms} (HPL)~\cite{Remiddi:1999ew}, up to the different sign convention,
\be
H(a_1,\ldots,a_m; z)=(-1)^p \,G(a_1,\ldots,a_m; z)\,,\quad a_i \in\{-1,0,1\}\,,
\ee
where $p$ counts how many $a_i$ are equal to +1. Since HPLs only depend on the outermost integration bound $z$ in eq.~\eqref{eq:Gdef}, their differentiation is trivial, and it is easy to show that the definitions~\eqref{dFPhi}-\eqref{symbol} specialize to
\begin{align}
\Delta_{m-1,1}H(a_1,\ldots,a_m; z)&= (-1)^{\text{sgn}({a_1})} H(a_2,\ldots,a_m; z)\otimes (z-a_1)\,,\label{eq:DeltaHPL}\\
\Delta_{m-2,1,1}H(a_1,\ldots,a_m; z)&=(-1)^{\text{sgn}({a_1})+\text{sgn}({a_2})} H(a_3,\ldots,a_m; z)\otimes (z-a_2)\otimes (z-a_1)\nonumber\,,\\
\vdots\nonumber\\
S\left[H(a_1,\ldots,a_m; z)\right]&=(-1)^p\left[(z-a_m)\otimes\ldots\otimes (z-a_1)\right]\,.
\end{align}
For later convenience let us also note that a more compact notation for all $a_i$ arguments can be adopted, whereby a string of subsequent zeros is replaced by
\be\label{HPL_a_to_m_notation}
\underbrace{0,0,\ldots0}_{m-1 \textrm{ times}},\pm1\to\pm m\,,
\ee
and the resulting, shorter string of arguments is placed as indices of the function. For example, classical logarithms, that are contained in HPLs, in this notation correspond to
\be
\text{Li}_m(z)=H_m(z)=H(\underbrace{0,\ldots,0,1}_{m-1\textrm{ times}};z)
\ee
and again for completeness their symbol will be (recalling that $d\log (z-1)=d\log (1-z)$)
\be
S\left[\text{Li}_m(z)\right]=-\big[(1-z)\otimes \underbrace{z\otimes\ldots\otimes z}_{m-1\textrm{ times}}\big]\,.
\ee
Finally, let us mention that products of $G$-functions with the same rightmost argument, and thus also HPLs, may be reexpressed as linear combinations thereof. Namely they form a (shuffle) algebra, as can be simply inferred from the definition~\eqref{eq:Gdef}, by appropriately splitting the integration range so that all dummy variables have a specific order.

In practice, by now there exist a variety of software tools that allow the evaluation of the functions presented in this subsection, as well as the application of their transformation properties, either in free computer algebra systems such as \texttt{GiNaC}~\cite{Bauer20021}, or in proprietary ones such as \texttt{Mathematica} and \texttt{Maple}, by SAGEX industry partners Wolfram Research and Maplesoft, respectively. These for example include the \texttt{Mathematica} paclages \texttt{HPL}~\cite{Maitre:2007kp} and \texttt{PolyLogTools}~\cite{Duhr:2019tlz}, as well as the native functionality of \texttt{GiNaC} (\texttt{Maple}~\cite{Frellesvig:2018lmm}) with respect to the numerical evaluation (and symbolic manipulation) of these functions.

\subsection{Cluster algebras and amplitude singularities}\label{sec:AlphabetCluster}

Once we have identified MPLs as the general class of functions that contain the amplitude, the next step is to clarify what the corresponding symbol alphabet is. For $n=6$, this came as a result of an explicit Feynman diagram computation~\cite{DelDuca:2009au,DelDuca:2010zg} at two loops, and was further supported by the analysis of closely related integrals~\cite{DelDuca:2011ne,Dixon:2011ng}. For general $n$, strong motivation is provided by the cluster algebra structure~\cite{Golden:2013xva} of the space of kinematics.

More precisely, in section \ref{sec:SymKin} we mentioned that the space of kinematics in terms of momentum twistors can be realized  as the quotient $Gr(4,n)/(\mathbb{C}^{*})^{n-1}$ of a Gra\ss mannian. It is the latter space that is naturally endowed with a cluster algebra structure~\cite{2003math.....11148S}, thus making it sensible to explore any implications this may have on the symbol alphabet. Before we spell that out, we will begin with a brief introduction on cluster algebras~\cite{1021.16017,1054.17024,CAIII,CAIV}, which have become a very active research area in contemporary mathematics since their inception in early 2000's. It is also worth noting that they have already found  applications in other areas of mathematical physics in the past, such as the proof of periodicicy of $Y$-systems and associated thermodynamic Bethe ans\"atze of certain integrable models \cite{Fomin:2001rc}, or the determination of BPS state spectra in supersymmetric field theories~\cite{Gaiotto:2010be,Alim:2011ae,Alim:2011kw}. In the realm of scattering amplitudes, their role was first appreciated at the level of the integrand, which also exhibits Gra\ss mannian structure~\cite{Arkani-Hamed:2012zlh}. For more recent work relating cluster algebras and tree-level amplitudes or loop integrands, see also~\cite{Arkani-Hamed:2017mur,Arkani-Hamed:2019vag}.

\subsubsection{Basics of cluster algebras.} With many excellent introductory articles on cluster algebras available in the literature,  as well as articles with detailed review sections on their connection to scattering amplitudes \cite{Vergu:2015svm,Drummond:2018dfd}, here will simply aim to highlight some of their features while mostly following one concrete example that will be relevant later on. 

The building blocks of cluster algebras are certain variables $a_i$, known as (Fomin-Zelevinsky) cluster $\cA$-coordinates, that are grouped into overlapping subsets $\{a_1,\ldots,a_d\}$ of \emph{rank} $d$,  the \emph{clusters}. Starting from an initial cluster, cluster algebras are constructively defined by a \emph{mutation} operation on the $\cA$-coordinates. They can also be generalized so as to contain \emph{frozen variables} or \emph{coefficients} $\{a_{d+1},\ldots,a_{d+m}\}$, whose main difference from the $\cA$-coordinates is that they do not mutate.

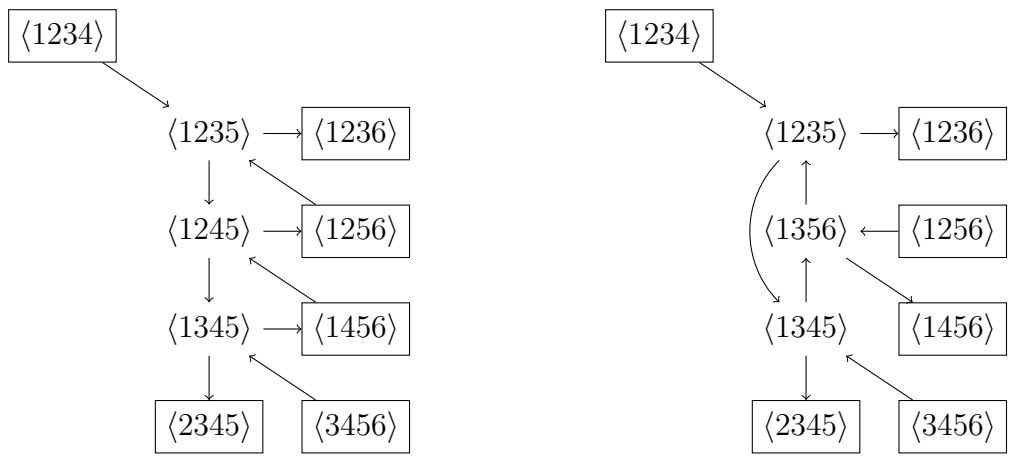
\begin{figure}
	\begin{minipage}{0.5\textwidth}
	\begin{center}
		\begin{tikzpicture}[scale=1.3]
		\node at (-1.5,4) [rectangle,draw] (a) {$\ab{1234}$};
		\node at (0,3) (b) {$\ab{1235}$};
		\node at (0,2) (c) {$\ab{1245}$};
		\node at (0,1) (b31) {$\ab{1345}$};
		\node at (0,0) [rectangle,draw] (b41) {$\ab{2345}$};
		\node at (1.5,3) [rectangle,draw] (b12) {$\ab{1236}$};
		\node at (1.5,2) [rectangle,draw] (b22) {$\ab{1256}$};
		\node at (1.5,1) [rectangle,draw] (b32) {$\ab{1456}$};
		\node at (1.5,0) [rectangle,draw] (b42) {$\ab{3456}$};
		\draw[->](a)--(b) ;
		\draw[->](b)--(c);
		\draw[->](c)--(b31);
		\draw[->](b31)--(b41);
		\draw[->](b32)--(c);
		\draw[->](c)--(b22);
		\draw[->](b22)--(b);
		\draw[->](b)--(b12);
		\draw[->](b31)--(b32);
		\draw[->](b42)--(b31);
	\end{tikzpicture}
		\end{center}
		\end{minipage}
		\begin{minipage}{0.5\textwidth}
		\begin{center}
			\begin{tikzpicture}[scale=1.3]
		\node at (-1.5,4) [rectangle,draw] (a) {$\ab{1234}$};
		\node at (0,3) (b) {$\ab{1235}$};
		\node at (0,2) (c) {$\ab{1356}$};
		\node at (0,1) (b31) {$\ab{1345}$};
		\node at (0,0) [rectangle,draw] (b41) {$\ab{2345}$};
		\node at (1.5,3) [rectangle,draw] (b12) {$\ab{1236}$};
		\node at (1.5,2) [rectangle,draw] (b22) {$\ab{1256}$};
		\node at (1.5,1) [rectangle,draw] (b32) {$\ab{1456}$};
		\node at (1.5,0) [rectangle,draw] (b42) {$\ab{3456}$};
		\draw[->](a)--(b) ;
		\draw[->](c)--(b);
		\draw[->](b31)--(c);
		\draw[->](b31)--(b41);
		\draw[->](c)--(b32);
		\draw[->](b22)--(c);
		\draw[->](b)--(b12);
		\draw[->](b42)--(b31);
		\draw[->](b) to[bend right=45]  (b31);
	\end{tikzpicture}
			\end{center}
			\end{minipage}
	\caption{Left: The quiver diagram for the $Gr(4,6)$ initial cluster. Right: The quiver that arises by mutating $\ab{1245}$ of the initial cluster, where the effect of the mutation is described in eqs.~\eqref{eq:mutation} and \eqref{eq:mutation1245} for the variable, and below eq.~\eqref{eq:PluckerRel} for the arrows of the quiver.}
\label{Gr46initial}
\end{figure}

In the simplest case, which is also sufficient for $\cN=4$ SYM amplitudes, cluster algebras can be described by directed graphs or quivers. On the left-hand side of figure \ref{Gr46initial}, the initial cluster of the $Gr(4,6)$ cluster algebra, relevant for six-particle scattering, is shown. Unboxed and boxed vertices of the quiver denote the $\cA$-coordinates and frozen variables, respectively, and we observe that they all correspond to the four-brackets (or equivalently Pl\"ucker coordinates) defined in subsection~\ref{sec:DCIKin}, namely $4\times 4$ minors of the $4\times n$ matrix realization of $Gr(4,n)$, here for $n=6$. The observant reader may however notice two differences between frozen and cluster variables, which turn out to hold more generally: First, that the former always have consecutive indices $\ab{i,i+1,i+2,i+3}$, modulo $n+i\sim i$ identifications, whereas this is not the case for the latter. And second, that the former are not allowed to have arrows between them.

The arrows of the quiver encode how its $\cA$-coordinates will transform under mutation. Concretely, if $a_k$ is a cluster $\cA$-coordinate, mutating it replaces it by
\be\label{eq:mutation}
a_k\to a^{\prime}_k=\frac{1}{a_k}\left(\prod_{\textrm{arrows }i\to k}a_i+\prod_{\textrm{arrows }k\to j}a_j\right)\,.
\ee
Let us see this in action in the $Gr(4,6)$ initial cluster, but for reasons that will become apparent very shortly, let us first switch to shorthand notation where each four-bracket is expressed in terms of the complement of the twistor labels it contains, for example $\ab{1235}=(46)$, $\ab{1345}=(26)$, with the sign convention chosen such that increasing order of labels on the left is mapped to increasing order of labels on the right (this is an instance of the more general $Gr(k,n)\simeq Gr(n-k,n)$ duality exchanging $k$- and $(n-k)$-planes). Then, the mutation of $\ab{1245}$ on the left-hand side of figure \ref{Gr46initial} yields
\be\label{eq:mutation1245}
\ab{1245}=(36)\to \frac{(46)(23)+(26)(34)}{(36)}=(24)\,,
\ee
where the shorthand notation allowed us to arrive at the last equality by using the familiar also in other contexts three-term Pl\"ucker relation or $SL(2,\mathbb{C})$ Schouten identity,
\be\label{eq:PluckerRel}
(ik)(jl)=(ij)(kl)+(il)(jk)\,,
\ee
for $i=2$, $j=3$, $k=4$ and $l=6$.

Except for $a_k\to a^{\prime}_k$, in the new quiver produced by this mutation, all the rest of the $\cA$-coordinates and coefficients remain unchanged. However, the arrows of the new quiver will differ, and may be obtained by those of the quiver before the mutation of $a_k$ by applying the following rules\footnote{It is very interesting to note that essentially the same quiver (but not cluster variable) mutation rules were independently proposed in the context of $\cN=1$ quiver gauge theories so as to describe their Seiberg duality~\cite{Feng:2001bn}, which generalizes the usual electric-magnetic duality of abelian gauge theory.}:
\begin{itemize}
\item For each path $i\to k\to j$ add an arrow $i\to j$, except if both $i$ and $j$ are frozen variables.
\item Reverse the direction of all arrows pointing to or originating from $k$.
\item Remove any pairs of arrows pointing in opposite directions, $\rightleftarrows$.
\end{itemize}
Going back to our example, we see that by virtue of these rules the mutation of $\ab{1245}$ in the $Gr(4,6)$ initial cluster leads to the new cluster shown on the right-hand side of figure \ref{Gr46initial}.

We have thus specified all the rules of the game, and obtaining the entire cluster algebra is a matter of applying them over and over at each vertex of every quiver we encounter.  While the graphical representation and rules we described so far are more accessible for a first exposure to cluster algebras, to this end it proves more efficient to exploit the fact that every quiver is in bijection with a skew-symmetric \emph{exchange matrix} matrix $B$ with elements
\be\label{eq:ExchMatrix}
b_{ij}=(\#\ \text{arrows}\,\, i\to j)-(\#\ \text{arrows}\,\, j\to i)\,.
\ee
In this manner, e.g. the exchange matrices $B,B'$ associated to the left- and right-hand side quivers of figure \ref{Gr46initial}, respectively, have nonzero elements with $i<j$ that are equal to
\be\label{eq:BBpElements}
\begin{aligned}
b_{12}&=b_{15}=b_{23}=b_{26}=b_{37}=b_{39}=-b_{14}=-b_{16}=-b_{27}=-b_{38}=1\,,\\
b'_{13}&=b'_{15}=b'_{27}=b'_{39}=-b'_{12}=-b'_{14}=-b'_{23}=-b'_{26}=-b'_{38}=-b'_{67}=1\,,
\end{aligned}
\ee
when ordering our $\cA$-coordinates and frozen variables as $\{\ab{1235},\ab{1245},\ab{1345},\ab{1234}$,
$\ab{1236},\ab{1256},\ab{1456},\ab{3456},\ab{2345}\}$ for $B$, and similarly with $
\ab{1245}\to \ab{1356}$ for $B'$. With this rearrangement of information, it can be shown that the $\cA$-coordinate mutation~\eqref{eq:mutation} becomes
\begin{equation}
	\label{equ:clusterMutation}
	a'_k = a_k^{-1}\left(\prod_{i=1}^{d+m}a_{i}^{\left[b_{ik}\right]_+} + \prod_{i=1}^{d+m}a_{i}^{\left[-b_{ik}\right]_+}\right) \,,
\end{equation}
with $\left[x\right]_+ = \max\left(0,x\right)$. Similarly, the rules we discussed below eq.~\eqref{eq:PluckerRel} for the transformation of the quiver translate into the following mutation rule for the exchange matrix,
\begin{align}\label{eq:Bmutation}
	b'_{ij} = 
		\begin{cases}
			-b_{ij}\,, \;\;  \text{if}\;\,i=k\,\text{or}\,j=k \,,\\
			b_{ij} + \left[-b_{ik}\right]_+b_{kj} + b_{ik}\left[b_{kj}\right]_+\;  \text{otherwise}\,,
		\end{cases}\
\end{align}
as can be readily verified in the example of eq.~\eqref{eq:BBpElements}.

Apart from the fact that the alternative definitions~\eqref{equ:clusterMutation}-\eqref{eq:Bmutation} are more amenable to computer implementation, they can be also generalized so as to describe cluster algebras with \emph{skew-symmetrizable} instead of skew-symmetric exchange matrices (more precisely, their \emph{principal} part with indices $i,j\le d$)\footnote{Alternatively, skew-symmetrizable cluster algebras may be defined by generalizing quivers to so-called valued quivers~\cite{2012arXiv1202.4161K}.}. In this more general setting, it is possible to prove that finite cluster algebras are classified by Dynkin diagrams, and that a skew-symmetric cluster algebra is finite if and only if one of its clusters takes the form of the associated Dynkin diagram when dropping arrow orientations as well as frozen variables and arrows from/to them~\cite{1054.17024}. Inspecting figure~\eqref{Gr46initial}, we can thereby infer that the $Gr(4,6)$ cluster algebra is of finite $A_3$ type.

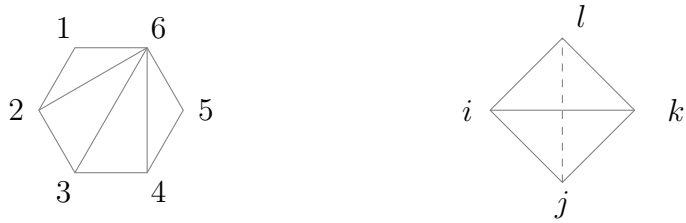
\begin{figure}
\begin{center}
\begin{tikzpicture}[scale=3]

    \foreach \i in {1,...,6}
   {
     \coordinate (a\i) at ($(60+60*\i:0.32)+(0,3.25)$);
   }
   \draw[gray,fill=white]
     (a1)
     \foreach \i in {2,...,6}
   {
     -- (a\i)
   } -- cycle;

   \draw[gray] (a2) -- (a6);
   \draw[gray] (a3) -- (a6);
   \draw[gray] (a4) -- (a6);

    \foreach \i in {1,...,6}
   {
     \node at ($(60+60*\i:0.42)+(0,3.25)$) {$\i$};
   }

       \foreach \i in {1,...,4}
   {
     \coordinate (a\i) at ($(90+90*\i:0.32)+(2,3.25)$);
   }
   \draw[gray,fill=white]
     (a1)
     \foreach \i in {1,...,4}
   {
     -- (a\i)
   } -- cycle;

   \draw[gray] (a1) -- (a3);
   \draw[gray, dashed] (a2) -- (a4);

     \node at ($(90+90*1:0.42)+(2,3.25)$) {$i$};
     \node at ($(90+90*2:0.42)+(2,3.25)$) {$j$};
ˇ    \node at ($(90+90*3:0.42)+(2,3.25)$) {$k$};
     \node at ($(90+90*4:0.42)+(2,3.25)$) {$l$};

\end{tikzpicture}
\end{center}
\caption{Left: Geometric interpretation of the $Gr(4,6)\simeq Gr(2,6)$ cluster shown in the left of figure~\ref{Gr46initial}, where $\ab{1234}$ corresponds to the edge $(56)$ etc, as a triangulation of a hexagon by non-crossing diagonals. Right: Mutations of this cluster algebra may be similarly geometrically interpreted as flips $(ik)\leftrightarrow(jl) $ of the diagonals of any quadrilateral subdiagram, see also eq.~\eqref{eq:PluckerRel}.}
\label{fig:HexTriangulation}
\end{figure}

Despite the relative simplicity of the definitions and properties of cluster algebras, the reader may perhaps be left wondering where they come from, if there is any physical or mathematical intuition behind them. To address this, drawing from~\cite{1021.16017} let us return to our $Gr(4,6)$ example, and note that the labels of the dual two-component brackets $(ij)$ can be interpreted as the vertices of a hexagon, such that the frozen variables correspond to its edges, whereas the $\cA$ coordinates of the clusters we have encountered so far to non-crossing diagonals. This observation in fact extends to the entire cluster algebra, since it can be shown that $\emph{all}$ mutations have the form of the three-term identity~\eqref{eq:PluckerRel} with $1\le i<j<k<l\le 6$, and hence they are geometrically equivalent to flipping the diagonal of a quadrilateral inside the hexagon, see figure \ref{fig:HexTriangulation}. Therefore the $Gr(4,6)$ cluster algebra is in natural bijection with all triangulations of the hexagon with non-crossing diagonals. Note that this bijection also includes the exchange matrix $B$ in eq.\eqref{eq:ExchMatrix}, or in other words the arrows of the quivers as seen e.g. in figure~\ref{Gr46initial}, which become arrows between adjacent sides of any triangle of the triangulation (not allowing arrows between two edges of the hexagon), with the orientation chosen in the anti-clockwise direction.  This geometric picture of cluster algebras as triangulations is in fact a very profound and universal one, as up to 18 exceptional cases, it has been shown to hold for all cluster algebras with a finite number of exchange matrices, even if they have an infinite number of clusters/variables~\cite{2011arXiv1111.3449F}.

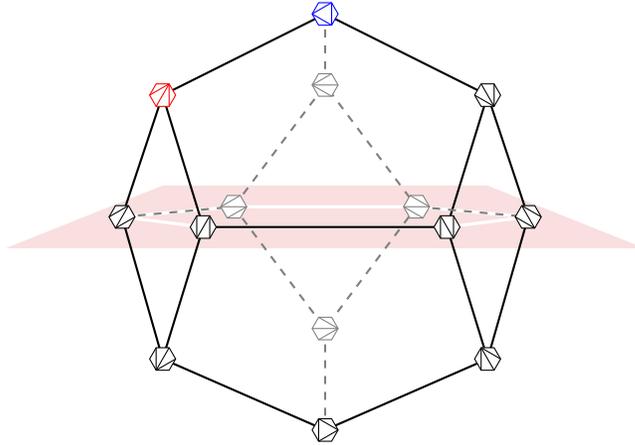
\begin{figure}
\begin{center}
\begin{tikzpicture}[scale=0.54]
\definecolor{bg}{RGB}{246,202,203}
\coordinate (A) at (1,0.8);
\coordinate (B) at (9,0.8);
\coordinate (C) at (13,-0.8);
\coordinate (D) at (-3,-0.8);
\draw[thick, color=white, fill=bg,fill opacity=0.6] (A) -- (B) -- (C) -- (D) -- cycle;

\draw[very thick,white] (0,0) -- (2.75,0.25) ;
\draw[very thick,white]  (7.25,0.25) -- (10,0);
\draw[very thick,white] (2,-0.25) -- (8,-0.25);
\draw[very thick,white] (0,0) -- (2,-0.25);
\draw[very thick,white] (8,-0.25) -- (10,0);
\draw[very thick,white] (2.75,0.25) -- (7.25,0.25);
  \draw[thick,fill=none] (0,0) -- (1,3) -- (2,-0.25) -- (1,-3.5)  -- cycle;
  \draw[thick,fill=none] (2,-0.25) -- (8,-0.25);
  \draw[thick,fill=none] (10,0) -- (9,3) -- (8,-0.25) -- (9,-3.5)  -- cycle;
  \draw[thick,fill=none] (1,3) -- (5,5) -- (9,3);
  \draw[thick,fill=none] (1,-3.5) -- (5,-5.25) -- (9,-3.5);
  \draw[thick,gray,dashed] (0,0) -- (2.75,0.25) ;
  \draw[thick,gray,dashed]  (7.25,0.25) -- (10,0);
  \draw[thick,gray,dashed]  (2.75,0.25) -- (5,3.25) -- (7.25,0.25) -- (5,-2.75) -- cycle;
  \draw[thick,gray,dashed]  (5,3.25) -- (5,5);
  \draw[thick,gray,dashed]  (5,-2.75) -- (5,-5.25);

\foreach \i in {1,...,6}
   {
     \coordinate (a\i) at ($(60+60*\i:0.32)+(0,0)$);
   }
   \draw[fill=white]
     (a1)
     \foreach \i in {2,...,6}
   {
     -- (a\i)
   } -- cycle;

   \draw (a6) -- (a2);
   \draw (a6) -- (a3);
   \draw (a3) -- (a5);

 \foreach \i in {1,...,6}
   {
     \coordinate (a\i) at ($(60+60*\i:0.32)+(1,3)$);
   }
   \draw[red,fill=white]
     (a1)
     \foreach \i in {2,...,6}
   {
     -- (a\i)
   } -- cycle;

   \draw[red] (a6) -- (a2);
   \draw[red] (a6) -- (a3);
   \draw[red] (a6) -- (a4);

 \foreach \i in {1,...,6}
   {
     \coordinate (a\i) at ($(60+60*\i:0.32)+(2,-0.25)$);
   }
   \draw[fill=white]
     (a1)
     \foreach \i in {2,...,6}
   {
     -- (a\i)
   } -- cycle;

   \draw (a1) -- (a3);
   \draw (a6) -- (a3);
   \draw (a6) -- (a4);

 \foreach \i in {1,...,6}
   {
     \coordinate (a\i) at ($(60+60*\i:0.32)+(1,-3.5)$);
   }
   \draw[fill=white]
     (a1)
     \foreach \i in {2,...,6}
   {
     -- (a\i)
   } -- cycle;

   \draw (a1) -- (a3);
   \draw (a6) -- (a3);
   \draw (a3) -- (a5);

 \foreach \i in {1,...,6}
   {
     \coordinate (a\i) at ($(60+60*\i:0.32)+(5,5)$);
   }
   \draw[blue,fill=white]
     (a1)
     \foreach \i in {2,...,6}
   {
     -- (a\i)
   } -- cycle;

   \draw[blue] (a2) -- (a4);
   \draw[blue] (a4) -- (a6);
   \draw[blue] (a6) -- (a2);

\foreach \i in {1,...,6}
   {
     \coordinate (a\i) at ($(60+60*\i:0.32)+(5,-5.25)$);
   }
   \draw[fill=white]
     (a1)
     \foreach \i in {2,...,6}
   {
     -- (a\i)
   } -- cycle;

   \draw (a1) -- (a3);
   \draw (a3) -- (a5);
   \draw (a5) -- (a1);

\foreach \i in {1,...,6}
   {
     \coordinate (a\i) at ($(60+60*\i:0.32)+(8,-0.25)$);
   }
   \draw[fill=white]
     (a1)
     \foreach \i in {2,...,6}
   {
     -- (a\i)
   } -- cycle;

   \draw (a1) -- (a3);
   \draw (a1) -- (a4);
   \draw (a4) -- (a6);

\foreach \i in {1,...,6}
   {
     \coordinate (a\i) at ($(60+60*\i:0.32)+(9,3)$);
   }
   \draw[fill=white]
     (a1)
     \foreach \i in {2,...,6}
   {
     -- (a\i)
   } -- cycle;

   \draw (a1) -- (a4);
   \draw (a2) -- (a4);
   \draw (a4) -- (a6);
   
\foreach \i in {1,...,6}
   {
     \coordinate (a\i) at ($(60+60*\i:0.32)+(10,0)$);
   }
   \draw[fill=white]
     (a1)
     \foreach \i in {2,...,6}
   {
     -- (a\i)
   } -- cycle;

   \draw (a1) -- (a4);
   \draw (a1) -- (a5);
   \draw (a2) -- (a4);
   
   \foreach \i in {1,...,6}
   {
     \coordinate (a\i) at ($(60+60*\i:0.32)+(9,-3.5)$);
   }
   \draw[fill=white]
     (a1)
     \foreach \i in {2,...,6}
   {
     -- (a\i)
   } -- cycle;

   \draw (a1) -- (a3);
   \draw (a1) -- (a4);
   \draw (a1) -- (a5);

   \foreach \i in {1,...,6}
   {
     \coordinate (a\i) at ($(60+60*\i:0.32)+(2.75,0.25)$);
   }
   \draw[gray,fill=white]
     (a1)
     \foreach \i in {2,...,6}
   {
     -- (a\i)
   } -- cycle;

   \draw[gray] (a2) -- (a5);
   \draw[gray] (a2) -- (a6);
   \draw[gray] (a3) -- (a5);

   \foreach \i in {1,...,6}
   {
     \coordinate (a\i) at ($(60+60*\i:0.32)+(5,3.25)$);
   }
   \draw[gray,fill=white]
     (a1)
     \foreach \i in {2,...,6}
   {
     -- (a\i)
   } -- cycle;

   \draw[gray] (a2) -- (a4);
   \draw[gray] (a2) -- (a5);
   \draw[gray] (a2) -- (a6);
   
      \foreach \i in {1,...,6}
   {
     \coordinate (a\i) at ($(60+60*\i:0.32)+(7.25,0.25)$);
   }
   \draw[gray,fill=white]
     (a1)
     \foreach \i in {2,...,6}
   {
     -- (a\i)
   } -- cycle;

   \draw[gray] (a1) -- (a5);
   \draw[gray] (a2) -- (a5);
   \draw[gray] (a2) -- (a4);
   
      \foreach \i in {1,...,6}
   {
     \coordinate (a\i) at ($(60+60*\i:0.32)+(5,-2.75)$);
   }
   \draw[gray,fill=white]
     (a1)
     \foreach \i in {2,...,6}
   {
     -- (a\i)
   } -- cycle;

   \draw[gray] (a1) -- (a5);
   \draw[gray] (a2) -- (a5);
   \draw[gray] (a3) -- (a5);   

 \end{tikzpicture}
 \end{center}
 \caption{The cluster polytope of the $Gr(4,6)\simeq A_3$ cluster algebra, with clusters representing the different triangulations of a hexagon, as discussed in the text and in figure~\ref{fig:HexTriangulation}, where also the vertex labels may be found. The initial and mutated cluster in the left- and right-hand side of figure \ref{Gr46initial} are color-coded in red and blue, respectively. The parity invariant plane is also drawn in pink. Adapted from refs.~\cite{Drummond:2018dfd,Chicherin:2020umh}.}
 \label{A3polytope}
 \end{figure}
Before concluding this introduction on cluster algebras, let us mention one further connection they have with geometry, that will be useful for us 
in what follows: Representing each cluster with a vertex, and each mutation with an edge yields the \emph{exchange graph} of  a finite rank-$d$ cluster algebra, which in fact defines a \emph{simple polytope}~\cite{Fomin:2001rc}, namely a geometric object with flat faces generalizing the polygon to higher dimensions, whose vertices are in addition adjacent to exactly $d$ edges. As an example, the cluster polytope of the $Gr(4,6)\simeq A_3$ cluster algebra is shown in figure \ref{A3polytope}. The bijection with triangulations discussed in the previous paragraph allows one to easily work this out, and infer that it has a total of 9 diagonals/cluster variables spread into 14 vertices/clusters, as well as 21 edges and 9 faces. 
In addition to its topological and combinatorial nature, this polytope also geometrically describes the compactification of the \emph{positive region} of $Gr(4,n)/(\mathbb{C}^{*})^{n-1}$, defined as the region where $\ab{ijkl}>0$ for $i<j<k<l$. In particular, each edge of the polytope can be assigned a (Fock-Goncharov) $\cX$-coordinate, related to the $\cA$-coordinates by~\cite{2003math.....11245F}
\begin{equation}
	\label{eq:xtoa}
	x_{i} \equiv \prod_{l=1}^{d+m}a_{l}^{b_{li}}\,,\quad i=1,\ldots d\,,
\end{equation}
such that each cluster provides a local coordinate chart describing this compactification, with the interior of the positive region corresponding to all $\infty>x_i>0$. The significance of the positive region will also be highlighted in Chapter 7~\cite{Herrmann:2022nkh} of the SAGEX Review~\cite{Travaglini:2022uwo}.

\subsubsection{Symbol letters from cluster variables.} 

\begin{figure}
	\begin{center}
		\begin{tikzpicture}[scale=1.3]
		\node at (-1.5,4) [rectangle,draw] (a) {$\ab{1234}$};
		\node at (0,3) (b11) {$\ab{1235}$};
		\node at (0,2) (b21) {$\ab{1245}$};
		\node at (0,1) (b31) {$\ab{1345}$};
		\node at (0,0) [rectangle,draw] (b41) {$\ab{2345}$};
		\node at (1.5,3)  (b12) {$\ab{1236}$};
		\node at (1.5,2)  (b22) {$\ab{1256}$};
		\node at (1.5,1)  (b32) {$\ab{1456}$};
		\node at (1.5,0) [rectangle,draw] (b42) {$\ab{3456}$};
		\node at (3,3)  (b13) {\ldots};
		\node at (3,2)  (b23) {\ldots};
		\node at (3,1)  (b33) {\ldots};
		\node at (3,0)  (b43) {\ldots};
		\node at (5,3)  (b14) {$\ab{123\scalebox{0.9}{$n-1 $}}$};
		\node at (5,2)  (b24) {$\ab{12\, \scalebox{0.9}{$n-2 $}\,\scalebox{0.9}{$n-1 $}}$};
		\node at (5,1)  (b34) {$\ab{1\, \scalebox{0.9}{$n-3 $}\,\scalebox{0.9}{$n-2 $}\,\scalebox{0.9}{$n-1 $}}$};
		\node at (5,0) [rectangle,draw] (b44) {$\ab{\scalebox{0.9}{$n-4 $}\,\scalebox{0.9}{$n-3 $}\,\scalebox{0.9}{$n-2 $}\,\scalebox{0.9}{$n-1 $}}$};
		\node at (8,3) [rectangle,draw] (b15) {$\ab{123\scalebox{0.9}{$n $}}$};
		\node at (8,2) [rectangle,draw] (b25) {$\ab{12\, \scalebox{0.9}{$n-1 $}\,\scalebox{0.9}{$n $}}$};
		\node at (8,1) [rectangle,draw] (b35) {$\ab{1\, \scalebox{0.9}{$n-2 $}\,\scalebox{0.9}{$n-1 $}\,\scalebox{0.9}{$n $}}$};
		\node at (8,0) [rectangle,draw] (b45) {$\ab{\scalebox{0.9}{$n-3 $}\,\scalebox{0.9}{$n-2 $}\,\scalebox{0.9}{$n-1 $}\,\scalebox{0.9}{$n $}}$};
		\draw[->](a)--(b11) ;
		\draw[->](b11)--(b21);
		\draw[->](b21)--(b31);
		\draw[->](b31)--(b41);
		\draw[->](b11)--(b12);
	    \draw[->](b21)--(b22);
	    	\draw[->](b31)--(b32);
        	\draw[->](b22)--(b11);
		\draw[->](b32)--(b21);
		\draw[->](b42)--(b31);
		\draw[->](b12)--(b22);
		\draw[->](b22)--(b32);
		\draw[->](b32)--(b42);
		\draw[->](b14)--(b24);
		\draw[->](b24)--(b34);
		\draw[->](b34)--(b44);
		\draw[->](b14)--(b15);
	    \draw[->](b24)--(b25);
	    	\draw[->](b34)--(b35);
        	\draw[->](b25)--(b14);
		\draw[->](b35)--(b24);
		\draw[->](b45)--(b34);

	\end{tikzpicture}
		\end{center}
	\caption{Quiver diagram for the initial $Gr(4,n)$ cluster.}
\label{Gr4Ninitial}
\end{figure}
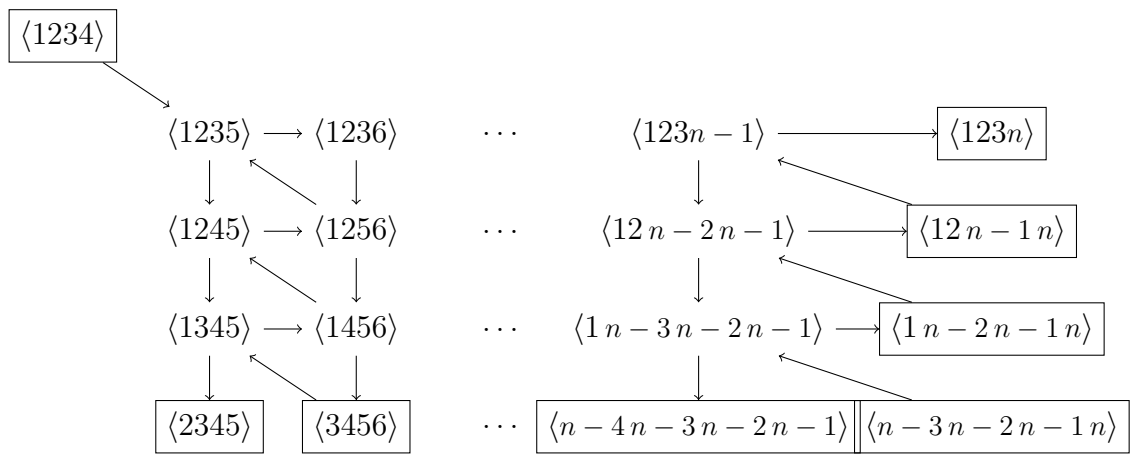

With this background knowledge on cluster algebras, we can now state their role in $\cN=4$ SYM amplitudes: In~\cite{Golden:2013xva} the remarkable observation was made, based on the then-known explicit computations, that

\vspace{6pt}
\noindent \fbox{\begin{minipage}{\textwidth}{
\emph{$Gr(4,n)$ cluster $\cA$-coordinates appear as symbol letters of
 the $n$-particle amplitude.}}
\end{minipage}}
\vspace{6pt}

That is, for the examples considered, the amplitude could be expressed in terms of polylogarithmic functions as defined in eq.~\eqref{dFPhi}, with $\phi_{\alpha_i}$ coinciding with $\cA$-coordinates (more precisely multiplicative combinations thereof, also including the frozen variables, that respect the scale and hence the dual conformal invariance of the theory).

For the six-particle amplitude, as we have seen the associated $Gr(4,6)$ cluster algebra consists of  9 $\cA$-coordinates, and these precisely coincide with the symbol alphabet of the two-loop correction to the amplitude~~\cite{DelDuca:2009au,DelDuca:2010zg,Goncharov:2010jf}. The six-particle or \emph{hexagon  bootstrap} was initiated in \cite{Dixon:2011pw} based on the assumption that this alphabet also remains stable at higher loops, and in this case its cluster algebraic structure may be considered as evidence backing this assumption. Explicitly, in the usual four-bracket notation a convenient choice for the six-particle alphabet reads~\cite{Caron-Huot:2019bsq}, \footnote{Note that any set of equal size, consisting of multiplicatively independent combinations of these letters, would make an equally valid choice. Indeed, in the original literature~\cite{Dixon:2011pw,Dixon:2011nj,Dixon:2013eka,Dixon:2014voa,Dixon:2014iba} the letters $u_i$ and $1-u_i$ were used. The relation with the presently used alphabet is $a_i=u_i/(u_{i-1} u_{i+1})$ and $m_i=1-1/u_i$.}
\be\label{eq:hexletters}
a_1=\frac{\textcolor{blue}{\ab{1245}}^2\ab{3456}^2\ab{6123}^2}{\ab{1234}\ab{2345}\ldots \ab{6123}}\,,\quad m_1=\frac{\textcolor{blue}{\ab{1356}\ab{2346}}}{\ab{1236}\ab{3456}}\,,\quad y_1=\frac{\textcolor{blue}{\ab{1345}\ab{2456}}\ab{1236}}{\textcolor{blue}{\ab{1235}\ab{1246}}\ab{3456}}\,,
\ee
with the cluster $\cA$-coordinates are color-coded in blue, together with two more cyclic transformations $l_1\to l_{1+i}$ with $l \in \{a,m,y\}$ induced by shifting $Z_m\to Z_{m-2i}$ on the right-hand side. The discrete parity and flip transformations of the letters may be inferred from the cluster polytope of figure \ref{A3polytope}, where they correspond to up-down and left-right reflection, respectively. For example, parity is equivalent to a $i\to i+3$ shift of momentum twistor labels, and thus  transforms $y_i\to 1/y_i$, while leaving $a_i,m_i$ invariant.

The initial quiver for generic $Gr(4,n)$, from which the next nontrivial $n=7$ case may be studied with the same set of rules we spelled out, is depicted in figure \ref{Gr4Ninitial}. While for the latter case the initial quiver does not have the topology of a Dynkin diagram, mutating $\ab{1256},\ab{1456}$ and $\ab{1345}$ does lead to an $E_6$-shaped cluster, and hence $Gr(4,7)$ is also a finite cluster algebra. In particular, one finds 42 different $\cA$-coordinates distributed in 833 distinct clusters (the order of the variables in each cluster does not matter). Again, these $\cA$-coordinates exactly match the symbol alphabet of the two-loop correction to the seven-particle amplitude~\cite{Caron-Huot:2011zgw}, and this observation was of central importance for generalizing the bootstrap program to higher multiplicity $n=7$ in~\cite{Drummond:2014ffa}. The choice for the corresponding symbol alphabet adopted in the latter reference is
\begin{align}
\label{eq:heptagonletters}
a_{11} &=
\frac{\ab{1234} \ab{1567} \textcolor{blue}{\ab{2367}}}{\ab{1237} \ab{1267} \ab{3456}}\,,
&
a_{41} &=
\frac{\textcolor{blue}{\ab{2457}} \ab{3456}}{\ab{2345} \ab{4567}}\,,
\nonumber
\\
a_{21} &=
\frac{\ab{1234} \textcolor{blue}{\ab{2567}}}{\ab{1267} \ab{2345}}\,,
&
a_{51} &=
\frac{\textcolor{blue}{\ab{1(23)(45)(67)}}}{\ab{1234} \ab{1567}}\,,
\\
a_{31} &=
\frac{\ab{1567} \textcolor{blue}{\ab{2347}}}{\ab{1237} \ab{4567}}\,,
&
a_{61} &=
\frac{\textcolor{blue}{\ab{1(34)(56)(72)}}}{\ab{1234} \ab{1567}}\,,
\nonumber
\end{align}
where we have again denoted the cluster $\cA$-coordinates in blue and defined
\be\label{brbilinear}
\ab{a(bc)(de)(fg)} \equiv
\ab{abde} \ab{acfg} - \ab{abfg} \ab{acde}\,,
\ee
together with the letters $a_{ij}$ obtained from $a_{i1}$ by cyclically relabeling the momentum twistors $Z_m \to Z_{m + j - 1}$. It is interesting to note that for $n=7$, and more generally when $n$ is odd, any individual cluster $\cA$-coordinate can be rendered invariant under rescalings $Z_i\to t Z_i$ by suitable products of powers of frozen variables. Here it is slightly more nontrivial to show that parity transformations map $a_{2i}\leftrightarrow a_{3,i-1}$ and  $a_{4i}\leftrightarrow a_{5i}$.

The appearance of not only Pl\"ucker variables but also homogeneous polynomials thereof~\eqref{brbilinear} as $\cA$-coordinates is a qualitatively new feature that persists for $Gr(4,n)$ cluster algebras with $n\ge 8$. However the relation of these cluster algebras with  $n$-particle alphabets is a significantly more subtle issue which we will address in subsection \ref{sec:TropSing}.

While not of direct relevance for this article, before closing let us briefly mention two more connections that have been established between cluster algebras and scattering amplitudes. The first one also pertains to planar $n$-particle amplitudes in $\cN=4$ SYM, and focuses on identifying appropriate cluster $\cX$-coordinates as arguments of the MPLs needed to describe them~\cite{Golden:2013lha,Golden:2014xqa,Parker:2015cia,Golden:2018gtk}, based on their Poisson structure~\cite{2002math......8033G}. This approach has been used very successfully in promoting symbols of two-loop MHV amplitudes~\cite{Caron-Huot:2011zgw} to functions~\cite{Golden:2014xqf,Golden:2014pua,Golden:2021ggj}, yet there is evidence that this is no longer possible at different MHV degree~\cite{Harrington:2015bdt}, and furthermore it is unclear how to generalize to higher loops. Finally, cluster algebras may also be used in order to define natural generalizations of string amplitudes~\cite{Arkani-Hamed:2019mrd,Arkani-Hamed:2019plo,Gates:2021tnp}.

\subsection{Cluster adjacency and extended Steinmann relations}\label{sec:SteinClus}

So far we have seen that cluster variables  dictate the singularities of $\cN=4$ SYM amplitudes. However cluster algebras have more structure than just the variables, for instance the clusters. It is thus natural to ask, do they also play a role in this context? Very interestingly, there is evidence that they do, in the form of~\cite{Drummond:2017ssj}

\vspace{6pt}

\noindent\fbox{\begin{minipage}{\textwidth}{
\emph{Cluster adjacency: In a symbol whose alphabet contains $\textrm{Gr}(4,n)$ cluster $\cA$-coordinates, two of them can appear consecutively	only if there exists a cluster where they both appear.}}
\end{minipage}}
\vspace{6pt}

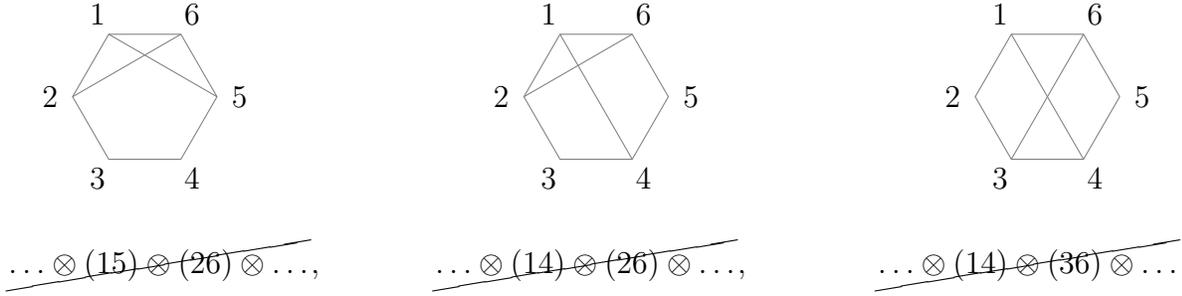
\begin{figure}
\begin{center}
\begin{tikzpicture}[scale=3]

    \foreach \i in {1,...,6}
   {
     \coordinate (a\i) at ($(60+60*\i:0.32)+(0,3.25)$);
   }
   \draw[gray,fill=white]
     (a1)
     \foreach \i in {2,...,6}
   {
     -- (a\i)
   } -- cycle;

   \draw[gray] (a1) -- (a5);
   \draw[gray] (a2) -- (a6);
    \foreach \i in {1,...,6}
   {
     \node at ($(60+60*\i:0.42)+(0,3.25)$) {$\i$};
   }
   
   \foreach \i in {1,...,6}
   {
     \coordinate (a\i) at ($(60+60*\i:0.32)+(2,3.25)$);
   }
   \draw[gray,fill=white]
     (a1)
     \foreach \i in {2,...,6}
   {
     -- (a\i)
   } -- cycle;

   \draw[gray] (a1) -- (a4);
   \draw[gray] (a2) -- (a6);
\foreach \i in {1,...,6}
   {
     \node at ($(60+60*\i:0.42)+(2,3.25)$) {$\i$};
   }   
   
   \foreach \i in {1,...,6}
   {
     \coordinate (a\i) at ($(60+60*\i:0.32)+(4,3.25)$);
   }
   \draw[gray,fill=white]
     (a1)
     \foreach \i in {2,...,6}
   {
     -- (a\i)
   } -- cycle;

   \draw[gray] (a1) -- (a4);
   \draw[gray] (a3) -- (a6);
    \foreach \i in {1,...,6}
   {
     \node at ($(60+60*\i:0.42)+(4,3.25)$) {$\i$};
   }
\end{tikzpicture}
\begin{flalign*}
&\cancel{\ldots \otimes (15)\otimes (26) \otimes \ldots}, \,\,\qquad\quad
\cancel{\ldots \otimes (14)\otimes (26) \otimes \ldots}, \,\,\qquad\quad\,\,\, \cancel{\ldots \otimes (14)\otimes (36) \otimes \ldots}\,&
\end{flalign*}
\end{center}
\caption{Pairs of $Gr(4,6)$ $\mathcal{A}$-coordinates not found in a cluster together, and hence not allowed to appear consecutively in the symbol of the six-particle amplitude, up to cyclic permutations and order reversal.}\label{fig:Adjacency}
\end{figure}

Let us distill the implications of cluster adjacency\footnote{Note that in the mathematics literature there exists a related notion of simultaneous inclusion in a cluster, which is known as `compatibility'~\cite{Fomin:2001rc}.} in our familiar $Gr(4,6)$ cluster algebra example. From our discussion of the geometric interpretation of the clusters as triangulations of a hexagon with non-crossing diagonals, it is evident that pairs of $\cA$-coordinates not found in a cluster together will in turn correspond to crossing diagonals, and are thus forbidden from appearing next to each other in the symbol of the six-particle amplitude, as shown in figure \ref{fig:Adjacency}.

The same information may be alternatively described by a \emph{neighbor set} \cite{Drummond:2018dfd} of an $\cA$-coordinate, that is the union of all clusters containing it, which in other words contains all the other variables (including the frozen ones), that can appear next to it in the symbol. For the $n=6$ case, reverting to the usual four-bracket notation these are
\begin{align}
    \ns{\ab{1245}} &= \{\ab{1245}, \ab{2456}, \ab{1345}, \ab{1246},  \ab{1235},\, \text{\& frozen variables.}\}\,,\label{eq:ns1245}\\
        \ns{\ab{1235}} &= \{\ab{1235}, \ab{2456}, \ab{2356}, \ab{1356}, \ab{1345}, \ab{1245}, \,\text{\& frozen variables.}\}\,,\label{eq:ns1235}
\end{align}
as well as their cyclic permutations, two for the first line and five for the second. Then, as a constraint on a polylogarithmic function $F$, and in the notation of eq.~\eqref{Delta11}, cluster adjacency may be formulated as
\be\label{eq:Nonadjacent_Constraint}
F^{\overline{\phi}_\beta,\phi_\beta}=0\,,
\ee
where $\overline{\phi}_\beta$ does not belong to the neighbor set of ${\phi}_\beta$, with the same condition also holding recursively for all left factors in the coproduct of $F$.

As we've discussed, actual symbol letters are dual  conformal invariant, namely products of the cluster and frozen variables that are invariant under rescalings of the twistors, i.e. homogeneous. It is thus more conventient to take this information into account on the right-hand side of eqs.~\eqref{eq:ns1245}-\eqref{eq:ns1235}, by defining the corresponding \emph{homogeneous neighbor sets},
\begin{align}
    \hns{\ab{1245}} &= \hns{a_1} = \{a_1,m_2,m_3,y_1, y_2 y_3\}\,,\label{eq:hns1245}\\
        \hns{\ab{1235}} &= \{a_1,a_2,\frac{m_1}{y_2},\frac{m_2}{y_2 y_3},m_3, y_1 y_2 y_3\}\,.\label{eq:hns1235}
\end{align}
again plus cyclic permutations. In the first line we could also promote the left-hand side to the conformally invariant letter $a_1$, since $\ab{1245}$ is the only cluster variable it depends on. This is not possible for the second line, relevant for the remaining letters $m_i,y_i$.

Moving on to the case $n=7$, all letters~\eqref{brbilinear} depend on a single $\cA$-coordinate, so we can directly focus on  the homogeneous neighbor sets. These are generated by
\begin{align}
    \hns{a_{11}} =\{ &a_{11}, a_{14}, a_{15}, a_{21}, a_{22}, a_{24}, a_{25}, a_{26}, a_{31}, a_{33}, a_{34}, a_{35}, a_{37}, a_{41},a_{43}, a_{46}, a_{51},\label{heptns}\\
     \qquad        &a_{53}, a_{56}, a_{62}, a_{67}\}\nonumber\\
    \hns{a_{21}} =\{&a_{11}, a_{13}, a_{14}, a_{15}, a_{17}, a_{21}, a_{23}, a_{24}, a_{25}, a_{26},a_{31}, a_{33}, a_{34}, a_{36},a_{37}, a_{41}, a_{43},\label{ns7first}\\
    \qquad        &a_{45}, a_{46}, a_{52}, a_{53}, a_{55}, a_{57}, a_{62}, a_{64}, a_{66}\}\nonumber\\
    \hns{a_{41}} =\{&a_{11}, a_{13}, a_{16}, a_{21}, a_{23}, a_{24}, a_{26}, a_{31}, a_{33}, a_{35}, a_{36}, a_{41}, a_{43}, a_{46},a_{51}, a_{62}, a_{67}\}\\
    \hns{a_{61}} =\{&a_{12}, a_{17}, a_{23}, a_{25}, a_{27}, a_{32}, a_{34}, a_{36}, a_{42}, a_{47}, a_{52}, a_{57}, a_{61}\}\,,\label{ns7last}
\end{align}
together with images under parity transformations and cyclic permutations. We will also comment on higher-multiplicity generalizations towards the end of this subsection.

\begin{figure}
\begin{center}
\includegraphics[width=0.7\textwidth]{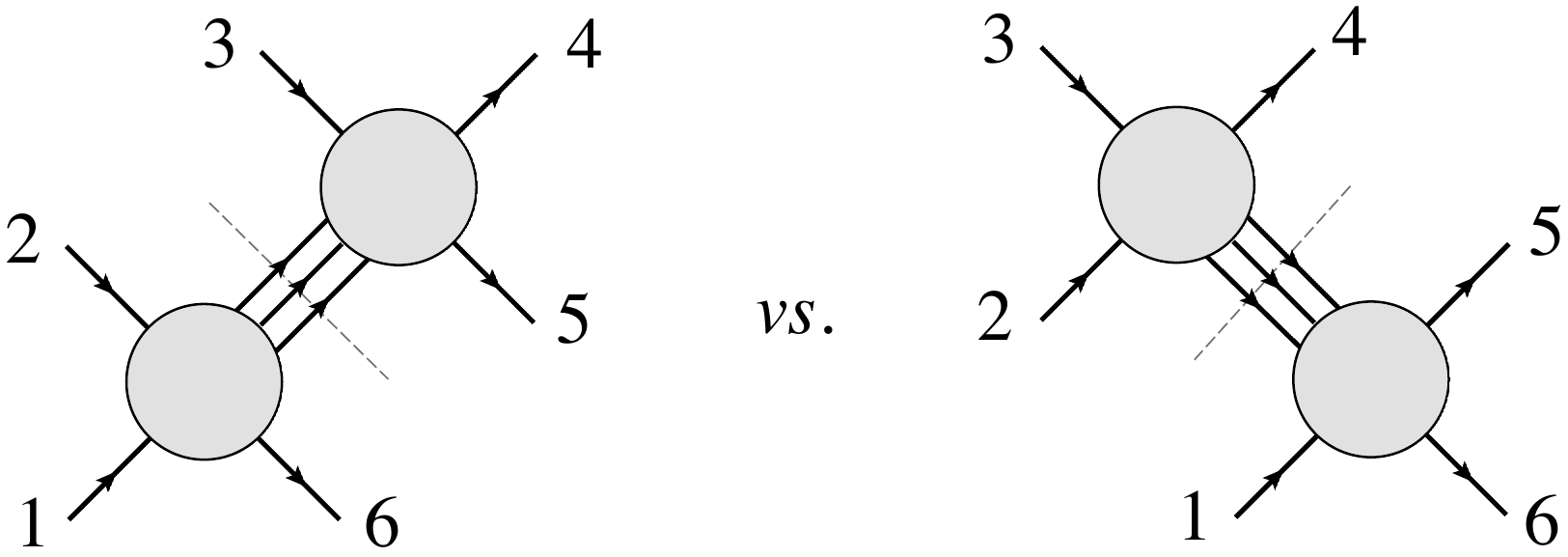}
\end{center}
\caption{Channels $s_{345}\propto \ab{2356}=(14)$ and $s_{234}\propto \ab{1245}=(36)$ for $3\to3$
kinematics. The Steinmann relations state that the discontinuity in one channel should not know about
the discontinuity in the other channel, and this yields the same constraint as the non-cluster adjacent pair on the right of figure \ref{fig:Adjacency}. Adapted from~\cite{Caron-Huot:2016owq}.}
\label{fig:stein}
\end{figure}

What about the physical interpretation of cluster adjacency? It turns out that some of its restrictions can be understood as the extended Steinmann relations~\cite{Caron-Huot:2018dsv,Caron-Huot:2019bsq}, as we will now explain. These are generalizations of  the usual Steinmann relations~\cite{Steinmann,Steinmann2,Cahill:1973qp}, which demand that the double discontinuities of any Feynman diagram (and thus of the amplitude they contribute to) vanish when taken in overlapping channels. In section \ref{sec:MPLs} we've seen that a discontinuity may be labeled by a Mandelstam invariant $s_{i,\ldots,j-1}$  which is analytically continued around its branch point. At the same time, by virtue of the Cutkosky rules \cite{Cutkosky:1960sp} this discontinuity may be obtained by placing on-shell the internal particles whose total energy equals $s_{i,\ldots,j-1}$, that is by replacing their propagators with delta functions. This is the notion of a cut, which splits the Feynman diagram into two parts, as seen in figure \ref{fig:stein}.

By the same logic, overlapping channels correspond to cut lines that intersect, in other words they divide the external particles of the Feynman diagram into four non-empty sets. In the example of the figure, these sets are $\{2\}$, $\{3,4\}$, $\{5\}$, and $\{6,1\}$. Focusing on three-particle Mandelstam invariants, but allowing the number of external particles $n$ to be arbitrary, the Steinmann relations at the level of the amplitude then imply
\be
\label{eq:disc_SteinmannA}
  \Disc_{s_{j,j+1,j+2}}\left(\Disc_{s_{i,i+1,i+2}}\left({A}_{n,k}\right) \right)=0\,,\quad \text{for}\,\,j=i\pm 1,i\pm 2\,,
\ee
with an obvious generalization to higher-particle Mandelstam invariants. Note that we refrain from considering two-particle invariants, since it is necessary for an invariant to be independent for the sake of analytic continuation. That is, no other Mandelstam invariant is allowed to change sign but the one we analytically continue, and this is generically not the case with two-particle invariants.

As eq.~\eqref{symbolDisc} reveals, at the level of the symbol a discontinuity around $\phi_\beta=0$ amounts to clipping off this particular letter from its first entry, with iterated discontinuities obtained by applying this procedure repeatedly. That is, if no other letter vanishes simultaneously, which can be ensured for rational alphabets such as the $n=6,7$ ones, due to the multiplicative independence of the letters. In this case, the Steinmann relations are therefore statements about which letters can appear next to each other in the first two entries of the symbol, and in the past they have been exploited in the amplitude bootstrap so as to simplify the construction of the function spaces containing the amplitude~\cite{Caron-Huot:2016owq,Dixon:2016nkn}. Particularly for $n=6$, we then notice that the unique, up to cyclic permutations, double discontinuity shown in figure \ref{fig:stein} precicely coincides with the non-cluster adjacent pair restriction shown on the right of figure~\ref{fig:Adjacency}.

Then, the analysis of a wealth of data obtained by the amplitude bootstrap, independently revealed that these restrictions on consecutive pairs of symbol letters apply not only in the first two slots, but to \emph{all} depths in the symbol, and they were thereby coined the \emph{extended Steinmann relations}\footnote{These results for $n=6$ were initially reported at {\it Amplitudes 2017}, in a talk by the author~\cite{YorgosAmps17}.}~\cite{Caron-Huot:2018dsv,Caron-Huot:2019bsq}. Given that in the $n=6,7$ alphabets \eqref{eq:hexletters}-\eqref{eq:heptagonletters} only $a_i$ and $a_{1i}$ are proportional to three-particle invariants, letting $F$ denote the appropriately normalized amplitudes, or any of their coproducts, or any finite integral within the space of MPLs with these alphabets, in this case they may thus concretely be expressed as\footnote{Imposing these equations in combination with the integrability conditions, to be discussed in the next section, automatically implies that they hold also with the order of letters reversed.}
\be\label{eq:FabExtStein}
\boxed{\text{Extended Steinmann Relations:}
\left\{\begin{array}{ll}
F^{a_i,a_{i+1}}=0\,,1\le i\le 3\,, &\text{for}\,\,n=6\,.\\
F^{a_{1i},a_{1i+\delta}}=0\,, \,\delta=1,2\,,\, 1\le i\le7\,,&\text{for}\,\,n=7 \,.
\end{array}\right.}
\ee
As physical quantities in perturbative quantum field theory are multivalued functions, whose different branches represent different kinematic regions, it is plausible that the extended Steinmann relations follow from the validity of the usual Steinmann relations in all branches. That is, since moving from one branch to another involves shifting functions by their discontinuities, and since this operation at the level of the symbol amounts to removing first entries, a condition between any pair of adjacent entries could be converted to the same one between the first two entries. In this manner, the bootstrap may point, for the first time, to a more general property of quantum field theory, at least in the planar limit. Indeed, the extended Steinmann relations have been confirmed to hold also for all planar two-loop five-point master integrals~\cite{Abreu:2020jxa}, though not for a family of non-planar integrals with the same external kinematics~\cite{Abreu:2021smk}.

Also in the form~\eqref{eq:FabExtStein}, it is clear that the extended Steinmann relations are contained in the cluster adjacency constraints~\eqref{eq:hns1245},\eqref{heptns} and \eqref{eq:Nonadjacent_Constraint}. What is less obvious but also true, is that for well-defined functions with physical branch cuts built out of the $n=6,7$ alphabets \eqref{eq:hexletters}-\eqref{eq:heptagonletters}, the extended Steinmann relations also automatically imply all remaining cluster adjacency restrictions. In other words there exists a quite nontrivial equivalence between the former and the latter with respect to the trancendental part of amplitudes. On the other hand, cluster adjacency also has important implications for the rational parts of amplitudes~\cite{Drummond:2018dfd,Mago:2019waa,Lippstreu:2019vug,Lukowski:2019sxw,Mago:2020eua}, as well as how these correlate with the transcendental parts, and this additional information has been very useful for bootstrapping $A^{(4)}_{7,1}$ \cite{Drummond:2018caf}. Note that the relation between symbol letters and rational parts of amplitudes is not only confined to $\cN=4$ SYM, as is has also been observed in five-gluon amplitudes in QCD~\cite{Abreu:2018zmy}.

\renewcommand{\arraystretch}{1.25}
\begin{table}[!t]
\centering
\resizebox{\textwidth}{!}{
\begin{tabular}[t]{l c c c c c c c c c c c c c c}
\hline\hline
weight $n$
& 0 & 1 & 2 & 3 & 4 &  5 &  6 &  7 &  8 &  9 & 10 & 11 & 12 & 13 \\
\hline\hline
First entry
& 1& 3 & 9 & 26 & 75 & 218 & 643 & 1929 & 5897 & ? & ? & ? & ? & ? 
\\\hline
Steinmann
& 1 & 3 & 6 & 13 & 29 & 63 & 134 & 277 & 562 & 1117 & 2192 & 4263 & 8240 & ?
\\\hline
Ext.~Stein.
& 1 & 3 & 6 & 13 & 26 & 51 & 98 & 184 & 340 & 613 & 1085 & 1887 & 3224 & 5431 
\\\hline\hline
\end{tabular}}
\caption{Dimensions of spaces of integrable symbols containing the six-particle amplitude, refined as one moves from top to bottom by successively imposing the analytic properties indicated on the left-hand side.}
\label{tab:HexSymbolDim}
\end{table}

\renewcommand{\arraystretch}{1.25}
\begin{table}[!t]
\centering
\begin{tabular}[t]{l c c c c c c c c}
\hline\hline
weight $n$ & 0 & 1 & 2 & 3 & 4 &  5 &  6 &  7  \\\hline\hline
First entry & 1 & 7 & 42 & 237 & 1288 & 6763 & ? & ?  \\\hline
Steinmann & 1 & 7 & 28 & 97 & 322 & 1030 & 3192 & 9570  \\\hline
Ext.~Stein. & 1 & 7 & 28 & 97 & 308 & 911 & 2555 & 6826 \\\hline\hline
\end{tabular}
\caption{Dimensions of spaces of integrable symbols containing the seven-particle amplitude, refined as one moves from top to bottom by successively imposing the analytic properties indicated on the left-hand side.}\label{tab:HepSymbolDim}
\end{table}

Apart from a mathematical curiosity or a formal physical property, it is also reasonable to ask what the practical significance of cluster adjacency/extended Steinmann relations is. The answer to this is that they drastically reduce the size of the function space containing the amplitude, thus making it far simpler to construct the former, and uniquely identify the latter. By now they have thus been incorporated in the construction of this function space, which we will detail in the next section, 
 but just to illustrate their power, in tables \ref{tab:HexSymbolDim} and \ref{tab:HepSymbolDim}, we provide the function space dimension as the weight increases, in comparison with the dimensions of the more redundant spaces used at earlier stages of the bootstrap.

Finally, let us  comment on the status of cluster adjacency and extended Steinmann relations at multiplicity $n\ge 8$. While cluster adjacency has been confirmed for all MHV amplitudes at $L=1,2$ with the help of the Sklyanin bracket~\cite{Golden:2019kks}, as we will see in subsection \ref{sec:TropSing} in general the symbol letters do not quite coincide with the cluster variables for $n\ge8$, and this creates several subtleties. For example a naive  application of cluster adjacency appears to be violated in certain integrals~\cite{He:2021mme}. As understood also in the latter reference, on this front it is the extended Steinmann relations that are on a firmer footing when expressed as a statement about multiple discontinuities generalizing eq.~\eqref{eq:disc_SteinmannA}, however in this case this does not simply translate into a statement about adjacent pairs in the symbol, analogous to eq.~\eqref{eq:FabExtStein}.

\section{The Steinmann Cluster Bootstrap for $\mathcal{N}=4$ SYM Amplitudes}\label{sec:HexHepFuns}

In the previous section, we described the essential characteristics of the cluster polylogarithmic function spaces containing six- and seven-particle amplitudes in $\cN=4$ SYM theory. Here, we will start subsection~\ref{sec:FirstLastEntry} by presenting certain additional analytic properties these spaces are endowed with, and in subsection~\ref{sec:FunctionConstruction} we will explain how to construct them iteratively in the weight. Subsection~\ref{sec:Limits} will then focus on how the amplitude may be singled out from this space with the help of independent information on its behavior in certain kinematic limits. Finally in subsection~\ref{sec:R62} we will apply all this knowledge in order to bootstrap the two-loop six-particle MHV amplitude $\cE^{(2}_{6,0})$. While the bootstrap has currently been applied to multiplicity $n=6,7$, whenever possible we will keep $n$ general in the presentation.

\subsection{Restrictions on the first and last symbol entries}\label{sec:FirstLastEntry}

In local perturbative quantum field theories, amplitudes can only have singularities when intermediate particles go on shell, corresponding to vanishing propagator denominators in contributing Feynman diagrams (vertices can either be constant or polynomial in the momenta). In massless theories, this is possible only when a Mandelstam invariant vanishes, and in the planar limit this is further restricted to the subset of Mandelstam invariants of cyclically adjacent momenta $s_{i,\ldots,j-1}=x^2_{ij}$ defined in eq.~\eqref{eq:Mandelstamtox}.  Given that the singularities of multiple polylogarithms are encoded in the first entry of their symbols, this implies that only the subset of letters formed exclusively out of products and ratios of the aforementioned Mandelstam variables is allowed to appear in the latter~\cite{Gaiotto:2011dt}. In $\cN=4$ SYM this subset precisely corresponds to the $n(n-5)/2$ conformal cross ratios defined in eq.~\eqref{u_def}, so concretely the \emph{first entry condition} becomes 
\be\label{eq:firstentries}
\text{First symbol entry of}\,\,\, A_{n,k}\in u_{ij}\,.
\ee
As a consequence of eq.~\eqref{xToZ}, in momentum twistor language the equivalent statement is that the first symbol entries are restricted to multiplicative dual conformal invariant combinations of the $\langle i-1 i j-1 j \rangle$ Pl\"ucker variables, and of no other Pl\"ucker variable or algebraic function thereof. In this manner, one can for example immediately identify that for the particular choices~\eqref{eq:hexletters} and \eqref{eq:heptagonletters} of $n=6,7$ alphabets we have made, \eqref{eq:firstentries} specializes to
\be\label{eq:firstentries67}
\text{First symbol entry of}\,\,\, A_{n,k}\in
\begin{cases}
a_i\,,\,\, i=1,\ldots, 3\,,& \text{for }n=6\,,\vspace{12pt}\\
a_{1i}\,,\,\, i=1,\ldots, 7\,,& \text{for }n=7\,.
\end{cases}
\ee

\setcounter{footnote}{0}

Similarly to the first entry, the last entry of the symbol of $\cN=4$ amplitudes is also constrained, this time by the $\bar Q$-equation~\cite{CaronHuot:2011kk}, which encodes how their Yangian symmetry is broken at loop level by infrared divergences. The precise form of the constraint depends on the helicity $k$, in line with the fact that beyond the $k=0$ or MHV case the trascendental functions associated to loop corrections to the amplitude are also multiplied by rational functions of helicities and momenta with tree-level origin (recall that in our normalized amplitude definition \eqref{eq:cE_def} we have divided out by the corresponding MHV rational factors). Hence, every linearly independent rational factor in the superamplitude, whose number for the first few helicity configurations has been found to be \cite{Elvang:2009wd}\footnote{Just to get a sense of these numbers, the NMHV amplitude has 5, 15, 35 and 70 NMHV components for $n=6,7,8,9$ and the N$^2$MHV one 105 and  490 components for $n=8,9$, respectively. The complete classification of these Yangian invariant rational functions, according to $n,k$ and cyclic class, has been carried out in~\cite{Arkani-Hamed:2012zlh}.},
\be
\begin{aligned}
\#\,\,&\text{linearly independent components of }A_{n,k}\,\,\text{for}\\
k=0:&\quad 1\,,\\
k=1:&\quad \binom{n-1}{4}\,,\\
k=2:&\quad \frac{(n - 5) (n - 4)^2 (n - 3)^2 (n - 2)^2 (n - 1)}{4!5!}\,,
\end{aligned}
\ee
will also come with a nontrivial transcendental function multiplying it. In general, it will be the final entries of these functions times the rational factors that are related to each other by the $\bar Q$-equation.

For simplicity, here we will thus only quote the \emph{MHV final entry condition},
\be\label{eq:finalentries}
\text{Final symbol entry of}\,\,\, A_{n,0}\in \langle i j-1 j j+1\rangle \,,
\ee
which we can again specialize to
\be\label{eq:finalentries67}
\text{Final symbol entry of}\,\,\, A_{n,0}\in
\begin{cases}
m_i,y_i\,,\,\, i=1,\ldots, 3\,,& \text{for }n=6\,,\vspace{12pt}\\
a_{2i},a_{3i}\,,\,\, i=1,\ldots, 7\,,& \text{for }n=7\,.
\end{cases}
\ee
in our choices of dual conformal invariant alphabets~\eqref{eq:hexletters} and \eqref{eq:heptagonletters}. The NMHV final entries entries may be found in~\cite{Dixon:2015iva}, \cite{Dixon:2016nkn} for $n=6,7$, and in~\cite{He:2020vob} for arbitrary $n$. 

Finally, it is worth noting that the $\bar Q$-equation in fact contains significantly more information than these final entry conditions, as it provides alternative representations for amplitudes, as integrals over a collinear limit of amplitudes with higher multiplicity and MHV degree, and lower loop order. In this manner, not only does it predict similar constraints deeper inside the symbol, such as MHV next-to-final-entry conditions~\cite{He:2021mme}, but it also offers an alternative route for the direct computation of amplitudes, as was successfully carried out in~\cite{He:2019jee,He:2020vob,Li:2021bwg}.

\subsection{Constructing the function space containing the $n$-particle amplitude}\label{sec:FunctionConstruction}
In the previous subsection we discussed the special restrictions holding at the two endpoints of the symbol, how about its remaining entries? One may be tempted to think that all possible combinations of letters or `words' of a given alphabet give rise to well-defined functions (so that for an alphabet of size $|\Phi|$ at weight $k$ we would obtain $|\Phi|^k$ such functions). However this is not true, since any well-defined function $F$ must satisfy the property that double derivatives with respect to two different independent variables $x_i, x_j$
commute,
\be
\frac{\partial^2F}{\partial x_i\partial x_j}\ -\
\frac{\partial^2F}{\partial x_j\partial x_i}=0 \,, \qquad i \neq j.
\label{eq:commdoublederiv}
\ee
When $F$ is an MPL, due to eqs.~\eqref{dFPhi} and~\eqref{dFPhiCop} this requirement implies the existence of linear relations between its double coproducts $F^{\phi_\alpha,\phi_\beta}$, 
\be
\sum_{\alpha,\beta=1}^{|\Phi|} D_{i\alpha\beta} \, F^{\phi_{\alpha},\phi_{\beta}}=0\,,\qquad i=1,2,\ldots,l\,,\label{DoubleCopMatrix}
\ee
where $D$ is a tensor with purely numeric entries, and $l$ counts the number of independent equations, which generally depends on the choice of alphabet. These equations are known as the \emph{integrability conditions}, and from eq.~\eqref{Delta11} we see that they are also equivalent to the fact that of all $|\Phi|^2$ combinations of letters, only a subset of weight-two symbols can appear at the last two coproduct slots.

Further focusing on the MPL function spaces containing $A_{n,k}$, in subsection~\ref{sec:DCIKin} we have already mentioned that the variables $x_i$ in eq.~\eqref{eq:commdoublederiv} can be chosen to be an algebraic independent subset of cross ratios, or more conveniently to coincide with the variables of any momentum twistor parametrization of the kinematics, such as the $Gr(4,n)$  $\cX$-coordinates of a given cluster. For the known six- and seven-particle alphabets, in this manner it can be shown that there exist $l=26$ equations for the $9^2=81$ double coproducts and $l=729$ equations for the $42^2=1764$ double coproducts, respectively~\cite{Dixon:2016nkn}. Their explicit form may be found in the ancillary file accompanying the \texttt{arXiv} submission of~\cite{Caron-Huot:2020bkp}, together with the extended Steinmann relations~\eqref{eq:FabExtStein}, which also assume the general form~\eqref{DoubleCopMatrix} and thus may be described by an enlarged matrix $D$.

With the knowledge of the $n$-particle alphabet as a starting point, the above properties allow us to recursively construct the space of  
(extended Steinmann/cluster adjacent) \emph{$n$-gon functions} containing the $L$-loop amplitude amplitude at weight $m=2L$, which we shall denote $\cH_{n,m}$, as follows: Given that a basis of functions on $\cH_{n,m-1}$ is known, we consider their $\{m-2,1\}$ coproduct representation \eqref{eq:Deltan1}, and attach another letter to them to the right in all possible ways. From this $\{m-2,1,1\}$ tensor product space that resembles eq.~\eqref{Delta11}, we then obtain $\cH_{n,m}$ by imposing the integrability and extended Steinmann relations \eqref{DoubleCopMatrix} on its elements. The procedure starts at  $m=1$ with the functions dictated by the first entry condition~\eqref{eq:firstentries}, and terminates at the desired weight $m=2L$, where also the final entry condition~\eqref{eq:finalentries} (for $k=0$) or its generalization (for $k>0$) may be imposed if one is interested in a particular helicity configuration.

Hence the bootstrap method simplifies amplitude computations by transforming them into linear algebra, and has so far successfully been applied to determine $A_{n,k}$ for $n=6,7$ to unprecedented loop orders, that would have been completely out of reach with traditional Feynman diagram methods~\cite{Dixon:2011pw,Dixon:2011nj,Dixon:2013eka,Dixon:2014voa,Dixon:2014iba,Drummond:2014ffa,Dixon:2015iva,Caron-Huot:2016owq,Dixon:2016nkn,Drummond:2018caf,Caron-Huot:2019vjl,Caron-Huot:2019bsq,Dixon:2020cnr}. 
The construction of $\cH_{n,m}$ is computationally the most challenging part of the bootstrap program as $m$ increases, but still the resulting systems of linear equations are many orders of magnitude smaller than, e.g. the integration-by-parts identities needed to determine the basis of master integrals at the same loop order. As first proposed and applied in~\cite{Drummond:2014ffa}, they can be most efficiently solved by finite field methods that avoid intermediate expression swell, implemented for example in software such as \texttt{IML} \cite{Chen:2005:BBC:1073884.1073899}, \texttt{SageMath} \cite{SageMath} or \texttt{SpaSM} \cite{spasm} has been used the past.  More recently, the \texttt{SymBuild}~\cite{Mitev:2018kie} package and the \texttt{FiniteFlow} framework~\cite{Peraro:2019svx}, with dedicated capabilities for constructing integrable symbols, have also been made available.

The procedure we have described works equally well for functions or for their symbols, and ensures that if $\cH_{n,m-1}$ has physical branch cuts, so will $\cH_{n,m}$, with the following exception for the case of functions: The space of solutions of the integrability conditions \eqref{DoubleCopMatrix} also contains functions such as
\be\label{eq:BadBranchFunction}
\zeta_{m-1}\log \phi_\alpha
\ee
where $\zeta_{m-1}$ is the Riemann zeta function and $\phi_\alpha$ is a letter that is not an allowed first entry \eqref{eq:firstentries}, which do not have physical branch cuts.

It is therefore necessary to eliminate such functions from our space, and one way to do so is by noting that the branch point at $\phi_\alpha=0$ also manifests itself as a pole in the derivative of the function. So ensuring that the function is analytic at $\phi_\alpha=0$ can be achieved by requiring that the corresponding residue of its derivative, or by virtue of eq.~\eqref{dFPhi} its left coproduct factor, vanishes as $\phi_\alpha\to 0$.

\setcounter{footnote}{0}

In practice, it is simpler to impose such \emph{branch cut conditions} on kinematic limits where more letters vanish simultaneously. For the purposes of this review, it will be sufficient to mention one such limit that has been used in the literature for $n=6$, the \emph{soft limit}. In the choice~\eqref{eq:hexletters} for the six-particle alphabet and in the $i$-th orientation it amounts to\footnote{This limit may also be expressed in terms of the more conventional cross ratios \eqref{eq:udef} as $u_i\to 1, u_{i-1},u_{i+1}\to0$ with $u_j/(1-u_i)$ held fixed for $j\ne i$.}
\be\label{eq:soft1}
\text{soft}_i: \,\,
\begin{gathered}
a_i\to \infty\,,\,\, a_{i-1}\to \frac{1}{a_{i+1}}\,,\\ m_{i+1}\to \frac{\sqrt{a_i}}{\sqrt{a_{i+1}}}\,,\,\, m_{i-1}\to \sqrt{a_i}{\sqrt{a_{i+1}}}\,,\,\, y_i\to 1\,,\\\text{with}\,\,\,a_{i+1}, y_{i+1}, y_{i-1}, {m_i}{\sqrt{a_i}}   \,\,\,\text{fixed}\,,\quad i=1,2,3\,,
\end{gathered}
\ee
with each of the soft limits corresponding to one square face, that also intersect the parity-even surface, of the $Gr(4,6)$ cluster polytope shown in figure \ref{A3polytope}. In these limits, functions in $\cH_{6,m}$ should additionally satisfy the following branch cut conditions \cite{Dixon:2013eka,Dixon:2015iva}
\be\label{eq:hexBranchCuts}
F^{m_i}\big|_{\text{soft}_i}=F^{y_{i-1}}\big|_{\text{soft}_i}=F^{y_{i+1}}\big|_{\text{soft}_i}=0\,,\quad i=1,2,3\,.
\ee
Analogous conditions for $n=7$ have been obtained in~\cite{Dixon:2020cnr}.

\subsection{Singling out the amplitude: Special kinematic limits}\label{sec:Limits}

Once the space of $n$-gon functions $\cH_{n,m}$ containing the $n$-particle amplitude has been constructed, the final step of the bootstrap method it to uniquely identify the latter from within this space, using information from kinematic limits where the behavior of the amplitude is already known. The simplest of these special kinematic configurations is the limit where the momenta of two consecutive external particles become collinear: Indeed, the BDS ansatz correctly captures not just the infrared singularity structure of the amplitude, but also its behavior under collinear factorization. As a consequence, in this limit the BDS-normalized amplitude smoothly reduces to the same amplitude with one leg less, for example for the MHV case this immediately carries over directly for the remainder function~\eqref{eq:R6},
\begin{equation}
\label{eq:collinearR}
\lim_{i{+}1 \parallel i} \cR^{(L)}_n = \cR^{(L)}_{n-1}\,,
\end{equation}
with the cascade terminating at $\cR^{(L)}_{5}=0$. As an example that will also be useful to us in the next subsection, let us quote here how the symbol alphabet behaves in the case of the $n=6$ collinear limit,
\be\label{eq:collinear}
\text{collinear}_i: 
\begin{gathered}
\,\,m_i\to \infty\,,\,\, m_{i-1}\to \frac{1}{m_{i+1}}\,,\\a_i\to\frac{(1+m_{i+1})^2}{m_i m_{i+1}}\,,\,\,  a_{i-1}\to {m_i}{m_{i+1}}\,,\,\,a_{i+1}\to \frac{m_i}{m_{i+1}}\,,\\
y_{i-1}\to 1\,,\quad y_{i+1}\to 1\,,\quad\text{with}\,\,\, m_{i+1}, y_{i}\,\,\,\text{fixed,}\quad i=1,2,3\,.
\end{gathered}
\ee
The three collinear limits correspond to the three edges of the $Gr(4,6)$ cluster polytope shown in figure \ref{A3polytope}, that lie on the parity-even surface.

Beyond the strict collinear limit, Mellin-Barnes-like integral representations for every term in the series expansion around an $(n-5)$-fold collinear limit may be predicted with the help of the integrability-based Wilson loop or Pentagon Operator Product Expansion (OPE)~\cite{Alday:2010ku,Basso:2013vsa,Basso:2013aha,Basso:2014koa,Basso:2014nra,Belitsky:2014sla,Belitsky:2014lta,Basso:2014hfa,Basso:2015rta,Basso:2015uxa,Belitsky:2016vyq}, to all loops.  These integral representations can then be systematically evaluated in closed form~\cite{Papathanasiou:2013uoa,Papathanasiou:2014yva,Caron-Huot:2019vjl}, thus providing direct input for the amplitude bootstrap, and in some cases the entire series expansion or certain well-defined subsector thereof may even be resummed so as to access more general kinematic configurations~\cite{Drummond:2015jea,Cordova:2016woh,Lam:2016rel,Belitsky:2017wdo,Belitsky:2017pbb,Bork:2019aud,Basso:2020xts}.

Last but not least, an excellent source of boundary kinematic data for the bootstrap is offered by the  high energy or multi-Regge kinematics (MRK), a very rich subject in its own right, which will be the focus of chapter 15~\cite{DelDuca:2022skz} of this review~\cite{Travaglini:2022uwo}. This owes to the development of an effective description of the latter by Balitsky, Fadin, Lipatov and Kuraev originally in QCD, that was later extended also to planar $\cN=4$ SYM~\cite{Bartels:2008sc,Fadin:2011we,Bartels:2011ge,Lipatov:2012gk,Dixon:2012yy,Bartels:2013jna,Dixon:2014iba,Drummond:2015jea,DelDuca:2016lad,DelDuca:2018hrv}. Interestingly, the dual conformal invariance of the theory renders it equivalent to the soft limit, so in order for the normalized amplitude to have nontrivial kinematic dependence there, it is necessary to first analytically continue away from the Euclidean region. At multiplicity $n=6$ we have already seen the soft/multi-Regge limit in eq.~\eqref{eq:soft1}, from where it becomes apparent that is natural to organize the weak coupling expansion of the amplitude also with respect to the order of the divergent logarithm, $\log^{L-p-1} a_1$, denoted as the (next-to)$^p$-leading-logarithmic (N$^p$LL) approximation. Remarkably,  this double expansion can be computed at any loop order and logarithmic approximation, not only for $n=6$~\cite{Basso:2014pla}, but also at arbitrary multiplicity~\cite{DelDuca:2019tur}, thanks to an analytic continuation connecting the multi-Regge with the near-collinear limit mentioned above. For $n=7$, these all-loop results have been recently checked against all available bootstrap data~\cite{Dixon:2021nzr}.

\subsection{A simple example: Bootstrapping the two-loop six-particle MHV amplitude}\label{sec:R62}

With all the bootstrap technology in place, let us now see it at work in a concrete example, the computation of the first nontrivial correction at $L=2$ loops (since the $L=1$ correction is by construction part of the BDS ansatz) to the six-particle amplitude in the MHV helicity configuration. This subsection is thus intended for the reader who is interested in learning how to perform actual bootstrap computations, and may otherwise be skipped.

In the first instance, we will construct the space of hexagon functions $\cH_{6,m}$ containing the amplitude and its derivatives, for $1\le m\le 2L=4$. At $m=1$, the first entry condition~\eqref{eq:firstentries67} requires this space to be
\be
\cH_{6,1}: \log a_i\,,\,\,i=1,2,3\,.
\ee
To go to higher weight, we first need to determine the integrability conditions in our choice~\eqref{eq:hexletters} for the six-particle alphabet. To this end, for a generic function $F \in \cH_{6,m}$ we compute the commutator of double derivatives with the help of the definitions~\eqref{dFPhi}-\eqref{dFPhiCop}, which generically takes the form
\begin{equation}
\sum_{\phi_\alpha,\phi_\beta\in{\Phi}} F^{\phi_\alpha,\phi_\beta} \left[\frac{\partial \log\phi_\alpha}{\partial x^i}\frac{\partial \log\phi_\beta}{\partial x^j}-\frac{\partial \log\phi_\alpha}{\partial x^j}\frac{\partial \log\phi_\beta}{\partial x^i}\right]\,,
\end{equation}
since terms with both derivatives acting on the letters $\phi$ automatically commute and cancel out. Next, we express the letters in terms of the three independent variables of any momentum twistor parametrization, such as the  $Gr(4,6)$ $\cX$-coordinates of the initial cluster, which we quoted in eq.~\ref{eq:Zweb6}, from which the term in brackets in the above equation is trivial to compute. This gives three equations for each $1\le i< j\le 3$, which should hold for any value of the $x_i$. We can therefore convert them into equations for the double coproducts $F^{\phi_\alpha,\phi_\beta}$ with purely numeric coefficients, either by collecting all terms under a common denominator and demanding that they hold separately for each coefficient of the polynomials in the numerator, or by evaluating them for sufficiently many values of the $x_i$. Explicitly, and in the shorthand notation $F^{[x,y]}=F^{x,y}-F^{y,x}$ the thus derived linear equations read
\be
\begin{aligned}
 F^{[a_1,a_2]} = F^{[a_1,m_1]}=F^{[a_1,y_1]}=F^{[a_1,y_2]}-F^{[a_1,y_3]}=F^{[m_1,y_2]}-F^{[m_1,y_3]}=0\,,\,\,\,i&=1,2,3\,,\\
  F^{[m_1,m_2]}+F^{[m_1,m_3]} =F^{[m_3,a_1]}+F^{[a_2,m_3]}+F^{[m_1,m_3]}+F^{[y_1,y_2]} =0\,,\,\,\,i&=1,2,\label{FabIntegrabilityMiddle}\\
   F^{[a_2,y_1]}+F^{[a_3,y_1]}+F^{[m_1,y_1]} = F^{[a_1,y_2]}+F^{[a_3,y_1]}+F^{[m_2,y_2]}& =0\,,\\
F^{[a_1,y_2]}+F^{[a_2,y_1]}+F^{[m_3,y_3]} = F^{[a_2,y_1]}-F^{[a_3,y_1]}-F^{[m_2,y_1]}+F^{[m_3,y_1]} &=0\,, \\
 F^{[a_1,y_2]}-F^{[a_3,y_1]}-F^{[m_1,y_2]}+F^{[m_3,y_1]} =F^{[m_2,a_1]}+F^{[a_3,m_2]}+F^{[m_1,m_2]}+F^{[y_1,y_3]}& =0\,, \\
F^{[m_1,m_2]}-F^{[y_1,y_2]}+F^{[y_1,y_3]}-F^{[y_2,y_3]} &=0\,.
\end{aligned}
\ee

\setcounter{footnote}{0}

With the six-particle integrability conditions~\eqref{FabIntegrabilityMiddle} at hand, constructing the weight-2 space of functions is now simply a matter of building an ansatz for the most general form of their $\Delta_{1,1}$ coproduct, obtained by adding another hexagon letter to $\cH_{6,1}$ in all possible ways (in compact symbol notation $\log\phi\to \phi$),
\be
\sum_{i=1}^3\sum_{j=1}^9 c_{ij} a_i \otimes \phi_j\,,
\ee
and imposing the integrability conditions simultaneously with the extended Steinmann relations~\eqref{eq:FabExtStein} on this ansatz.\footnote{Note that cluster adjacency may be exploited so as to reduce the number of initial unknowns and equations, by attaching a letter $\phi_j$ only to those functions whose final entries belong to the neighbor set \eqref{eq:hns1245}-\eqref{eq:hns1235} of $\phi_j$ \cite{Drummond:2018dfd,Drummond:2018caf}. While this increases the efficiency of the method, for the sake of simplicity we will refrain from applying it here.} In total these are 29 homogeneous linear equations, but when applied to the above ansatz only 21 of them are linearly independent, and  fix an equal number of the 27 unknowns $c_{ij}$. Thus the coefficients of the remaining six unknowns will span the allowed weight-two space, and  explicitly we find these to be
\be\label{eq:H62}
\cH_{6,2}:
\begin{cases}
-\left[(a_{i+1} a_{i-1})\otimes m_{i}\right]=-\left[(1+m_i)^2\otimes m_i\right] \to -2H_{-2}(m_i)\,,\\
a_i\otimes a_i\to \frac{1}{2}\log^2{a_i}\,,\\
\end{cases}\quad i=1,2,3\,.
\ee
In the first line we chose to express the letters using their subset $m_i$ as the independent variables,
\be
a_i=\frac{(1+m_{i+1}) (1+m_{i-1})}{1+m_i}\,.
\ee
The simple form of the $\Delta_{1,1}$ coproducts also allows us to immediately identify the corresponding functions, see in particular eq.~\eqref{eq:DeltaHPL} for the first line, indicated with arrows in the above equation. More precisely, the $\Delta_{1,1}$ coproduct is equivalent to the total differential of a function and thus specifies it up to a constant, which we are free to choose for example so as to  simplify the functional expression, as is done here. This choice is tantamount to the choice of base point for the integration of the total differential, which is usually chosen as a potentially singular point of the functions. So if we wish our basis to be independent of this choice of singular base point, it is natural to also include the transcentental constants the above functions evaluate to at these points, in this case $\zeta_2=H_{2}(1)$.

\setcounter{footnote}{0}

Moving on to weight three, to find a basis of functions we similarly form an ansatz of the six functions of eq.~\eqref{eq:H62} tensored with the nine hexagon letters, giving rise to a total of 54 unknowns. We do not need to include $\zeta_2 \log\phi_j$ in our ansatz since these functions have identically vanishing double coproducts, and hence correspond to trivial solutions of the integrability conditions and extended Steinmann relations, which we know beforehand. We then apply the aforementioned constraints on the right two slots of the $\Delta_{1,1,1}$ coproduct of the ansatz, in the first instance obtaining $29\times 3=87$ equations, since these should hold separately for each of the leftmost coproduct slots,  which are algebraically independent. However these reduce to 41 linearly independent equations, whose solution space is then found to be,\footnote{From this point on, we will employ the compact symbol notation only to the right factor of the $\Delta_{m-1,1}$ coproduct, and otherwise retain the complete functional form of the left factor.}
\be\label{eq:H63}
\cH_{6,3}:
\begin{cases}
2H_{-2}(m_i)\otimes m_i \to 2H_{-3}(m_i)\,,\\
-2H_{-2}(m_i)\otimes  (1+m_i)^2 \to -4H_{-1,-2}(m_i)\,,\\
\frac{1}{2}\log^2{a_i}\otimes a_i\to \frac{1}{6}\log^3 a_i\,,\\
-H_{-2}(m_i)\otimes \frac{a_{i-1}}{a_{i+1}}-\frac{1}{2}(\log^2 a_{i-1}-\log^2 a_{i+1})\otimes m_i \to -H_{-2}(m_i)\log\frac{a_{i-1}}{a_{i+1}}\,,\\
\tilde F\equiv {\displaystyle \sum_{i=1}^3}\left[\frac{1}{2}(\log^2 a_{i-1}+\log^2 a_{i+1})-2H_{-2}(m_1)-2H_{-2}(m_2)-2H_{-2}(m_3)\right]\otimes y_i\,,
\end{cases}.
\ee
with $i=1,2,3$. Notice that unlike all functions encountered thus far, $\tilde F$ is parity odd, owing to the appearance of the  parity-odd letters $y_i$ for the first time. Of the remaining even functions, only the last one is not immediately identifiable with the help of the usual HPL definitions, and requires expressing the difference of the squares of logarithms as a product thereof, using the $m_i$ as the independent variables, and applying the shuffle algebra relations mentioned in subsection~\ref{sec:MPLs}.

At this stage all functions are defined modulo $\zeta_2 \log\phi_j$, the trivial solution we chose not to include in our ansatz. This ambiguity may be fixed by further imposing the branch cut conditions \eqref{eq:hexBranchCuts}, which were automatically satisfied at lower weight since no analogous ambiguity existed. While it is easy to check that all even $\cH_{6,3}$ functions in eq.~\eqref{eq:H63} satisfy them, for $\tilde F$ we find that in the soft limits~\eqref{eq:soft1}
\be
\tilde F^{y_{i-1}}\big|_{\text{soft}_i}=\tilde F^{y_{i+1}}\big|_{\text{soft}_i}=-4\zeta_2\,.
\ee
The only slightly nontrivial step needed to show this, and more generally to evaluate all kinematic limits we will consider later in this subsection, are HPL $x\to 1/x$ argument inversion identities, which for example can be obtained with the package \texttt{HPL}~\cite{Maitre:2007kp}. From the above equation, it is clear that in order to ensure that $\tilde F$ has good branch cuts, we need to redefine it as
\be
\tilde F\equiv\sum_{i=1}^3\left[\frac{1}{2}(\log^2 a_{i-1}+\log^2 a_{i+1})-2H_{-2}(m_1)-2H_{-2}(m_2)-2H_{-2}(m_3)+4\zeta_2\right]\otimes y_i\,,
\ee
and now the only ambiguity remaining in its definition is its value at a point. This may then be fixed by picking this point anywhere on the parity-invariant surface, where by definition any parity-odd function vanishes. In our choice of basis, $\tilde F$ is in fact equal to twice the transcendental part of the six-dimensional hexagon integral studied in~\cite{Dixon:2011ng}, where also an explicit expression of the latter in terms of classical polylogarithms may be found. Finally, in $\cH_{6,3}$ we may additionally include the part of the trivial solution which obviously also satisfies the branch cut conditions, $\zeta_2 \log a_i$,  as well as $\zeta_3$, by the same reasoning that led to the inclusion of $\zeta_2$ at one weight lower.

Arriving at weight 4, since our goal here is MHV amplitude, which is parity even and obeys the final entry condition~\eqref{eq:finalentries67}, we may simplify the calculation by incorporating these constraints directly in our initial ansatz. Namely we tensor the even $\cH_{6,2}$ functions with $m_i$, and $\tilde F $ with $y_i$. For this subspace of $\cH_{6,4}$, which we shall denote $\cH^{+,\text{MHV}}_{6,4}$, we find the basis
\be\label{weight4}
\cH^{+,\text{MHV}}_{6,4}:
\begin{cases}
-2H_{-3}(m_i)\otimes m_i \to -2H_{-4}(m_i)\,,\\
4H_{-1,-2}(m_i)\otimes  m_i \to 4H_{-2,-2}(m_i)\,,\\
\hat\Omega^{(2)}_i\,,\\
-2\zeta_2 H_{-2}(m_i)\,,
\end{cases}
\quad i=1,2,3\,,
\ee
where 
\begin{align}
\Delta_{3,1}\hat\Omega^{(2)}_i=&\hat\Omega^{(2),m_i}_{i}\otimes m_i+\hat\Omega^{(2),m_{i+1}}_{i}\otimes m_{i+1}-\tilde F\otimes (y_i y_{i+1})\,,\label{Omega2}\\
\hat \Omega^{(2),m_{i}}_{i}\equiv& -4 H_{-3}\left(m_i\right)+2\left[ H_{-1,-2}\left(m_i\right)+ H_{-1,-2}\left(m_{i+1}\right)- H_{-1,-2}\left(m_{i-1}\right)\right]+\frac{1}{3} \log ^3a_{i+1}
\\
&- H_{-2}\left(m_{i-1}\right)\log \tfrac{a_{i+1}}{a_i}+ H_{-2}\left(m_i\right)\log \tfrac{a_{i-1}}{a_{i+1}}- H_{-2}\left(m_{i+1}\right)\log \tfrac{a_i}{a_{i-1}}+4 \zeta_2 \log a_{i+1}\nonumber\\
\hat \Omega^{(2),m_{i+1}}_{i}=&\hat\Omega^{(2),m_{i}}_{i}\Big|_{\substack{m_j\leftrightarrow m_{i+1}\\ a_i\leftrightarrow a_{i+1}}}\,.
\end{align}
Clearly, $\hat \Omega^{(2)}_i$ is symmetric under exchange of letters with indices $i\leftrightarrow (i+1)$. Recalling that $\tilde F$ vanishes in the parity-invariant surface containing (intersecting) the collinear (soft) limit, and using HPL argument inversion identities as before, it's a straightforward exercise to show that the above functions already satisfy the branch cut conditions~\eqref{eq:hexBranchCuts}, and thus require no further modification. As with lower weights, we also include the constant $\zeta_4$ in $\cH^{+,\text{MHV}}_{6,4}$.

Therefore the only task remaining in order to fully specify our $\cH^{+,\text{MHV}}_{6,4}$ basis is the value of $\hat \Omega^{(2)}_i$ at a point. We can do this by noting that its $\Delta_{3,1}$ coproduct vanishes in the $m_{i-1}\to\infty$ orientation of the collinear limit~\eqref{eq:collinear}, therefore it reduces to a constant which is natural to also set to zero. This choice in fact renders $\hat \Omega^{(2)}_i$ four times the double pentagon integral $\Omega^{(2)}$, which we will encounter again in section~\ref{sec:IntegralBootstrap},
in its three possible orientations.

Having fully specified $\hat \Omega^{(2)}_i$ in this manner, it is not difficult to similarly obtain its other two nontrivial collinear limits, which we will need later on, from the coproduct representation \eqref{Omega2}. In the $m_{i}\to\infty$ orientation in particular we find,
\begin{align}
\Delta_{3,1}\hat\Omega^{(2)}_i\to&4 \Big[\left(H_{-3}\left(m_{i+1}\right)+H_{-2,0}\left(m_{i+1}\right)-2 H_{-2,-1}\left(m_{i+1}\right)+\log m_i\, H_{-2}\left(m_{i+1}\right)\right)\otimes m_i\nonumber\\
& +  \Big(H_{-2,0}\left(m_{i+1}\right) -2 H_{-2,-1}\left(m_{i+1}\right)-H_{-1,-2}\left(m_{i+1}\right)+4 H_{-1,-1,-1}\left(m_{i+1}\right)\nonumber\\
&-2 H_{-1,-1,0}\left(m_{i+1}\right)-2 \log m_i (H_{-1,-1}\left(m_{i+1}\right)+ H_{-1,0}\left(m_{i+1}\right)+ H_{-2}\left(m_{i+1}\right))\nonumber\\
&+ \left(\tfrac{1}{2}\log ^2m_i+\zeta_2\right) H_{-1}\left(m_{i+1}\right)\Big)\otimes m_{i+1}\Big]\,.
\end{align}
This expression can be trivially integrated to yield $\hat\Omega^{(2)}_i$ up to a constant, since it is equivalent to an ordinary differential equation  for the function with respect to the only surviving finite variable $m_{i+1}$. That is, focusing on the coproduct component in question, eq.~\eqref{eq:DeltaHPL} allows us to replace
\be
H_{\vec l}(m_{i+1})\otimes m_{i+1}\to H_{0,\vec l}(m_{i+1})+c\,.
\ee
The integration constant is then fixed by the fact that $\hat\Omega^{(2)}_i$ should vanish at the $m_{i+1}\to0$ endpoint of the $m_{i}\to\infty$ collinear line: Indeed, this point is an overlap with the $m_{i-1},m_{i}\to\infty$, $m_{i+1}\to 0$ soft limit, and the latter in turn also overlaps with the $m_{i-1}\to\infty$ collinear limit, where as we have seen  the function vanishes. In this manner, we finally obtain
\begin{align}
\frac{\hat\Omega^{(2)}_i}{4}\xrightarrow[m_{i}\to\infty]{\text{coll.}}&  H_{-3,0}\left(m_{i+1}\right) -2 H_{-3,-1}\left(m_{i+1}\right)-H_{-2,-2}\left(m_{i+1}\right)+4 H_{-2,-1,-1}\left(m_{i+1}\right)\nonumber\\
&-2 H_{-2,-1,0}\left(m_{i+1}\right)-2 \log m_i \left[H_{-2,-1}\left(m_{i+1}\right)+ H_{-2,0}\left(m_{i+1}\right)+ H_{-3}\left(m_{i+1}\right)\right]\nonumber\\
&+ \left(\tfrac{1}{2}\log ^2m_i+\zeta_2\right) H_{-2}\left(m_{i+1}\right)\,,\label{OmegaColl1}
\end{align}
whereas the third collinear limit orientation follows for free by exploiting the flip symmetry of the function so as to replace $m_{i}\leftrightarrow m_{i+1}$ in the above formula.

Now that the hard part of constructing the function space containing the amplitude is over, all that is left to determine the latter is to form an ansatz from all the basis functions, and determine the coefficients by comparing it to special kinematic limits where we have independent information on the behavior of the amplitude. Taking into account the dihedral symmetry of the BDS-like normalized six-particle amplitude, our initial ansatz contains merely five unknowns,
\be\label{eq:E62ansatz}
\cE_{6,0}^{(2)}=c_1\sum_{i=1}^3 [-2H_{-4}(m_i)]+c_2\sum_{i=1}^3 4H_{-2,-2}(m_i)+c_3\sum_{i=1}^3 \hat \Omega^{(2)}_i+c_4\, \sum_{i=1}^3 [-2\zeta_2 H_{-2}(m_i)]+c_5 \zeta_4\,,
\ee
where we remind the reader that the function $\hat \Omega^{(2)}_i$ is defined by its $\Delta_{3,1}$ coproduct~\eqref{Omega2} and the fact that it vanishes in the $w_{i-1}\to\infty$ collinear limit~\eqref{eq:collinear}.

In order to fix the coefficients of the ansatz we will also consider the collinear limit, where eqs.\eqref{eq:R6},\eqref{Gcusp} and \eqref{eq:collinearR} imply that at this loop order the amplitude has the simple behavior
\be\label{eq:E62Coll}
\cE_{6,0}^{(2)}\xrightarrow{\text{collinear}} \frac{1}{2}\left(\cE_{6,0}^{(1)}\right)^2\,.
\ee
More precisely, given that our ansatz has already taken dihedral symmetry into account, we may focus on a single orientation, say $m_3\to \infty$, where the one-loop correction~\eqref{E61def} reduces to
\be
\cE_{6,0}^{(1)}\xrightarrow[m_{3}\to\infty]{\text{collinear}}-\frac{1}{2} \log ^2m_1-\frac{1}{2} \log ^2 m_3-2 \zeta_2\,.
\ee
The final step is to also evaluate our ansatz in the limit. For $\hat \Omega^{(2)}_{i}$ this has already been done in eq.~\eqref{OmegaColl1}, and for the other functions we proceed similarly. Recalling also the MPL identity~\eqref{eq:Gzeros}, after the dust settles the difference of the right-hand sides of eqs.~\eqref{eq:E62Coll} and \eqref{eq:E62ansatz}, evaluated on the $m_3\to\infty$ collinear limit will be a sum of functions
\be
\log^i (m_3) \,\zeta_j H_{-l_1,-l_2,\ldots -l_k}(m_1)\,,
\ee 
of total weight four, with coefficients depending on the unknowns $c_i$. Since each of these functions is algebraically independent, their coefficients should vanish separately, and solving this set of equations yields the unique solution
\be\label{eq:E63cSol}
c_1= -2\,,\quad c_2=-\frac{1}{4}\,,\quad c_3=\frac{1}{4}\,,\quad c_4= 1\,, \quad c_5= 8\,.
\ee
Congratulations, you have just bootstrapped the two-loop correction to the planar six-particle amplitude, or equivalently lightlike hexagon Wilson loop in planar $\cN=4$ SYM theory! Its initial computation in terms of Feynman diagrams required a rather nontrivial effort, and resulted in a 17-page long sum of multiple polylogarithms, that is equivalent to the expressions \eqref{eq:E62ansatz},\eqref{eq:E63cSol}.

\renewcommand{\arraystretch}{1.25}
\begin{table}[!t]
\centering
\def\sp{@{{\,}}}
\begin{tabular}[t]{l\sp\sp c\sp\sp c\sp\sp c\sp\sp c\sp\sp c\sp\sp c\sp\sp c\sp\sp c}
\hline
Constraint                      & $L=1$\, & $L=2$\, & $L=3$\, & $L=4$\, & $L=5$\, & $L=6$
\\\hline
1. $\cH_6$ & 6 & 27 & 105 & 372 & 1214 & 3692?\\\hline\hline
2. Symmetry       & (2,4) & (7,16) & (22,56) &  (66,190) & (197,602)& (567,1795?)\\\hline
3. Final-entry       & (1,1) & (4,3) & (11,6) &  (30,16) & (85,39) & (236,102)\\\hline
4. Collinear       & (0,0) & (0,0) & $(0^*,0^*)$ &  $(0^*,2^*)$ & $(1^{*3},5^{*3})$ & $(6^{*2},17^{*2})$ \\\hline
5. LL MRK       & (0,0) & (0,0) & (0,0) &  (0,0) & $(0^*,0^*)$ & $(1^{*2}$,$2^{*2})$ \\\hline
6. NLL MRK       & (0,0) & (0,0) & (0,0) &  (0,0) & $(0^*,0^*)$ & $(1^*,0^{*2})$ \\\hline
7. NNLL MRK       & (0,0) & (0,0) & (0,0) &  (0,0) & (0,0) & $(1,0^*)$ \\\hline
8. N$^3$LL MRK       & (0,0) & (0,0) & (0,0) &  (0,0) & (0,0) & (1,0) \\\hline
9. Full MRK       & (0,0) & (0,0) & (0,0) &  (0,0) & (0,0) & (1,0) \\\hline
10. $T^1$ OPE       & (0,0) & (0,0) & (0,0) &  (0,0) & (0,0) & (1,0) \\\hline
11. $T^2$ OPE       & (0,0) & (0,0) & (0,0) &  (0,0) & (0,0) & (0,0) \\\hline
\end{tabular}
\caption{Remaining parameters in the ans\"{a}tze for
the (MHV, NMHV) amplitude after each constraint is applied,
at each loop order. The superscript ``$*$'' (``$*n$'') denotes an additional
ambiguity ($n$ ambiguities) which arises only due to lack of knowledge
of the cosmic normalization constant $\rho$ at the given stage. The ``$?$'' indicates an ambiguity about
the number of weight 12 odd functions that are ``dropouts''; they
are allowed at symbol level but not function level. From ref.~\cite{Caron-Huot:2019vjl}.}
\label{tab:full}
\end{table}

Let us close this section with a few remarks on how the bootstrap ideas we have presented in this example apply more generally.
\begin{itemize}
\item For the hexagon bootstrap we have considered here, it is possible to further reduce the size of $\cH_{6,m}$, and thus to facilitate the identitication of the amplitude, by only including the constants $\zeta_{2n}$ with $n\ge 2$ as independent functions. This requires a further modification of the amplitude normalization by a coupling-dependent constant, dictated by what is known as a (cosmic Galois) coaction principle~\cite{Caron-Huot:2019bsq}, see also the review~\cite{Caron-Huot:2020bkp}, initially carried out order by order in perturbation theory, and later conjectured to all loops in~\cite{Basso:2020xts}. The number of functions remaining after applying consecutive constraints, so as to uniquely determine the amplitude together with this `cosmic' normalization through six loops, is summarized in table~\ref{tab:full}. The difference between our count of 5 unknowns in our ansatz~\eqref{eq:E62ansatz} and the 4 unknowns quoted for the MHV case in the $L=2$ column and final-entry row, is precisely due to our redundant inclusion of $\zeta_2$ as an independent constant. Though it does not appear in the table, $\cE_{6,0}^{(7)}$ has also been determined in~\cite{Caron-Huot:2019vjl}, and $\cE_{6,1}^{(7)}$ is also known~\cite{DDunpublished}.

\item Similarly to what we did for the more complicated functions encountered in our example, in general it proves more economical to recursively represent each weight-$m$ MPL in terms of its differential  or $\Delta_{m-1,1}$ coproduct, together with its value at a point~\cite{Dixon:2013eka}. As is discussed in chapter 3~\cite{Abreu:2022mfk} of the SAGEX review~\cite{Travaglini:2022uwo}, this is not a restriction however, since explicit $G$-function representations may be found algorithmically when there exists a choice of variables such that the symbol letters are rational functions thereof. This is even simpler when these functions are further restricted to be linear, where algorithmic integration via \emph{fibration bases} \cite{Brown:2009qja,Bogner:2014mha}, see also~\cite{Ablinger:2014yaa,Anastasiou:2013srw}, has been implemented in the software packages \texttt{HyperInt}~\cite{Panzer:2014caa}, \texttt{MPL}~\cite{Bogner:2015nda} and \texttt{PolyLogTools}~\cite{Duhr:2019tlz} (with the first of the three in fact based on a further refinement of this algorithm).

\item Building on the aforementioned coproduct representation,  a significant efficiency upgrade that becomes necessary at higher loops is to encode all the information on integrable symbols and functions, as well as of the equations needed to construct them, in terms of tensors with purely numeric entries~\cite{Dixon:2016nkn}, see also~\cite{Caron-Huot:2019bsq}. The latter tensor has already appeared when expressing the integrability and Steinmann relations as we did in eq.~\eqref{DoubleCopMatrix}, and the former tensor simply relates a function basis element at weight $m$ to a basis element at weight $m-1$ as well as to the position of a letter of an ordered alphabet. The advantage of this approach is that it not only provides the most compact way of storing all function data, but most importantly that it reduces the iterative construction of the function space exclusively to matrix operations. At the level of the symbol, it has been implemented in the package~\texttt{SymBuild}\cite{Mitev:2018kie}.
\end{itemize}

\section{New Frontiers}\label{sec:NewFrontiers}

\subsection{$n>7$ singularities from tropical Gra\ss mannians}\label{sec:TropSing}
In the previous sections, we have seen the spectacular success of the bootstrap program in determining scattering amplitudes in planar $\cN=4$ SYM theory at multiplicity $n=6,7$. A key idea is that given a finite set of singularities, or more precisely symbol letters, the space of polylogarithmic functions containing the amplitude is also finite at each loop order, and thus one can efficiently construct it and single out the actual amplitude. For the aforementioned multiplicities, this set of symbol letters  exactly matches the variables of a $Gr(4,n)$ cluster algebra, thus lending support to the expectation that it should remain stable as the loop order increases.

\begin{figure}
    \centering
    \begin{tikzpicture}[scale=1.5,baseline={([yshift=-.5ex]current bounding box.center)}]
        \node[fill=black,circle,draw=black, inner sep=0pt,minimum size=7pt] at (-2,0) {};
        \node[fill=black,circle,draw=black, inner sep=0pt,minimum size=7pt] at (-2,-1) {};
        \node[fill=black,circle,draw=black, inner sep=0pt,minimum size=7pt] at (-1,-1) {};
        \node[fill=black,circle,draw=black, inner sep=0pt,minimum size=7pt] at (-1,0) {};
        \draw[line width=0.4mm] (-2,0) -- (-2,-1) -- (-1,-1) -- (-1,0) -- cycle;
        \draw[line width=0.4mm] (-2.6,-1.2) -- (-2,-1);
        \draw[line width=0.4mm] (-2,-1) -- (-2.2,-1.6);
        \draw[line width=0.4mm] (-1,-1) -- (-0.4,-1.2);
        \draw[line width=0.4mm] (-1,-1) -- (-0.8,-1.6);
        \draw[line width=0.4mm] (-2.6,0.2) -- (-2,0);
        \draw[line width=0.4mm] (-2.2,0.6) -- (-2,0);
        \draw[line width=0.4mm] (-0.8,0.6) -- (-1,0);
        \draw[line width=0.4mm] (-1,0) -- (-0.4,0.2);
        \node at (-0.2,0.2) {$j$};
        \node at (-0.8,-1.8) {$k$};
        \node at (-2.8,-1.2) {$l$};
        \node at (-2.2,0.8) {$i$};
         \draw[loosely dotted,thick] (-2.4,-1.2) arc (197:254:0.35);
         \draw[loosely dotted,thick] (-0.8,-1.4) arc (287:344:0.35);
         \draw[loosely dotted,thick] (-2.2,0.4) arc (107:164:0.35);
         \draw[loosely dotted,thick] (-0.6,0.2) arc (17:74:0.35);
    \end{tikzpicture}
    \caption{The one-loop four-mass box formed by $n\ge 8$ cyclically ordered momentum twistors, distributed in the four corners as indicated by four of their labels.}
    \label{fig:1L4mBox}
\end{figure}
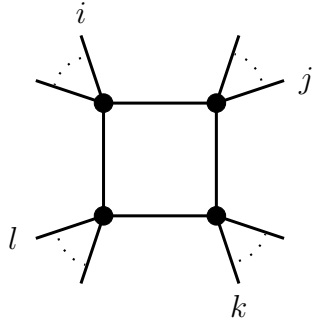

However even in the ideal setting of the simplest interacting gauge theory, the following significant conceptual and practical challenges prevented the application of the bootstrap in order to efficiently compute amplitudes at higher multiplicity $n$ in general kinematics:
\begin{enumerate}
\item $Gr(4,n)$ cluster algebras with $n\ge 8$ become infinite~\cite{2003math.....11148S}, and thus provide no predictability on what the symbol alphabet should be.
\item By construction, cluster $\cA$-coordinates are rational functions of the Pl\"ucker coordinates $\ab{ijkl}$.\footnote{This is a direct consequence of the mutation rule~\eqref{eq:mutation} or~\eqref{equ:clusterMutation}, although in fact all denominators cancel, and in the end $\cA$-coordinates simplify to homogeneous polynomials of the four-brackets or Pl\"ucker coordinates, as seen e.g. in eq.\eqref{brbilinear}.} Yet for $n\ge 8$, symbol letters that also contain square roots thereof are known to appear, and can hence not be captured by cluster algebras.
\end{enumerate}
A prototypical example of an integral yielding square-root letters is the one-loop four-mass box depicted in figure \ref{fig:1L4mBox}, which in particular contains $\sqrt{\Delta_{ijkl}}$ with
\be\label{Delta4mBox}
\Delta_{ijkl}\equiv(f_{ij} f_{kl} - f_{ik} f_{jl} + f_{il} f_{jk})^2-4 f_{ij} f_{jk} f_{kl} f_{il}\,,\quad f_{ij} \equiv \langle i\,i{+}1\,j\,j{+}1\rangle\,,
\ee
see for example~\cite{Bourjaily:2013mma}. While one could hope that individual Feynman diagram contributions would cancel out so as to yield a simpler result for the amplitude, it can be shown that this is the only diagram contributing to a particular component of $A^{(1)}_{8,2}$~\cite{Henn:2018cdp}.

\begin{figure}
\begin{center}
\includegraphics[width=0.7\textwidth]{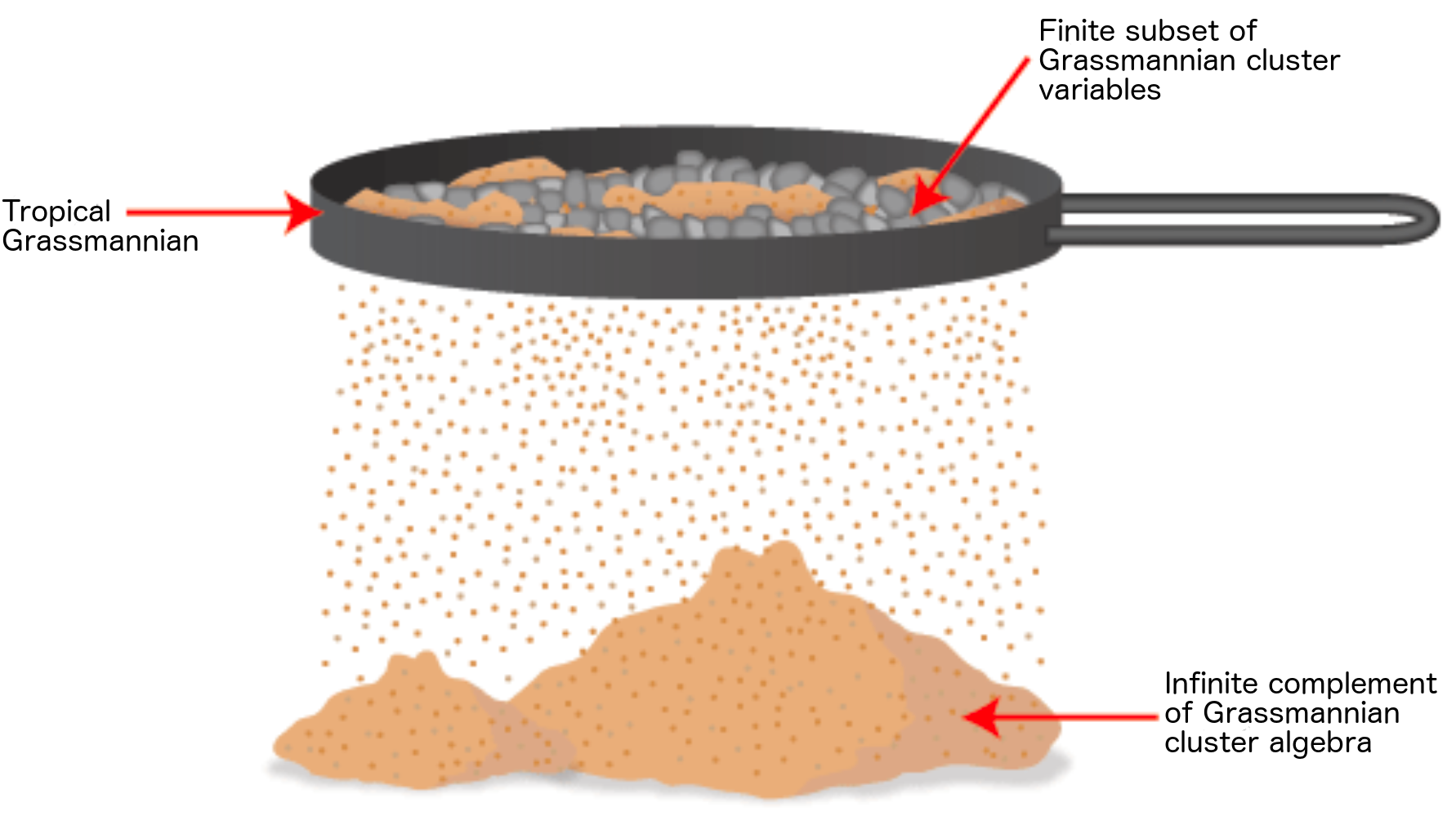}
\end{center}
\caption{Rough sketch of how a finite subset of variables of $Gr(k,n)$ cluster algebras may be selected with the help of their associated positive tropical Gra\ss mannians.} 
\label{fig:sifting}
\end{figure}

Very similar natural resolutions of these longstanding problems were simultaneously proposed in~\cite{Drummond:2019cxm,Arkani-Hamed:2019rds,Henke:2019hve} based on the relation of cluster algebras with geometric objects known as positive tropical Gra\ss mannians~\cite{2003math......4218S,Speyer2005}, or equivalently their duals as constructed by `stringy canonical form' integrals~\cite{Arkani-Hamed:2019mrd}. In the first instance, this resolution, which we will describe in more detail in the next subsections, may be pictorially represented as in figure \ref{fig:sifting}. It boils down to explicit predictions for the $n=8$ alphabet, and more recently it has been generalized in principle to any $n$, and in practice to $n=9$~\cite{Henke:2021ity}. These predictions are in agreement with all currently known data for amplitudes at these multiplicities~\cite{He:2019jee,He:2020vob,Li:2021bwg} and for $n=8$ they are also backed by a related but distinct approach based on scattering diagrams and wall-crossing~\cite{Herderschee:2021dez}\footnote{It has been observed that an alternative means for reproducing the square-root letters found in the known data, is by  solving polynomial equations associated to certain plabic graphs~\cite{Mago:2020kmp,He:2020uhb,Mago:2021luw}. As soon as one attempts to also incorporate rational letters in this approach, however, non-plabic graphs are required as well~\cite{Mago:2020nuv}. In this case the solution space includes all cluster variables of $G(4,n)$, that is the alphabet becomes infinite again.}. They thus pave the way for bootstrapping new results.

\subsubsection{The positive tropical Gra\ss mannian.}

\begin{figure}
\begin{center}
\begin{minipage}{0.45\textwidth}
{\includegraphics[width=0.6\textwidth]{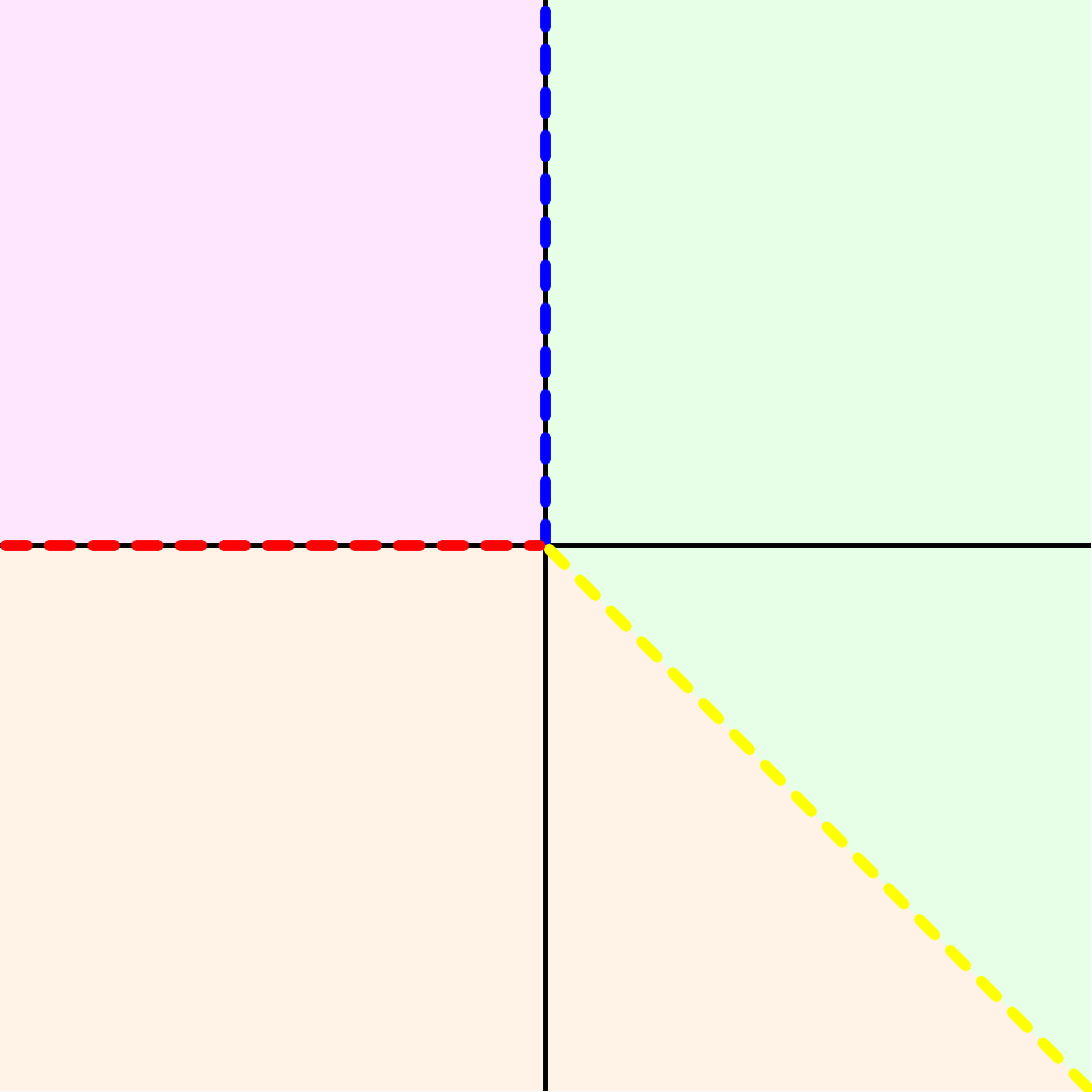}}
\end{minipage}
\begin{minipage}{0.45\textwidth}
{\includegraphics[width=0.7\textwidth]{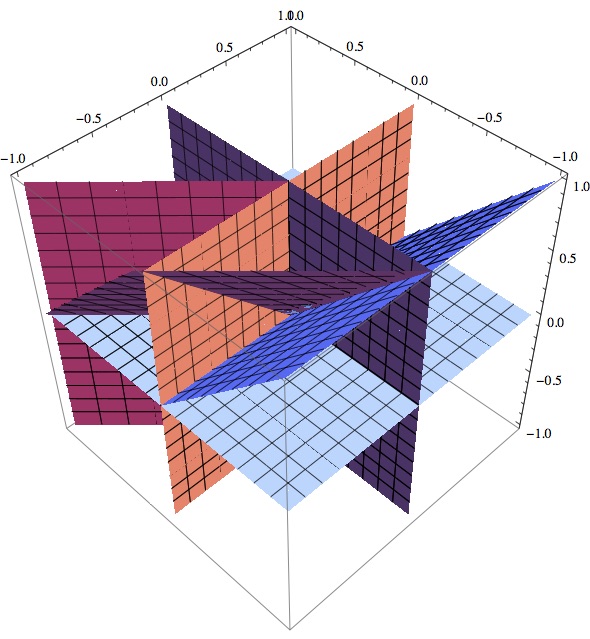}}
\end{minipage}
\end{center}
\caption{Left: The $\ab{1346}$ tropical hypersurface, given by the half-planes (in the transverse $x_3$ direction)  $(x_1=0, x_2>0)$, $(x_1=-x_2, x_2<0)$, $(x_2=0, x_1<0)$ where $\min(0,x_1,x_1+x_2)$ is attained twice simultaneously. From ref.\cite{Henke:2019hve}. Right: The $Tr_+(4,6)$ fan.}
\label{fig:ab1346Fan}
\end{figure}

Let us begin by defining this object, before explaining how it leads to finite alphabet predictions. The simplest way to define the positive part~\cite{Speyer2005} of the tropical Gra\ss mannian~\cite{2003math......4218S} $Tr_+(k,n)$, is by first expressing all $Gr(k,n)$ Pl\"ucker variables in terms of the $\cX$-coordinates of the initial cluster of the corresponding cluster algebra, which can be constructed with the web algorithm presented in the former paper. Then, one \emph{tropicalizes} this parametrization of the Pl\"ucker variables, which practically means one replaces
\be
\text{Tropicalization}:\quad \begin{tabular}{rcl}
addition& $\longrightarrow$& minimum\\
multiplication& $\longrightarrow$& addition\\
$\mathbb{C}^*$ constants & $\longrightarrow$& 0\\
0& $\longrightarrow$& $\infty$
\end{tabular}
\ee
We have already seen the $Gr(4,6)$ $\cX$-coordinate parametrization in eq.\eqref{eq:Zweb6}, so to be concrete let's tropicalize the Pl\"ucker variable shown in eq.\eqref{eq:ab1346},
\be
\ab{1346}=1+x_1+x_1 x_2\to \min(0,x_1,x_1+x_2)\,.
\ee
The \emph{tropical hypersurface} for any such tropicalized polynomial is the $(d-1)$-dimensional surface in $\mathbb{R}^d$ where the minimum is attained twice simultaneously\footnote{For $Gr(k,n)$, $d=(k-1)(n-k-1)$, but from this point onwards we will specialize to $k=4$.}, see the left-hand side of figure \ref{fig:ab1346Fan} for the case of $\ab{1346}$. Since the (in)equalities enforcing these are invariant under rescaling, we may equivalently describe the hypersurface with integer-valued vectors
\be\label{eq:ab1346Rays}
g=(-1,0,0)\,\quad g'=(0,1,0)\,,\quad g''=(1,-1,0)\,.
\ee

The \emph{positive tropical Gra\ss mannian} $Tr_+(4,n)$ is then defined as the union of tropical hypersurfaces for all $\ab{ijkl}$. Being a solution of linear (in)equalities, it is inherently finite-dimensional. Its building blocks are 1-dimensional intersections of tropical hypersurfaces emanating from the origin, known as \emph{rays}. A positive span of certain sets of rays then yields the regions in $\mathbb{R}^d$ bounded by the tropical hypersurfaces, where all tropicalized $\ab{ijkl}$ are continuous. These are the \emph{cones}, and the set of all cones then forms a (tropical) \emph{fan}. The full $Tr_+(4,6)$ fan is illustrated on the right-hand side of figure \ref{fig:ab1346Fan}.

One may also consider further generalizations of $Tr_+(4,n)$ where any subset of $\ab{ijkl}$ (or in fact any $\cA$-coordinate of the corresponding cluster algebra) is tropicalized, leading to different fans but with similar properties with what we have described above. Since $Tr_+(4,n)$ is not invariant under parity, which is a symmetry of the MHV amplitudes, a particularly natural choice is to tropicalize the maximal parity-invariant subset of Pl\"ucker variables $\ab{i-1i j-1j}, \ab{ij-1jj+1}$. It is therefore this choice of partial tropicalization  of the positive Gra\ss mannian that we will adopt from now on, and we will denote it as $pTr_+(4,n)$. Further choices and their implications are discussed in~\cite{Drummond:2020kqg}.

\subsubsection{A tropical sieve for rational letters.}
Now that we have defined $(p)Tr_+(4,n)$, we can describe their relation to cluster algebras. Starting with the $pTr_+(4,6)=Tr_+(4,6)$ fan to the right of figure \ref{fig:ab1346Fan}, careful observation reveals that it is dual to the $Gr(4,6)$ cluster polytope shown in figure \ref{A3polytope}! This can be seen by drawing a vertex inside each cone, and connecting them with lines if they are separated by a plane. More rigorously, it can be proven that any cluster $\cA$-coordinate $a$ may be uniquely written as~\cite{CAIV}
\begin{equation}
	\label{eq:AVarRayForm}
	a = \prod_{i=1}^d a_{i}^{g_i} \cdot \frac{F\left(x_{1},\dots,x_{d}\right)}{F_{\mathbb{T}}\left(y_{1},\dots,y_{d}\right)}\,,
\end{equation}
where $a_i, x_i$ are the $\cA$-coordinates and $\cX$-coordinates of the initial cluster, and the precise definition of the so called coefficients $y_i$ and (tropical) $F_{(\mathbb{T})}$-polynomial will not be important for our purposes. What matters in the above formula is that each $\cA$-coordinate is in one-to-one correspondence with an integer vector $(g_1,\ldots,g_d)$ that defines a ray similarly to eq.~\eqref{eq:ab1346Rays}, each cluster then defines a cone spanned by the rays of its $\cA$-coordinates, and finally the entire cluster algebra defines a \emph{cluster fan}.

\begin{figure}
	\centering
	\begin{minipage}{0.3\textwidth}
		\includegraphics[width=1\textwidth]{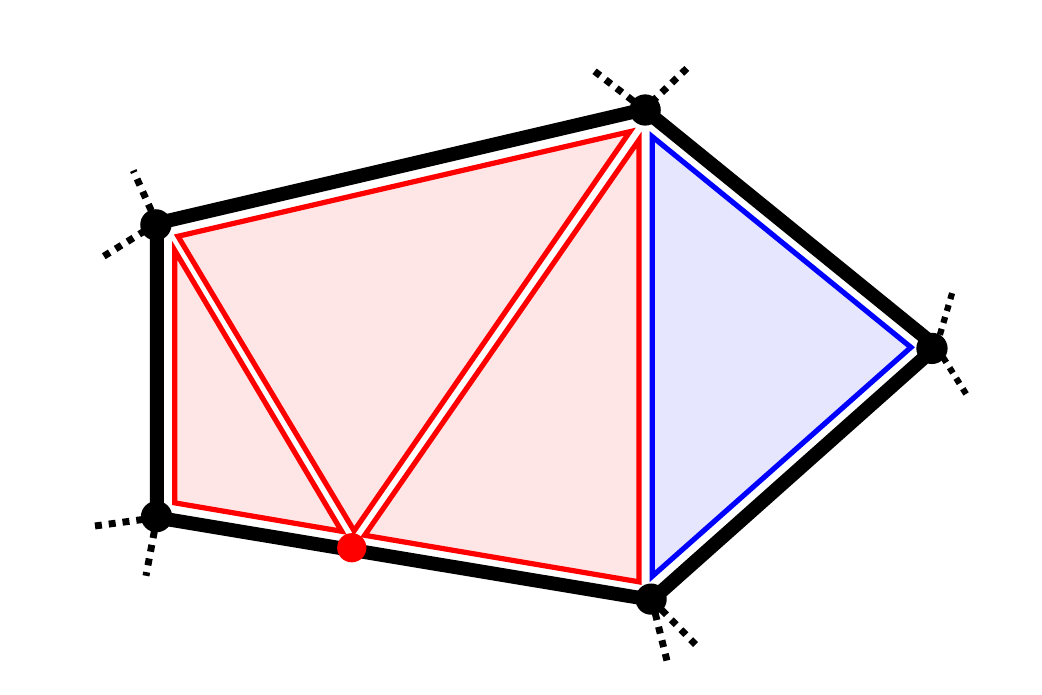}
	\end{minipage}
	\hskip 32pt
	\begin{minipage}{0.3\textwidth}
		\includegraphics[width=1\textwidth]{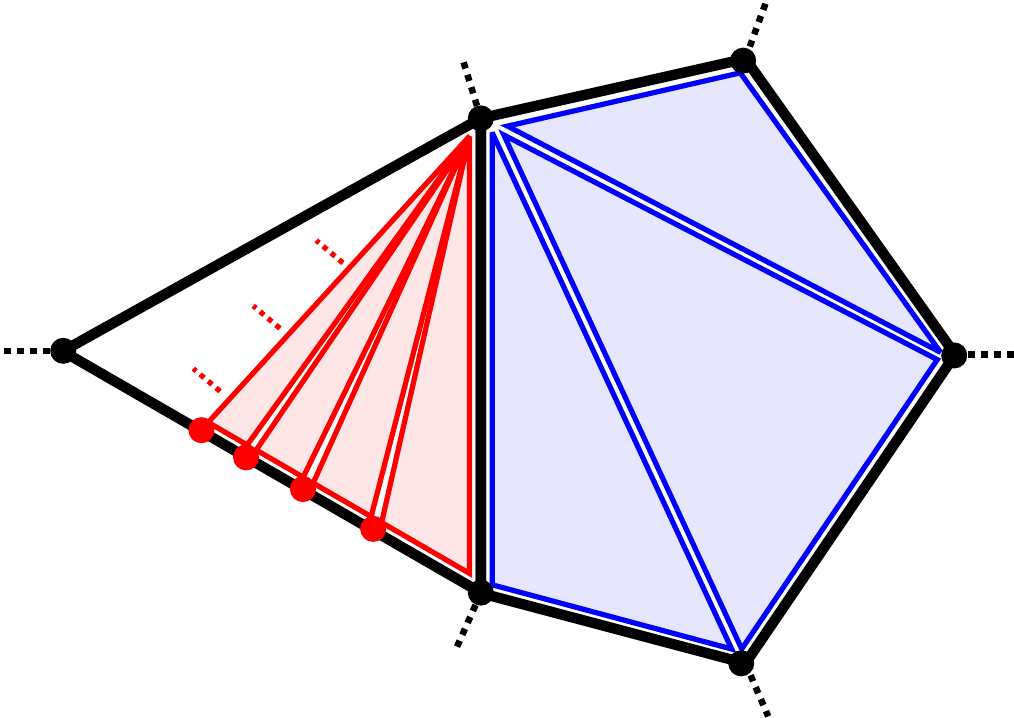}		
	\end{minipage}
	\caption{Illustrative examples of the (redundant) triangulation of a tropical fan by a (a) finite and (b) infinite cluster algebra. Each of the figures depicts two cones of a $3$-dimensional fan intersected with the unit sphere $S^2$ in black. The cones and the redundant rays from the redundant triangulation are drawn in red, those from the non-redundant triangulation in blue. Adapted from ref.~\cite{Henke:2021ity}.}
	\label{fig:infiniteRedundantTriangulation}
\end{figure}

So in the $Gr(4,6)$ example the cluster fan coincides with the tropical fan, and more generally the former \emph{triangulates} the latter\footnote{More precisely, due to different choice of conventions it is the fan of the \emph{dual} cluster algebra, obtained by transposing its exchange matrices or inverting the arrow direction in its quivers~\cite{2011arXiv1101.3736N}, that triangulates the tropical fan. Combinatorially, the cluster fan and its dual are equivalent.}~\cite{Speyer2005}. This fact has already been used to compute tree-level amplitudes of generalized biadjoint scalar theory~\cite{Cachazo:2019ngv}, defined as a natural extension of the Cachazo-He-Yuan formulation~\cite{Cachazo:2013hca,Cachazo:2013iea} for the corresponing amplitudes in ordinary biadjoint scalar theory, which are essentially given by the volume of the tropical Gra\ss mannian~\cite{Drummond:2019qjk}. There, it was also pointed out that the fan of finite cluster algebras may contain not only additional boundaries compared to the tropical fan, but also additional or \emph{redundant rays}, as shown on the left of figure \ref{fig:infiniteRedundantTriangulation}. 

A key insight behind the works~\cite{Drummond:2019cxm,Arkani-Hamed:2019rds,Henke:2019hve} is to turn this logic around in the case of infinite cluster algebras, where it is natural to expect that their infinities are due to inifitely redundant triangulations, as shown on the right of figure \ref{fig:infiniteRedundantTriangulation}. Therefore we may prevent this from happening by selecting the \emph{finite subset} of cluster variables whose rays coincide with the tropical rays!  Concretely, starting from the initial cluster of the $Gr(4,n)$ cluster algebra, where all cluster variables have non-redundant rays, after every mutation we compare the ray of the resulting cluster variable to the set of all $pTr(4,n)$ rays, which may be computed independently e.g. with the program \texttt{polymake}~\cite{polymake:2000}. We then stop mutating whenever a redundant ray is reached. It is in this sense that $pTr(4,n)$ acts as a sieve as shown in figure \ref{fig:sifting}.

\begin{figure}
\begin{center}
\includegraphics[width=\textwidth]{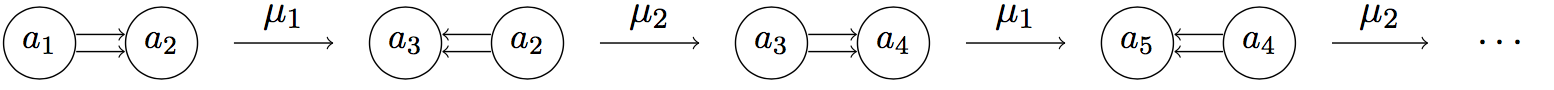}
\caption{The $A_1^{(1)}$ cluster algebra, where up to orientation each quiver is mapped back to itself after one mutation. Adapted from ref.~\cite{Henke:2019hve}.}
\label{fig:A11}
\end{center}
\end{figure}

\subsubsection{Square-root letters from infinite mutation sequences.}
At this point, we have resolved the first challenge we have discussed at the beginning of this section. How about the second challenge, associated to the appearance of square-root letters in the alphabet of the amplitude? To this end, an important observation comes from existing studies in the mathematical literature, of sequences of cluster mutations that map a quiver back to itself, and thereby lead to recursion relations between the $\cA$-coordinates. In figure \ref{fig:A11} we present the simplest example of an infinite cluster algebra where this occurs, of affine rank-2, or $A_1^{(1)}$ in the extended Dynkin diagram classification, type. Very interestingly, the limit of consecutive $\cA$-coordinates along this infinite mutation sequence becomes~\cite{Canakci2018},
\be
\lim_{i\to \infty}\frac{a_i}{a_{i-1}}=\frac{a_2}{2a_1}\left({1+x_1+x_1 x_2+\sqrt{(1+x_1+x_1 x_2)^2-4 x_1 x_2}}\right)\label{eq:A11Limit}
\ee
with $x_1=1/a_2^2$, $x_1=a_1^2$. 

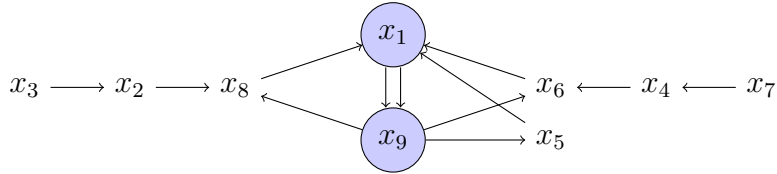
\begin{figure}
\begin{center}
	\begin{tikzpicture}[scale=1.4]
		\node at (0,0.5) (6) {$x_3$};
		\node at (1,0.5) (9) {$x_2$};
		\node at (2,0.5) (2) {$x_8$};
		\node[blueNode] at (3.5,1) (5) {$x_1$};
		\node at (5,0.5) (1) {$x_6$};
		\node at (6,0.5) (7) {$x_4$};
		\node at (7,0.5) (8) {$x_7$};
		
		\node[blueNode] at (3.5,0) (3) {$x_9$};
		\node at (5,0) (4) {$x_5$};
		
		\draw[->] (6) -- (9);
		\draw[->] (9) -- (2);
		\draw[->] (2) -- (5);
		\draw[->] (1) -- (5);
		\draw[->] (7) -- (1);
		\draw[->] (8) -- (7);
		
		\draw[->] (3) -- (2);
		\draw[->] (3) -- (1);
		\draw[->] (4) -- (5);
		\draw[->] (3) -- (4);
		
		\draw[->] ([xshift=2pt]5.south) -- ([xshift=2pt]3.north);
		\draw[->] ([xshift=-2pt]5.south) -- ([xshift=-2pt]3.north);
	\end{tikzpicture}
\end{center}
\caption{Quiver of a $Gr(4,8)$ cluster containing an $A_1^{(1)}$ subalgebra highlighted by the blue nodes. The $x_i$ labels denote the $\mathcal{X}$-coordinates of the cluster. Adapted from ref.~\cite{Henke:2019hve}.}
\label{fig:Gr48A11Cluster}
\end{figure}

The second main idea of refs.~\cite{Drummond:2019cxm,Arkani-Hamed:2019rds,Henke:2019hve} was to thus also consider infinite mutation sequences, so as to obtain generalized cluster variables of the form~\eqref{eq:A11Limit}, which should correspond to square-root symbol letters of amplitudes! Indeed, $Gr(4,n)$ cluster algebras with $n\ge 8$ do contain $A_1^{(1)}$ subalgebras\footnote{A subalgebra of a cluster algebra is obtained by freezing, i.e. not mutating certain variables in one of its clusters. For the $A_1^{(1)}$ subalgebra of figure \ref{fig:Gr48A11Cluster}, this would be all variables but $x_1, x_9$.}, as shown in figure \ref{fig:Gr48A11Cluster}. The fine print is that the analogous to eq.~\eqref{eq:A11Limit} limit value also depends on all the cluster variables held frozen in the infinite mutation sequence, hence in the language of frozen variables one would have to separately analyze these for every cluster of $Gr(4,n)$ containing an $A_1^{(1)}$ subalgebra.

Nevertheless there exists a framework for simultaneously describing any choice of frozen variables~\cite{CAIV}: It involves grouping them into \emph{coefficients},
\be
\quad y_i=\frac{\displaystyle\prod_{\textrm{arrows }\fbox{$\scriptstyle j$}\to i}a_j}{\displaystyle\prod_{\textrm{arrows }\fbox{$\scriptstyle j$}\leftarrow i}a_j}\,,
\ee
associated to the unfrozen variable they connect to, and defining mutation rules for them analogous to those of the $\cA$- or $\cX$ coordinates, independently of their constituent frozen coordinates. The simplest case of \emph{principal coefficients} amounts to $y_i=a_{d+i}$, $i=1,\ldots d$, namely one frozen variable attached to every unfrozen one.

In~\cite{Henke:2019hve}, infinite $A_1^{(1)}$ mutation sequences with principal coefficients were analyzed as a proof of concept, see also~\cite{2018arXiv180605094R}. References~\cite{Drummond:2019cxm} and \cite{Arkani-Hamed:2019rds} additionally found the generating functional of the mutation sequences for the particular case of frozen variables needed to embed them inside the $Gr(4,8)$ cluster algebra.  In~\cite{Drummond:2019cxm} it was furthermore noticed that when embedding  $A_1^{(1)}$ in a larger cluster algebra, it is possible to also take the direction of approach to the limit ray  into account, so as to associate many square-root letters to each limit ray\footnote{As with the selection of cluster variables/rational letters, also the limit ray must coincide with a $pTr(4,n)$ ray so as not to discard the associated square-root letters.}. It was also proposed how to do this in a particular fashion, that was subsequently supported by a complementary approach based on scattering diagrams~\cite{Herderschee:2021dez}. In this manner, one obtains a candidate eight-particle alphabet consisting of 272 rational and 18 square-root dual conformal invariant letters, which as a highly nontrivial check contains all those found in explicit computations of $A^{(2)}_{8,1}$~\cite{He:2019jee} and more recently $A^{(3)}_{8,0}$~\cite{Li:2021bwg}.

Finally, infinite higher-rank $A_1^{(m)}$ mutation sequences with general coefficients were worked out in~\cite{Henke:2021ity}, and this provides the missing link for predicting finite symbol alphabets in principle at any multiplicity $n$. In the latter reference these general results were also specialized to the $n=9$ case, yielding 3078 rational and 2349 square-root letters expected to appear in the amplitude. Support for the correctness of this proposal comes again from the fact that it contains the alphabet appearing in the independent determination of $A^{(2)}_{9,1}$\cite{He:2020vob}, as well as from the agreement of the entire rational part  with an alternative proposal based on tensor diagrams~\cite{Ren:2021ztg}.  The $n=8$ and $n=9$ alphabets are too length too quote here, but they may be found in the ancillary files of~\cite{Henke:2021ity}.

While resolving the longstanding issues we presented at the beginning of this subsection, the line of research we have described has also led to very interesting open questions that deserve further inquiry. For example, at $n=9$ we have for the first time the appearance of square-root letters whose radicand does not correspond to that of the one-loop box~\eqref{Delta4mBox}, and it would thus be worthwhile to identify specific Feynman integrals giving rise to them. Perhaps more importantly, a new qualitative feature starting at this multiplicity is the existence of $pTr(4,n)$ rays which are inaccessible from the $Gr(4,n)$ cluster algebra even when enlarging the latter so as to also include limits of infinite mutation sequences. It is currently unclear if the missing rays are associated to more intricate algebraic letters beyond square roots, or point towards the need for more complicated, elliptic generalizations of MPLs starting to contribute at $n=9$.  Indeed, while it is known that such functions certainly appear at $n=10$~\cite{CaronHuot:2012ab}, the possibility that these also appear at lower multiplicity cannot be excluded at the moment. 

Even in the latter case, as we have stressed the key prerequisite for the bootstrap approach is the finiteness of the expected space of functions at each loop order, which is not necessarily restricted to MPLs. For example, the symbol calculus has been developed also for elliptic generalizations of MPLs~\cite{Broedel:2018iwv}, and its application on the elliptic double box in fact reveals that it is more similar to the non-elliptic case than previously expected~\cite{Kristensson:2021ani}. The coaction and hence also the symbol contained in it has in fact been defined for even more general classes of periods and Feynman integrals~\cite{Brown:2015fyf,2015arXiv151206410B}, see additionally the recent review~\cite{Bourjaily:2022bwx} on Feynman integrals involving special functions beyond MPLs. Hence, also for quantities expressible in terms of these types of functions, what is needed to render them amenable to the bootstrap is a principle dictating a finite set of integration kernels contributing to them. In light of this, it would be very exciting to find a means to associate this kind of generalized symbol letters to the $pTr(4,n)$ missing rays, and explore their relevance for scattering amplitudes and Feynman integrals.

\subsection{Bootstrapping Feynman integrals}\label{sec:IntegralBootstrap}

\begin{figure}
\begin{center}
\begin{minipage}{0.45\textwidth}
\begin{tikzpicture}[baseline={([yshift=-.5ex]current bounding box.center)},scale=0.18]
        \draw[black,thick] (0,0)--(0,5)--(4.76,6.55)--(7.69,2.5)--(4.76,-1.55)--cycle;
        \draw[black,thick] (-15,5)--(-19.76,6.55)--(-22.69,2.5)--(-19.76,-1.55)--(-15,0);
        \draw[decorate, decoration=snake, segment length=12pt, segment amplitude=2pt, black,thick] (4.76,6.55)--(4.76,-1.55);
        \draw[decorate, decoration=snake, segment length=12pt, segment amplitude=2pt, black,thick] (-19.76,6.55)--(-19.76,-1.55);
         \draw[black,thick] (9.43,2.5)--(7.69,2.5);
        \draw[black,thick] (4.76,6.55)--(5.37,8.45);
        \draw[black,thick] (4.76,-1.55)--(5.37,-3.45);
        \draw[black,thick] (0,5)--(-5,5)--(-5,0)--(0,0);
        \draw[black,thick,densely dashed] (-5,5)--(-10,5);
        \draw[black,thick,densely dashed] (-5,0)--(-10,0);
        \draw[black,thick] (-10,0)--(-10,5)--(-15,5)--(-15,0)--cycle;
        \draw[black,thick] (-19.76,6.55)--(-20.37,8.45);
        \draw[black,thick] (-19.76,-1.55)--(-20.37,-3.45);
        \draw[black,thick] (-24.69,2.5)--(-22.69,2.5);
\filldraw[black] (2.3,2.5) node {{$L$}};
\filldraw[black] (-17.7,2.5) node {{$1$}};
\filldraw[black] (-12.5,2.5) node {{$2$}};
\filldraw[black] (-7.5,2.5) node {{$\cdots$}};
\end{tikzpicture}
\end{minipage}
\begin{minipage}{0.15\textwidth}
    \begin{tikzpicture}[baseline={([yshift=-.5ex]current bounding box.center)},scale=0.15]
        \draw[black,thick] (0,0)--(5,0)--(6.55,4.76)--(2.50,7.69)--(-1.55,4.76)--cycle;
        \draw[black,thick] (1.5,9.43)--(2.5,7.69)--(3.5,9.43);
        \draw[black,thick] (-0.21,-1.99)--(0,0)--(-1.83,-0.81);
        \draw[black,thick] (6.83,-0.81)--(5,0)--(5.21,-1.99);
        \draw[black,thick] (2.5,0)--(2.5,3.5)--(6.55,4.76);
        \draw[black,thick] (2.5,3.5)--(-1.55,4.76);
         \draw[black,thick] (6.55,4.76)--(8.45,6.37);
        \draw[black,thick] (-3.55,6.37)--(-1.55,4.76);
    \end{tikzpicture}
\end{minipage}
\begin{minipage}{0.6\textwidth}
\includegraphics[width=\textwidth]{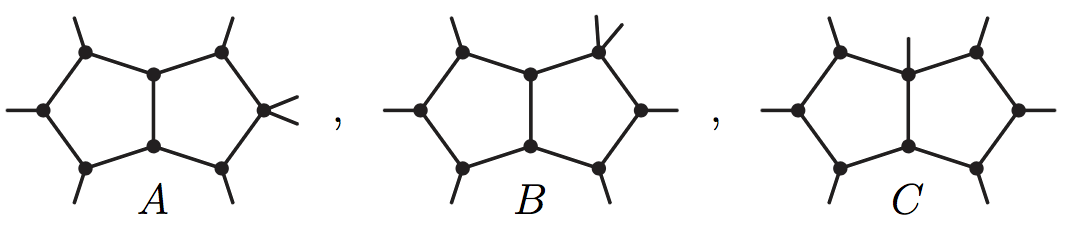}
\end{minipage}

\end{center}
\caption{Examples of planar, four-dimensional finite dual conformal integrals that have been bootstrapped. Top left: Six-point double pentagon ladder $\Omega^{(L)}$~\cite{Dixon:2011nj,Caron-Huot:2018dsv} (see also~\cite{He:2021non} for its generalization to massive left- and right-most legs through $L=4$). Bottom: Seven-point double pentagons \cite{Drummond:2017ssj}(B,C, the latter also at $L=3$), \cite{Henn:2018cdp} (A,C). Top right: Eight-point $L=3$ wheel~\cite{He:2021eec}. Adapted from refs.~\cite{Bourjaily:2018aeq,He:2021non,He:2021eec}}.
\label{fig:omega_L}
\end{figure}
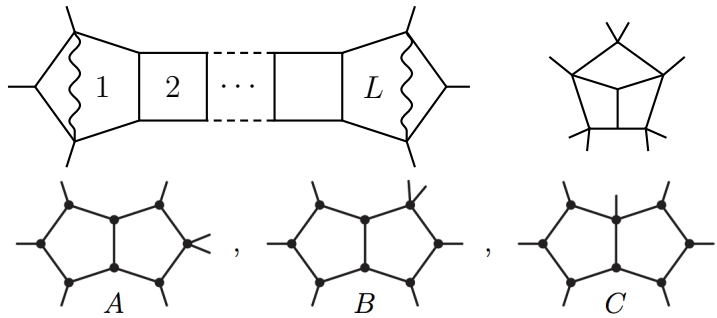

While the analytic bootstrap approach to perturbative quantum field theory has been initiated and more extensively developed in the context of $\cN=4$ SYM amplitudes, that we have presented so far, the same methodology is applicable in many other situations as well. This in particular includes individual Feynman integrals that  belong to the class of multiple polylogarithms defined in section~\ref{sec:MPLs}, especially if the are \emph{pure}, meaning that they have uniform weight, and their rational coefficients do not depend on the kinematics. Identifying such integrals is possible at the level of the integrand by examining their leading singularities~\cite{Cachazo:2008vp}, and if necessary modifying the integrals by taking out any overall factors, or by choosing the numerators appropriately, such that these leading singularities are constant~\cite{ArkaniHamed:2010gh}.

\subsubsection{Survey of explicit results.}
The constant leading singularity criterion was originally understood in the realm of $\cN=4$ SYM as well, and so the first pure weight integrals that were thus identified, and later on bootstrapped, were planar, finite and dual conformal in strictly four dimensions. These most notably contain the six-point double pentagon integral $\Omega^{(2)}$, that has already appeared in the construction of our hexagon function space in subsection~\ref{sec:R62}, as well as its generalizations to higher loops or legs, some of which are depicted in figure~\ref{fig:omega_L}.\footnote{Many of these integrals may be also computed by direct methods, see for example~\cite{McLeod:2020dxg} for $\Omega^{(L)}$ through $L=10$, and \cite{Bourjaily:2018aeq} for the type A,B seven-point double pentagon ladders through $L=4$, as well as their eight-point generalizations through $L=3$ and $L=2$, respectively.} The wavy lines denote numerators carefully chosen so as to render the integrals pure, see also chapter 7~\cite{Herrmann:2022nkh} of the SAGEX review~\cite{Travaglini:2022uwo} for a concrete one-loop example. The fact that at consecutive loop orders these integrals are related by differential equations~\cite{Drummond:2010cz} or equivalently recursive integral representations~\cite{He:2020uxy}, allows one to easily locate them inside the expected function space. Combined with integrability, bootstrap methods have also led to a closed form expression for a doubly infinite class of four-point fishnet integrals~\cite{Basso:2017jwq}.

\begin{figure}
\begin{center}
\begin{minipage}{0.3\textwidth}
\includegraphics[width=0.9\textwidth]{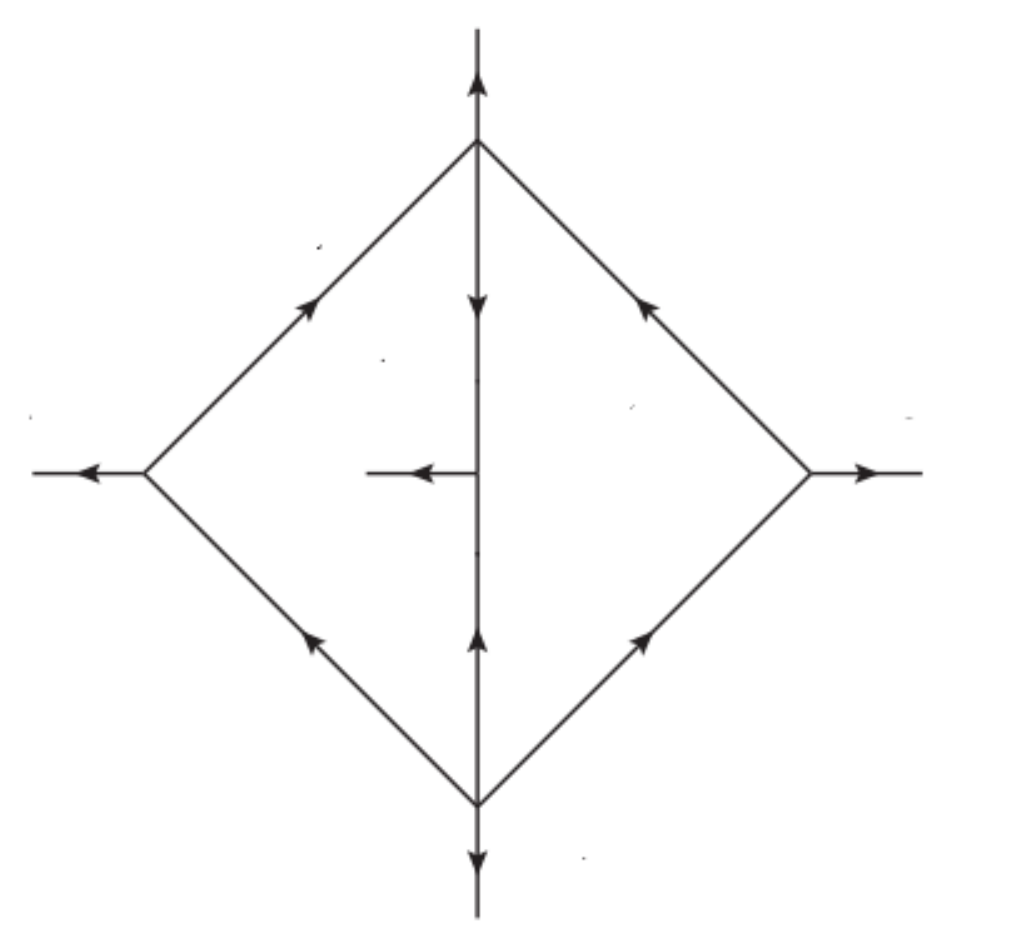}
\end{minipage}
\begin{minipage}{0.3\textwidth}
\includegraphics[width=\textwidth]{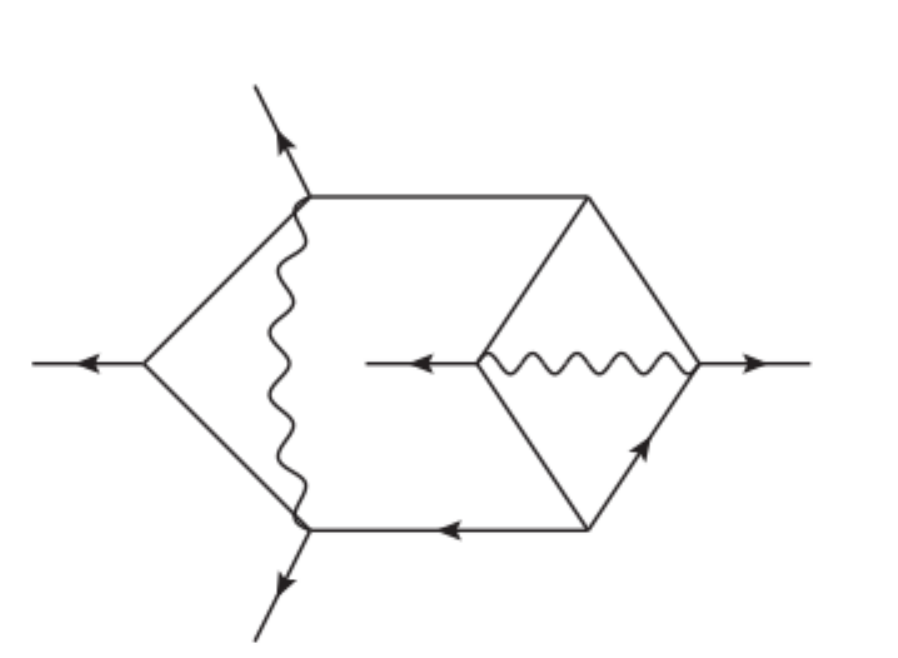}
\end{minipage}
\end{center}
\caption{Nonplanar two-loop five-point massless integrals that have been bootstrapped up to $\mathcal{O}(\epsilon^0)$ in dimensional regularization. Adapted from ref.~\cite{Chicherin:2017dob}.}.
\label{fig:NonPlanarPentIntegrals}
\end{figure}

As with many computational tools first developed in the laboratory of the simplest gauge theory, the leading singularity analysis for finding pure weight integrals is generally applicable, and for integrands only having simple poles in the integration varibles~\cite{Arkani-Hamed:2014via}, it has in fact been automated in the \texttt{DlogBasis} package~\cite{Henn:2020lye}. With the help of this analysis, the Feynman integral bootstrap has also been applied to the nonplanar cases relevant for massless five-point scattering shown in figure~\ref{fig:NonPlanarPentIntegrals}~\cite{Chicherin:2017dob}, under the additional assumption that the corresponding symbol alphabet may be obtained from permutations of the known planar two-loop alphabet~\cite{Gehrmann:2015bfy}. In order to fix an ansatz for the integral, additional information may be generically obtained by taking limits where it reduces to other simpler integrals that are already known or can be computed simply, either exactly or as series expansions, or by taking discontinuities that decrease the weight and hence also the complexity.

Finally, it is worth mentioning another type of bootstrap that is closely related to the differential equations obeyed by a basis of master integrals contributing to a particular process, and represented by the vector ${\bf f}$, with respect to the independent kinematic variables $\vec{x}$. If the integrals in question are expressible in terms of MPLs, then finding the transformation that brings the differential equations to canonical form~\cite{Henn:2013pwa}\footnote{The notion of master integrals and the (canonical) differential equations they obey are discussed in more detail in chapter 3~\cite{Abreu:2022mfk} of the SAGEX review~\cite{Travaglini:2022uwo}. Different strategies for transforming differential equations to canonical form have been implemented in publicly available software such as \texttt{epsilon}~\cite{Prausa:2017ltv}, \texttt{Fuchsia}~\cite{Gituliar:2017vzm}, \texttt{Canonica}~\cite{Meyer:2017joq}, \texttt{Initial}~\cite{Dlapa:2020cwj} and \texttt{Libra}~\cite{Lee:2020zfb}.},
\begin{align}\label{canonicalDE}
d\, {\bf f}(\vec{x};\eps)  = \eps  \left[ \sum\nolimits_i  {\bf A}_{i} d \log \phi_i(\vec{x}) \right] {\bf f}(\vec{x};\eps) \,,
\end{align}
where $d= \sum_j dx_j  \partial_{x_j}$, $\phi_i$ are the letters and ${\bf A}_{i}$ are constant matrices, is a very powerful method for algorithnmically evaluating them in closed form to arbitrary order in the dimensional regulator $\epsilon=(4-D)/2$. That is, when supplemented with boundary conditions, and  provided that the letters are rational functions of the kinematic variables. If algebraic letters such as the square roots we saw in the previous section appear, under certain conditions variable transformations that rationalize them may also be found algorithmically~\cite{Besier:2018jen}, and have been implemented in the package~\texttt{RationalizeRoots}~\cite{Besier:2019kco}. Given that this however doesn't always work, an alternative presented in~\cite{Heller:2019gkq,Heller:2021gun}, as also reviewed in the textbook~\cite{Weinzierl:2022eaz}, is to match ${\bf f}$ and its derivatives to an ansatz built out of MPLs with the known alphabet, constructed by searching for admissible arguments of these MPLs. In particular, this search is guided by the requirement that the letters of the candidate MPLs should factorize over the known alphabet, and builds on analogous methods previously developed for rational alphabets~\cite{Duhr:2011zq}.

\subsubsection{The role of cluster algebras.}
\begin{figure}
\begin{minipage}{0.55\textwidth}
  \includegraphics[width=0.9\textwidth]{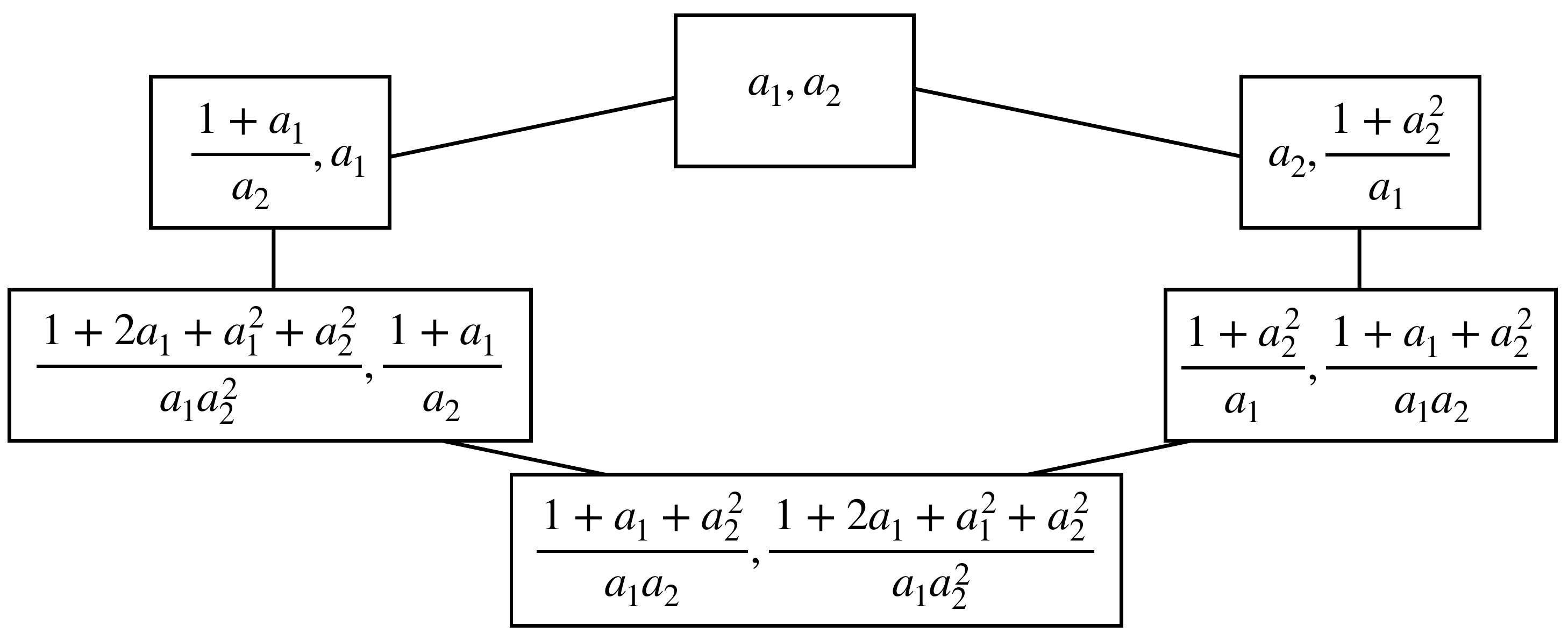}
\end{minipage}
\begin{minipage}{0.4\textwidth}
   \includegraphics[width=\textwidth]{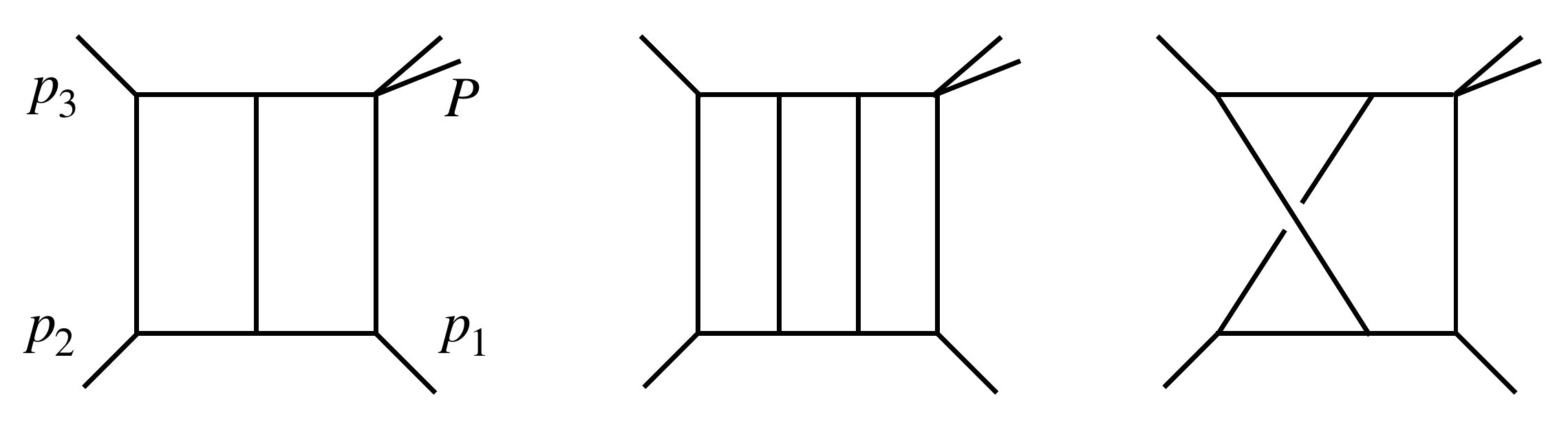}
   \end{minipage}
\caption{Left: The exchange graph of the $C_2$ cluster algebra, with cluster coordinates ordered as $a_i,a_{i+1}$. Right: Examples of four-point one-mass integrals whose alphabet is given by the $C_2$ cluster coordinates. Adapted from ref.~\cite{Chicherin:2020umh}.}
\label{fig:2Loop1MassBoxes}
\end{figure}

It is important to note that a prior knowledge (or educated guess) of the symbol alphabet is either strictly necessary for the amplitude or integral bootstrap, or tremendously helpful for bringing the differential equations of master integrals to canonical form, as the remaining dependence of~\eqref{canonicalDE} on the purely numeric matrices ${\bf A}_{i}$ can be determined much more easily. We have seen in subsections~\ref{sec:SteinClus} and \ref{sec:TropSing} that cluster algebras and their generalizations may give strong clues about the right alphabet, however until recently their appearance was confined to the realm of $\cN=4$ SYM theory. Excitingly, this changed with a recent publication~\cite{Chicherin:2020umh}, where it was discovered that cluster algebras underlie the analytic structure of a host of Feynman integrals in dimensional regularization. This most notably includes all known four-point integrals with one off-shell (or equivalently massive) leg, two-loop planar and nonplanar~\cite{Gehrmann:2000zt,Gehrmann:2001ck}, $L$-loop ladders~\cite{DiVita:2014pza,Panzer:2015ida}, and more recently the three-loop tennis court~\cite{Canko:2021hvh,Canko:2021xmn}, see also figure \ref{fig:2Loop1MassBoxes} for some examples. Specifically, the alphabet of this entire class of integrals is described by a $C_2$ cluster algebra.

In the remainder of this subsection, let us discuss how the latter connection was established, and what it implies. The $C_2$ cluster algebra is the first non-simply laced case we encounter, so given that it has no standard quiver representation, we will define it directly in terms of the exchange matrix $B$ of its initial cluster. The  elements $b_{ij}$ of (the principal part of) $B$ are essentially given by those of the associated Cartan matrix $C$ by~\cite{1054.17024}
\be
b_{ij}=(-1)^{\text{sign}(i-j)}\left(c_{ij}-\delta_{ij}c_{ii}\right)\,,
\ee
therefore for the $C_2$ case without frozen variables we have,
\be
C=\left(
\begin{array}{cc}
2&-1\\
-2&2
\end{array}
\right)
\quad \Rightarrow \quad B=\left(
\begin{array}{cc}
0&1\\
-2&0
\end{array}\right)\,.
\ee
Mutating the initial $\cA$-coordinates $\{a_1,a_2\}$ according to the rules~\eqref{equ:clusterMutation}-\eqref{eq:Bmutation}, then yields in total six $\cA$-coordinates, arranged in six clusters as depicted in the corresponding exchange graph on the left of figure \ref{fig:2Loop1MassBoxes}.

One the other hand, the aforementioned integrals are expressible in terms of the well-studied class of two-dimensional harmonic polylogarithms (2dHPLs)~\cite{Gehrmann:2000zt,Duhr:2012fh}, which have the alphabet, in the dimensionless variables $z_1 =2( p_1 \cdot p_2) / P^2$ and $ z_2=  2( p_2 \cdot p_3)/ P^2$  where the momentum labels as shown in the above figure, 
\begin{align}\label{eq:2dHPL}
\Phi_{\text{2dHPL}}=\{z_1,z_2, z_3, 1-z_1,1-z_2,1-z_3\}\,,
\end{align}
with $z_1 + z_2 + z_3 =1$. Remarkably, the 2dHPL alphabet (\ref{eq:2dHPL}) is equivalent to the $C_{2}$ alphabet given in the left of figure \ref{fig:2Loop1MassBoxes}, as can be readily verified by applying the variable transformation 
\begin{equation}
z_1 = - \frac{a_2^2}{1+a_1} \,, \quad z_2= - \frac{1+a_1+a_2^2}{a_1 (1+a_1)}  \,. \label{z1z2a1a2}
\end{equation}
Transformations of this type between equivalent alphabets may be searched for systematically in a fashion analogous to the search for admissible MPL arguments, discussed  a few paragraphs above. For example, the fact that $z_1, z_2$ are both variables and symbol letters implies that their logarithm should be a linear combination of logarithms of the $\cA$-coordinate alphabet of figure \ref{fig:2Loop1MassBoxes}, and that $1-z_i$ should factorize over this alphabet.

\setcounter{footnote}{0}

It is very interesting to note that this cluster algebra connection also extends to a variety of processes in Quantum Chromodynamics, where the aforementioned integrals contribute as master integrals at two loops. These include for example three-jet production in electron-positron annihilation~\cite{Garland:2002ak}, as well as vector boson plus jet~\cite{Gehrmann:2013vga}, or perhaps more importantly,  Higgs boson plus jet production~\cite{Gehrmann:2011aa} in proton-proton collisions (in the massless and the heavy-top limit, respectively). As is also reviewed in chapter 1~\cite{Brandhuber:2022qbk} of the SAGEX review~\cite{Travaglini:2022uwo}, the latter amplitude is in fact a three-particle \emph{form factor}. These quantities are defined as vacuum expectation values of local operators between the vacuum and an $n$-particle external state, so they are between the fully on-shell amplitudes and the fully off-shell correlators.\footnote{Protected correlators, their connection to scattering amplitudes in $\cN=4$ SYM, and related bootstrap approaches for computing them are also discussed in chapter 8\cite{Heslop:2022xgp} of the SAGEX review~\cite{Travaglini:2022uwo}.}. In the case at hand, the operator  $H\Tr F^2$, where $H$ is the Higgs boson and $F$ the gauge field strength, is the leading effective vertex in the effective field theory that arises when integrating out the top mass. The analogous quantity in $\cN=4$ SYM, which will be the focus of the next subsection, replaces this operator with any component of the stress-tensor multiplet of the theory, can also be expressed in terms of the same set of master integrals, and was first computed in~\cite{Brandhuber:2012vm}.

Finally, we turn to the implications of the $C_2$ cluster algebra for the 2dHPL master integrals and the physical quantities they compute. We have seen in subsection~\ref{sec:SteinClus} that in $\cN=4$ SYM theory cluster algebras additionally restrict which symbol letters can appear next to each other, could we hope for something similar here as well? Surprisingly, it turns out that the following subset of cluster adjacency restrictions hold,
\be\label{eq:C2adjacency}
\hspace{48pt}\cancel{\ldots \otimes a_{2l+1}\otimes a_{2m+1} \otimes \ldots} \,\,\quad\Leftrightarrow\quad
\cancel{\ldots \otimes 1-z_l \otimes 1-z_m \otimes \ldots}\,,\quad l \ne m \,,
\ee
to all orders in $\epsilon$! The same restrictions were also independently observed when bootstrapping the $\cN=4$ SYM three-particle form factor~\cite{Dixon:2020bbt}, and as in the case of amplitudes they considerably reduce the size of the corresponding function space. 

Can we understand why only this subset and not all of the adjacencies occur? To this end, it is very suggestive that $C_2$ is the parity-invariant surface of the $A_3$ cluster algebra, relevant for six-particle scattering: This readily follows by inspecting their cluster polytopes, shown in figures \ref{fig:2Loop1MassBoxes} and \ref{A3polytope}.\footnote{This is an instance of a more general folding procedure, which carries over from Dynkin diagrams to cluster algebras, and allows one to embed $B_n,C_n,F_4$ and $G_2$ inside $A_{2n-1},D_{n+1},E_6$ and $D_4$, respectively~\cite{Fomin:2001rc}.}. In more detail, with the help of $\cX$-coordinates it can be shown that the six nonvanishing $A_3$ letters of eq.~\eqref{eq:hexletters} (to avoid clash of notation, in the latter equation we switch $a_1\to a, a_2\to b, a_3\to c$), are related to the $C_2$ cluster variables as 
\be
\begin{aligned}\label{eq:B2toA3Letters}
a_1=\sqrt a\,,\quad a_3=\sqrt c\,,\quad a_5=\sqrt b\,,\quad  a_{2i}=\sqrt {m_{3-i}}\,,\quad i=1,2,3\,.\\
\end{aligned}
\ee
With this identification, we observe that the adjacency restrictions~\eqref{eq:C2adjacency} precisely become the extended Steinmann relations for six-particle massless scattering~\eqref{eq:FabExtStein}! This observation may point towards the right formulation of the (extended) Steinmann relations at lower multiplicity $n<6$, which is currently not well understood.\footnote{We emphasize however that the adjacency conditions~\eqref{eq:C2adjacency} do not correspond to discontinuities with respect to two-particle Mandelstam invariants.} Further aspects of cluster-algebraic structures and their generalizations to Feynman integrals have been discussed in~\cite{He:2021esx,He:2021fwf,He:2021non,He:2021mme,He:2021eec}, and providing a first-principle derivation of their presence would be a very interesting goal for the future.

\subsection{Bootstrapping a three-particle form factor}
In the previous subsection, we saw that two-loop three-particle form factors are expressible in terms of the 2dHPL alphabet~\eqref{eq:2dHPL}, whose equivalent description in terms of a $C_2$ cluster algebra makes plausible that the same holds true also at higher loops. In $\cN=4$ SYM theory, while the computation of the two-loop form factor of the stress tensor multiplet was carried out by unitarity methods, at the same time it was also shown that its symbol may also be uniquely determined by a bootstrap approach very much alike the one we have described for scattering amplitudes~\cite{Brandhuber:2012vm}. However what prevented the application of this approach at higher loops was the absence of enough independent information on the behavior of the form factor in kinematics limits, in order to uniquely identify it inside the expected space of functions. This limitation has been recently overcome, with the extension of the integrability-based Pentagon OPE for predicting the near-collinear limit expansion of amplitudes and Wilson loops, mentioned in subsection~\ref{sec:Limits}, also to form factors~\cite{Sever:2020jjx,Sever:2021nsq,Sever:2021xga}. As a result, the form factor in question has been computed through five loops in~\cite{Dixon:2020bbt}, and results through eight loops also appeared very recently~\cite{Dixon:2022rse}.

While a detailed exposition of the three-particle form factor bootstrap would be out of the scope of this review, let us present some of the main features, to also illustrate the close similarity to the amplitudes case. The quantity of interest is the infrared-finite part of the form factor, which was originally obtained by factoring out the exponentiated one-loop form factor~\cite{Brandhuber:2012vm}, in other words the corresponding BDS ansatz,
\be
 \mathcal{F}_{3} = \mathcal{F}_3^{\text{BDS}} \exp[ R ] \,.
\ee
In this normalization, by convention the finite part was chosen to be represented by its logarithm, or remainder function $R$. However it turns out to be advantageous to pull out the finite kinematic-dependent part of the one-loop form factor, so as to define the BDS-like normalized form-factor $E$~\cite{Dixon:2020bbt}\footnote{Further refinement of the normalization so as to reduce the number of independent constants is possible, but we will not describe this here.},
\be
E = e^{\frac{1}{4} \Gcusp E^{(1)} + R }\,,
\ee
where its one-loop contribution is given by
\be
E^{(1)} = -\sum_{i=1}^3\left[2  \Li_2(1-z_i) + \ln z_{i} \ln z_{i+1} \right]+ 4 \zeta_2\,, 
\ee
in terms of the kinematic variables we have defined above eq.~\eqref{eq:2dHPL}. Then, at weight 1 the space of functions $F_{1}$ containing $E$ and its derivatives is dictated by the first entry condition,
\be
F_1=\{\log z_i\}\,,\quad i=1,2,3\,,
\ee
and at higher weight it is constructed from the alphabet~\eqref{eq:2dHPL}, ensuring that it obeys the integrability condition~\eqref{eq:commdoublederiv} or \eqref{DoubleCopMatrix}, the adjacency condition~\eqref{eq:C2adjacency}, which may be equivalently formulated as
\be
F^{1-z_l,1-z_m}=0\,,\quad l\ne m\,,
\ee
as well a branch cut condition analogous to eq.~\eqref{eq:soft1},
\be
F^{1-z_i}\Big|_{\stackrel{z_i\to 1}{z_{i+1}\to0}} = 0,,\quad i=1,2,3\,,
\ee
An important difference between amplitudes and form factors, however, is that in the latter case the branch cut condition also removes symbol-level functions.

Once the function space at the desired weight has been constructed, then an ansatz for $E$ is uniquely fixed by imposing its full $z_i\to z_j$ permutation symmetry, the restriction on the
\be\label{eq:firstentriesFF}
\text{final symbol entry of}\,\,\, E\in \frac{1-z_i}{z_i}\,,
\ee
the fact that in the strict collinear limit in one orientation, e.g.
\be
\lim_{z_2\to 0}R=0\,,
\ee
and finally constraints coming from the aforementioned near-collinear OPE.

A few final remarks are in order: First, in~\cite{Brandhuber:2012vm} the surprising observation was made, that when normalized in the same fashion, the maximal transcendental part of the leading-color term of the two-loop Higgs amplitude~\cite{Gehrmann:2011aa} discussed in the previous subsection, coincides with $R^{(2)}$ at symbol level (but not quite at function level~\cite{Duhr:2012fh}). It would be very interesting to find out if this agreement persists also at higher loops. Second, in the previous subsection we saw that the alphabet of the six-particle amplitude reduces to the alphabet of the form factor in the parity invariant surface. Quite remarkably, this kinematic relation also extends to the level of dynamics: From the existing data on both sides, it was observed that the MHV six-particle amplitude on the parity-even surface coincides with the form factor, up to certain variable substitution and the reversal of the order of letters in its symbol~\cite{Dixon:2021tdw}! More precisely, the latter symbol-level order reversal also extends to functions up to factors of $\pi$, and is captured by the antipode operation on the Hopf algebra structure of MPLs, as is reviewed e.g. in~\cite{Duhr:2012fh}. It would be very interesting to understand the physical origin of this correspondence, and explore whether it persists for amplitudes and form factors with more legs or different MHV degree.

\section*{Acknowledgments}

This work  was supported  by the European Union's Horizon 2020 research and innovation programme under the Marie Sk\l{}odowska-Curie grant agreement No.~764850 {\it ``\href{https://sagex.org}{SAGEX}''}. We also acknowledge support from the Deutsche Forschungsgemeinschaft under
Germany's Excellence Strategy EXC 2121 Quantum Universe 390833306.
  \\ 
 
\bibliography{refs}
\end{document}